\documentclass[mathleft
]{an}
\usepackage{graphicx}
\usepackage{rotating}
\usepackage{amssymb} 
\usepackage{lscape}
\usepackage{longtable}
\usepackage{times}
\usepackage{fge}
\begin{document}

% The following seven commands are intended for editorial usage and should be ignored by
% the author(s).
\Pagespan{789}{}% Document's page range.
% If second parameter is left empty, the last page is computed automatically.
\Yearpublication{2011}%
\Yearsubmission{2010}%
\Month{11}%
\Volume{999}%
\Issue{88}%
% \DOI{This.is/not.aDOI}%

\sloppy

\title{Sunspot numbers based on historic records in the 1610s -- \\
early telescopic observations by Simon Marius and others}

\author{R. Neuh\"auser\inst{1} \thanks{Corresponding author: \email{rne@astro.uni-jena.de}}
\and D.L. Neuh\"auser\inst{2}
}

\titlerunning{Sunspot observations by Marius}
\authorrunning{Neuh\"auser \& Neuh\"auser}

\institute{
Astrophysikalisches Institut und Universit\"ats-Sternwarte, FSU Jena,
Schillerg\"a\ss chen 2-3, 07745 Jena, Germany (e-mail: rne@astro.uni-jena.de)
\and
Schillbachstra\ss e 42, 07743 Jena, Germany
}

\received{2015 Sep 24}
\accepted{2016 Mar 9}


\abstract{
Hoyt \& Schatten (1998) claim that 
Simon Marius would have observed the sun from 1617 Jun 7 to 1618 Dec 31 
(Gregorian calendar) all days, except three short gaps in 1618, 
but would never have detected a sunspot --
based on a quotation from Marius in Wolf (1857), but mis-interpreted by Hoyt \& Schatten.
Marius himself specified in early 1619 that
{\em for one and a half year ... rather few or more often no spots
could be detected ... which was never observed before} (Marius 1619).
The generic statement by Marius can be interpreted such that the active day fraction
was below 0.5 (but not zero) from fall 1617 to spring 1619 and that it was 1 before fall 1617
(since August 1611).
Hoyt \& Schatten cite Zinner (1952), who referred to Zinner (1942), 
where observing dates by Marius since 1611 are given, but which were not used by Hoyt \& Schatten.
We present all relevant texts from Marius where he clearly stated
that he observed many spots 
in different form on and since
1611 Aug 3 (Julian) = Aug 13 (Greg.)
(on the first day together with Ahasverus Schmidnerus);
14 spots on 1612 May 30 (Julian) = Jun 9 (Greg.),
which is consistent 
with drawings by Galilei
and Jungius for that day, the latter is shown here for the first time; 
at least one spot on 
1611 Oct 3 and/or 11 (Julian), i.e. Oct 13 and/or 21 (Greg.), 
when he changed his sunspot observing technique; 
he also mentioned that he has drawn sunspots for 
1611 Nov 17 (Julian) = Nov 27 (Greg.);
in addition to those clearly datable detections,
there is evidence in the texts for regular observations.
For all the information that can be compared to other observers, 
the data from Marius could be confirmed, so that his texts are highly credible.
We also correct several 
shortcomings or apparent errors
in the database by Hoyt \& Schatten (1998)
regarding 1612 (Harriot), 1615 (Saxonius, Tard\'e), 1616 (Tard\'e), 1617-1619 (Marius, Riccioli/Argoli),
and Malapert (for 1618, 1620, and 1621).
Furthermore, Schmidnerus, Cysat, David \& Johann Fabricius, Tanner, Perovius, Argoli, and Wely
are not mentioned as observers for 1611, 1612, 1618, 1620, and 1621 in Hoyt \& Schatten.
Marius and Schmidnerus are among the earliest datable telescopic sunspot observers 
(1611 Aug 3, Julian), namely
after Harriot, the two Fabricius (father and son), Scheiner, and Cysat.
Sunspots records by Malapert from 1618 to 1621 show that the last low-latitude spot
was seen in Dec 1620, while the first high-latitude spots were noticed in June and Oct 1620,
so that the Schwabe cycle turnover (minimum) took place around that time, 
which is also consistent with the sunspot trend mentioned by Marius 
and with naked-eye spots and likely true aurorae.
We consider discrepancies in the Hoyt \& Schatten (1998) systematics, 
we compile the active day fractions for the 1610s, and we critically 
discuss very recent publications on Marius, which include the following Maunder Minimum.
Our work should be seen as a call to go back to the historical sources.
}

\maketitle

\section{Introduction}

The reconstruction of past solar activity is essential to understand the internal physics
of the Sun and (sun-like) stars as well as to possibly predict future solar activity
and space weather. 
Sunspots are used as proxy for solar activity
and have been observed 
for
millennia by the unaided eye and since 1610 also with the telescope.

R. Wolf studied solar activity and introduced the sunspot number:
The daily Wolf or Z\"urich sunspot number R$_{\rm Z}$ for an individual observer is defined as follows: 
\begin{equation}
R_{Z} = k \cdot (10 \cdot g + n)
\end{equation}
with the total number of individual sunspots $n$, the number of sunspot groups $g$,
and the individual correction factor $k$ of the respective observer. 
The {\em international sunspot number} is available for the time since 
1700 at, e.g., www.sidc.be/silso/datafiles (see Clette et al. 2014 and also Hathaway 2010)
and shows the Schwabe cycle with maxima and minima 
every $\sim 10$ yr according to sunspot observations (Schwabe 1843),
or $131 \pm 14$ months (Hathaway \& Wilson 2004, Hathaway 2010).

Hoyt \& Schatten (1998, henceforth HS98) have then defined the daily 
{\em group sunspot number} R$_{\rm G}$ as follows:
\begin{equation}
R_{G} = \frac{12.08}{N} \cdot \sum_{k_{i}^{\prime}=1}^{N} (k_{i}^{\prime} \cdot G_{i})
\end{equation}
with the correction factor $k_{i}^{\prime}$ and group sunspot number $G_{i}$
of the {\em i}-th observer
(single spots which are not part of another group, are considered a group here),
and N being the number of observers used for the
daily mean (only observers with $k_{i}^{\prime}$ between 0.6 and 1.4 were used
for the time since 1848). From the daily means, HS98 also derived
monthly and yearly values. In some months and years, though, in particular
in the 17th century, there are only
very few days with observations or even no observations at all.

Instead of the {\em group sunspot number}, one can also just list the
{\em group number}, which would be simply the number of groups seen by a certain observer.
Such a number would not include the factor $12.08$ from Equ. (2).

In HS98, dates of telescopic sunspot observations are listed 
together with the name of the observer, the place, and
the number of sunspot groups observed by that observer on each day.\footnote{See
ftp://ftp.ngdc.noaa.gov/STP/space-weather/solar-data/solar-indices/sunspot-numbers/group/daily-input-data/}
If an observer detected only one spot, 
HS98 consider them as one sunspot {\em group},
but groups of course can also include one or more large spots or numerous (small) spots.

A possible estimate of solar activity is also the
fraction of active days (F$_{\rm a}$): the number of days (N$_{\rm a}$) 
with at least one sunspot divided by the number (N) of observing days in a given period
(e.g. Maunder 1922, Kovaltsov et al. 2004), see Sect. 6.

Here, we discuss the sunspot observations from Simon Marius from 
Ansbach, Germany (not far from Nuremberg), who has observed sunspots from 1611 until at least 1619.
We present a few clearly dated sunspot detections, which were not 
considered in HS98 -- with some impact to the
daily, monthly, and yearly group sunspot numbers compared to HS98.
The texts from Marius also deliver qualitative and quantitative input,
which can hardly be considered with Equ. (2).
The period 1611 to 1619 is of particular importance and relevance,
because it is shortly after the invention of the telescope and 
shortly before the start of the Maunder Grand Minimum.
The duration and depth of the Maunder Grand 
Minimum (first noted by Sp\"orer 1887, then amplified by Maunder 1890 and Eddy 1976)
has received much attention since then (e.g., Ribes \& Nesme-Ribes 1993, 
Usoskin et al. 2007,
Vaquero et al. 2011, Vaquero 2012, Vaquero \& Trigo 2014, 2015, 
Clette et al. 2014, Zolotova \& Ponyavin 2015, and Usoskin et al. 2015);
it is usually dated from 1645 to 1715.
Vaquero \& Trigo (2015) argue that what they call the {\em extended Maunder Minimum}
would have started in 1618 
during or around a Schwabe cycle minimum around that time.

In Sect. 2, we introduce Simon Marius and the other observers studied here.
We will first discuss the time from 1617 to 1619, because this includes the
years for which HS98 list Marius (1617 and 1618) -- extended to 1619
following Zinner (1952), who noticed that Marius observed until at least 1619;
we will discuss Marius and compare him to other observers in those years,
see Sect. 3.
Then, in Sect. 4, we present the other relevant reports from Marius (and some others in relation to Marius)
from which we deduce dates for his observations (1611--1615).
We discuss some shortcomings of the group sunspot number system in Sect. 5.
And we calculate the active day fractions in Sect. 6.
The dating of the Schwabe cycle minimum around 1620 is discussed in Sect. 7
together with a critical reflection of a recent work by Zolotova \& Ponyavin (2015)
regarding the reports from Marius and the timing of this minimum -- 
we present drawings by Malapert used to date the Schwabe cycle minimum.  
We finish with a summary in Sect. 8.

Some parts of this paper were presented before in a proceedings paper for
a workshop on Marius in German language 
(Neuh\"auser \& Neuh\"auser 2016).

\section{The observers and their time}

Here, we introduce Simon Marius, the other observers discussed, as well as their time
and calendar issues.

\subsection{Simon Marius}

Simon Mayr (Lat. Marius) was born 1573 Jan 10 (Julian calendar) in Gunzenhausen, 
in Bavaria in Germany.
In 
1586 and 1589-1601, Marius studied -- with interruptions -- at the academy in Heilsbronn,
Germany, where he recorded the local weather. In May 1601, with a scholarship, he went
to Prague, Czech Republic, 
to meet and study with Tycho Brahe and Johannes Kepler. While Marius did not meet with Brahe
personally due to Brahe's illness, he did meet David Fabricius and observed
with Brahe's instruments. He left Prague a few months later in 1601 and started to study medicine 
in Padua, Italy, in December, where he may have met with Galilei (Prickard 1917). 
He went back home in 1605 and started to work as {\em Court Astronomer} for Joachim Ernst, 
Margrave of Brandenburg-Ansbach, in nearby Ansbach 
beginning in
1606.
We show an image of Marius in Fig. 1.

\begin{figure}
\begin{center}
{\includegraphics[angle=0,width=8cm]{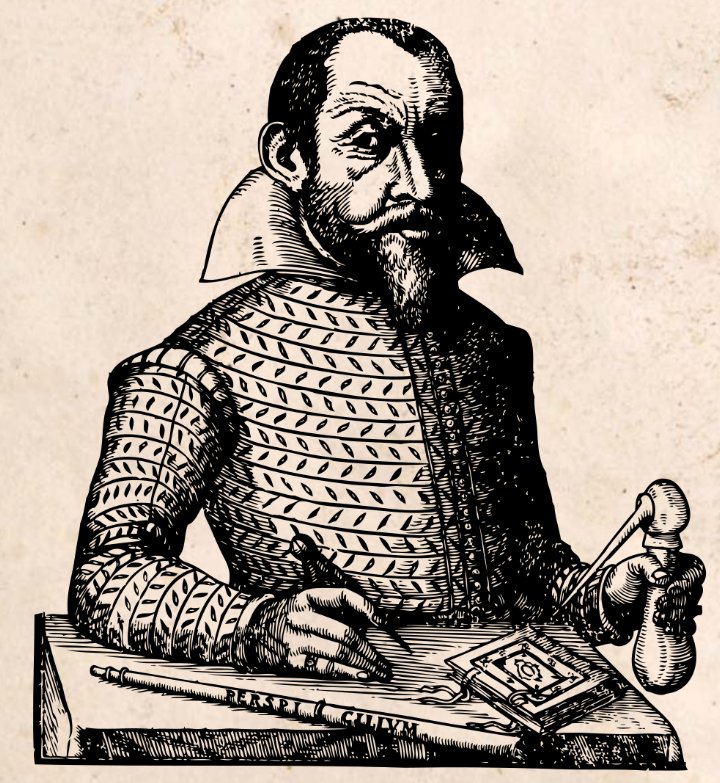}}
\end{center}
\caption{This engraved image of Simon Marius (1573-1624) 
is part of a figure shown in his book {\em Mundus Iovialis} (Marius 1614)
and in his Prognosticon for 1622, a small telescope is seen.}
\end{figure}

Marius published calendars with astronomical and astrological data as well as
{\em Prognostica} (yearly forecasts with astrological and astronomical texts) since 1601
at Lauer Publishing in Nuremberg. He and his wife, Felicitas Lauer, the daughter
of his publisher, had seven children.
Marius suffered from strong headaches since the Padua time and died on 1624 Dec 26 
(old Julian style) 
in Ansbach,
i.e. during the 30-year long civil war in central Europe.

See, e.g., Prickard (1917), Zinner (1942), Wolfschmidt (2012ab), and Gaab (2016) for more details on his life and work,
mostly in Ansbach.
Zinner (1886-1970), an astronomer from nearby Bamberg, who also worked on the history of astronomy, 
studied Marius in detail and noticed most of his sunspot observations (Zinner 1942, 1952).

Marius started celestial observations in Ansbach with the newly invented telescope 
in summer 1609 (Marius 1614).
He discovered the four large moons around Jupiter simultaneous and
independent of Galilei (e.g. Bosscha 1907, Prickard 1917),\footnote{Galilei observed the Jupiter
moons for the first time on AD 1610 Jan 7-11 (Gregorian), while Marius observed them for the
first time in 1609 Nov (Julian) and wrote first data records on 1609 Dec 29 (Julian),
i.e. on 1610 Jan 8 (Gregorian), e.g. Prickard 1917; that Marius (1614) used the Julian calendar
is written by him on his page 96 (in the edition of Schl\"or 1988).} 
and their current names were suggested by Marius (1614).
Marius also observed the phases of Mercury, sunspots, eclipses, 
comets, the Andromeda galaxy, and the (super-)nova of 1604.
His sunspot observations are discussed in this paper.
He exchanged letters with David Fabricius, Kepler, Maestlin (Kepler's teacher), and others.
Only a few of his letters have been found to date (e.g. Zinner 1942).

The work by Marius, partly in Latin and partly in German language,
as it is available in digital form
on the Marius portal (www.simon-marius.net) has been consulted:
we read his two main works, on the Iovian system ({\em Mundus Iovialis}, Marius 1614) including the
1615 appendix, and on the comet of 1618 (Marius 1619), as well as
his Prognosticon for 1613 (Marius 1612);
these three works contain his information on sunspots; 
all works on the Marius portal were checked 
by H. Gaab (priv. comm.); we also consulted
Klug (1904) and Zinner (1942), who bring extended quotations from the
Prognostica 
for 1601-1603, 1606-1616, 1618-1622, and 1625
with relevant astronomical content
(some of them published posthume).
In addition, the Calendars for 1615, 1619, and 1625 
have been consulted by
H. Gaab and K. Matth\"aus (priv. comm.) at the Staatsarchiv N\"urnberg
(not yet available in electronic form),
where no information on spots were found.
Some further Calendars were not yet consulted, or not even found.
Marius' original observational diaries (or log books) are considered lost.
% Cal 1607, 1608, 1610, 1613, 1614, 1619, 1622, 1625, 1626, not yet

\subsection{The other observers}

We also introduce briefly the other sunspot observers discussed 
here in comparison with Marius (sorted in alphabetic order),
and give the present name of relevant towns, where they lived,
with the present name of the respective country.

{\bf Andrea Argoli} (1570--1657).
Argoli was born in Tagliacozzo, Italy, and worked as astronomer and 
lawyer 
at U La Sapienza in Rome, Italy;
his main work {\em Pandosion sphaericum} from 1644 on the geocentric model has two pages on sunspots,
partly cited by Riccioli (see Sect. 3.4), but not listed in HS98.
As far as the year 1634 is concerned, 
not studied 
by us, Argoli was already mentioned as
additional sunspot observer by Vaquero (2003). We show below that Argoli also reports about observations
in 1618, but no 
detections (while comets were seen).

{\bf Johann Baptist Cysat SJ} 
(1586--1657).
Cysat joined Scheiner for sunspot observations in Ingolstadt, Germany, in March 1611
(see Sect. 4.5), which is not listed in HS98.
Cysat was born in 1586 and 
died in 1657, 
both in Luzern, Switzerland.
He was a Jesuit and studied mathematics and astronomy with Scheiner at U Ingolstadt.
After Scheiner left for Rome in 1618, Cysat became professor 
of mathematics at U Ingolstadt.
From 1624 to 1627 he was director of the Jesuit college at Luzern, then be became rector of U Innsbruck,
Austria, in 1637. Malapert (1633) mentions sunspot observations made 
at Ingolstadt in Bavaria, Germany,
(in Latin: {\em Ingolstadii in Bavaria}) by an observer, whose name he does not 
specify; Malapert shows those observations as 
a drawing 
for 1618 Mar (see Sect. 7.2); according to Vaquero \& Vazquez (2009), Malapert meant Cysat.

{\bf David Fabricius} (1564--1617).
David Faber (or Goldschmidt or Goldsmid, Lat.: Fabricius) from Dornum in East Frisia,
north-western Germany, studied at U Helm\-stedt, Germany, and worked 
as a pastor in Frisia;
he had visited Tycho Brahe in Prague (where he met with Marius);
among his most important observations were the naked-eye observations of the (super-)nova of 1604/1605
(now called SN 1604 or Kepler's supernova), where he obtained very precise positional data like
Kepler, which led to the identification of the SN remnant (see, e.g., Baade 1943, Stephenson \& Green 2002); 
among his notable telescopic observations are those of sunspots (Sect. 4.3)
together with his son in early 1611, published by his son (Fabricius 1611),
but not listed in HS98;
see also Reeves \& Van Helden (2010).

{\bf Johann Fabricius} (1587--1616).
Johann(es) Faber (or Goldschmidt or Goldsmid, Lat.: Fabricius), like his father David, also from Dornum in East Friesland,
north-western Germany, studied at U Helmstedt and Wittenberg in Germany as well as U Leiden in The Netherlands,
where he bought a telescope; he was the first to publish about telescopic sunspots (Fabricius 1611),
he is not listed in HS98;
Johann Fabricius noticed that spots near the center of the solar disk move faster than those at the edge
(where they also appear smaller due to foreshortening), so that he concluded that the spots
were on the surface of the Sun, and not due to orbiting objects;
see, e.g., Casanovas (1997) and Reeves \& Van Helden (2010) for more details; see Sect. 4.3.

{\bf Galileo Galilei} (1564--1642).
Galilei was born in Pisa, Italy, where he studied and started to teach. 
Later, he taught astronomy at U Padua until 1610.
In 1609, he heard about the newly 
invented telescope, built one, 
and used it for celestial observations.
He recognized craters on the moon, phases of Venus, moons around Jupiter, 
rings atound Saturn, and individual stars in the Milky Way. 
In a letter to Maffeo Barberini (later Pope Urban VIII) dated 1612 Jun 2,
Galilei mentioned to have started sunspot observations in Dec 1610.
Drawings are available since 1612 Feb 12.
See also Wallace (1984),
Vaquero \& Vazquez (2009), and Reeves \& Van Helden (2010).
He supported the Copernican system and was involved in several major controversies. 
Since he was supported by the Medici, he suggested to call the
large Jovian moons after them. See Sect. 4.6.

{\bf Thomas Harriot} (ca. 1560--1621).
Harriot studied in Oxford, UK, 1577--1590 and then worked on astronomical and mathematical problems;
among his astronomical achievements are the 
earliest datable
telescopic observations of the Sun
in 1610 Dec (see e.g. Herr 1978 and our Sect. 7.1) and a first telescopic drawing of the Moon;
Harriot found 
the law of refraction
before Snellius (but after Ibn Sahl);
he was employed for many years by Henry Percy, 9th Earl of Northumberland,
he did not publish much, many of his writings were found long after his death 
in 1786 by von Zach 
(e.g. Chapman 2008);
he is also known for the intervention of mathematical signs for {\em larger than} 
and {\em smaller than} and for introducing the potato to Europe (Shirley 1983).
See Sect. 4.6 for an error in HS98 on Harriot's observations in June 1612.

{\bf Joachim Jungius} (1587--1657).
Jungius is mostly known as mathematician and philosopher, but also obtained a lot of astronomical observations;
he was born in L\"ubeck, Germany, studied at U Rostock, Germany, and U Padua, Italy, 
and became professor for mathematics at U Giessen, 
Germany, at
the age of 22 years
(later, he moved back to Rostock);
he was in charge of the observatory tower at the newly founded U Giessen from 1611 to 1614 (Harden 2014),
the likely location for his telescopic sunspot observations in 1611 and 1612;
see Harden (2014) for a recent summary of the life and astronomical work of Jungius.
HS98 give Hamburg as the site of his observations, but it was Giessen, while his manuscripts 
now lie at the U Hamburg, Germany, library (we obtained his sunspot observations from them in digital form).
We show one of his observations in Sect. 4.6.

{\bf Charles Malapert SJ} (1581--1630).
Malapert was a Belgian Jesuit and astronomer;
he held a chair in the Jesuit college in Kalisz, Poland, from 1613 to 1617;
in 1630, he started a journey to Madrid, Spain, to occupy a chair at the Jesuit college, but died on the way
(Birkenmajer 1967).
His sunspot observations are found in Malapert (1620, 1633) with drawings, see Sect. 7.2.

{\bf Simon Perovius SJ} 
(1580 or 1586 to 1656).
Perovius is given 
as a sunspot observer 
for Kalisz, Poland (Latin: {\em Callissii in Polonia}) by Malapert (1633),
see Sect. 7.2.
According to Malapert (1633), Pero\-vius was a Jesuit and professor for Mathematics in Kalisz, 
which is known as the oldest town in Poland; Perovius was born in 1580 or 1586, he took over the Kalisz observatory directorship
after Malapert left in 1617, see Birkenmajer (1967) for the observations by Malapert and Perovius
(joined also by Alexius Sylvius Polonius).
Perovius died on 1656 Apr 26 in Cracow, Poland, and is sometimes also spelled Szymon and/or Petrovius
(see www.sjweb.info/arsi/documents/Defuncti$\_$1640-1740$\_$vol$\_$IIII$\_$N$\_$R.pdf).
Perovius' drawings are available in Malapert (1633) for 1618 Mar and July.
Vaquero \& Vazquez (2009) also mention Simon Perovius in Kalisz in connection to Malapert.
Perovius is not listed in HS98.

{\bf Giovanni Battista Riccioli SJ} (1598--1671).
Riccioli was born in Ferrara, Italy, and later became a Jesuit and astronomer
with his most important book being the {\em Almagestum Novum} published 1651,
where he preferred the Tychonian geo-heliocentric model over others
and where he displays, among other material, a telescope directed towards the Sun (on the title page);
he observed sunspots until 1661 (see Sect. 3.4 for 1618); see Graney (2015) for details.

{\bf Petrus Saxonius} (1591--1625).
Saxonius (Saxo) was born in Husum in northern Germany and 
studied at L\"ubeck, Leipzig, Altdorf, Leiden, T\"ubingen, and Wittenberg, 
all in Germany and The Netherlands;
he was travelling in southern Germany in 1614, visiting Scheiner in Ingolstadt 
and Maestlin in T\"ubingen, among others;
Saxonius visited Marius on (or since) 1615 Jul 4/14;
according to Wolf (1857) and HS98, 
Saxonius observed 
sunspots in Feb and Mar 1616
as obtained from 
his cupper plate entitled 
{\em Maculae solares ex selectis observationibus}
published at U Altdorf near Nuremberg 
(but see Sects. 4.7 and 4.8);
he worked since September 1617 as professor for mathematics at U Altdorf (Gaab 2011).
Because
Saxonius was from protestant northern Germany, where his father worked as Archidiakon in Husum, 
i.e. for the protestant church, 
and since Petrus Saxonius worked in the protestant area of Nuremberg, like Marius, 
he used the old Julian calendar in his writings.

{\bf Christoph Scheiner SJ} (1573 or 1575 to 1650).
Scheiner was born in Markt Wald in Swabia, Germany, and became a Jesuit and physicist:
he studied and taught at U Ingolstadt, Germany, later in Rome, Italy;
Scheiner and his assistant Cysat saw sunspots in March 1611, namely through clouds at U Ingolstadt (see Sect. 4.5).
Later he observed sunspots with the telescope since fall 1611 and published about them in 1612 
in letters sent under a pseudonym (Apelles) to Marcus Welser in Augsburg, Germany (Scheiner 1612);
many of his sunspot drawings are published in his book {\em Rosa Ursina sive Sol}.
Scheiner disputed about the discovery and nature of spots with Galilei (e.g. Braunm\"uhl 1891, Shea 1970, Biagioli 2002).

{\bf Ahasverus Schmidnerus} 
(ca. 1580--1634).
Schmidnerus studied medicine first in Wittenberg,
Germany (graduation October 1610), then in Basel, Switzerland (graduation 1612),
he visited Marius on his route between Wittenberg and Basel, 
see Sect. 4.1;
Schmidnerus later worked as medical doctor in K\"onigsberg
(formerly in Germany, now located in Russia), and 
died after 1634 (Komorowski 2008).
He has shown sunspots to Marius, but is not listed in HS98.

{\bf Adam Tanner SJ} (1572--1632).
Tanner was born in Innsbruck, Austria, and died in Unken near Salzburg, Austria;
he was a Jesuit and professor for theology in Munich, Ingolstadt, Dillingen, all in Germany,
Prague, Czech Republic, and Vienna, Austria (e.g. Rahner 1983).
Tanner is an additional, new sunspot observer for 1611 Oct 21 on until the end of 1612, not listed in HS98.
His observations were mentioned in Scheiner's {\em Rosa Ursina}.
Tanner's main work is {\em Universa Theologica scholastica}, where he reported about his spot observations
(Reeves \& Van Helden 2010, Sharratt 1996). See Sect. 4.5.

{\bf Jean Tard\'e} (1561--1636).
Tard\'e was born in Gascony, studied law at U Cahors, and worked as canon at the cathedral of Sarlat, all in France; 
during his visit to Italy in 1614, he met with Galilei in Florence;
he went back to Sarlat in 1615, started to observe sunspots, 
and advocated the transit theory for sunspots, see Baumgartner (1987).
He published his findings with drawings in Tard\'e (1620). Some of his data from 1615 to 1617 were 
listed in Wolf (1859) and used by HS98
(who gave {\em Farlat} as his town, but correct is {\em Sarlat}, now {\em Sarlat-la-Caneda}),
see Sect. 4.8.

{\bf Guilielmus Wely}
(born ca. 1600).
This person is given as sunspot observer for Coimbra, Portugal, (Latin: {\em Conimbrice in Lusitania}) 
by Malapert (1633) and Riccioli (1651), in the latter as Gulielmus Velius.
According to Malapert (1633), Wely was a Jesuit and professor for Mathematics in Coimbra, 
one of the oldest universities worldwide.
Wely's measurements for separations of spots from the solar limb are available in Malapert (1633) for
1620 Oct and Dec as well as 1621 Sep.
In the manuscript volume ARSI, Lusitania 39 fo.117v and Lusitania 44 II, fo.393v, 
in a document entitled {\em Catalogus brevis} for 1621 and 1622,
for the Jesuit college at Coimbra, there is a reference to {\em Guilhelmo Velli} (1621) 
and {\em Guilielmo Vielli} (1622),
who is listed among the Jesuits students for the 2nd year theology course 
(B. Mac Cuarta, priv. comm.).
The year of his death is not known to us (and he did not die as a Jesuit, B. Mac Cuarta, priv. comm.).
Vaquero \& Vazquez (2009) also mention Wely from Coimbra in connection to Malapert.
We discuss his observations in Sect. 7.2, he is not listed in HS98.
(Guilielmus Wely should not be confused with Gulielmini from Bologna, Italy,
who is listed in HS98 for observations 1675-1696.)

Argoli, Cysat, D. and J. Fabricius, Perovius, Schmidnerus, Tanner, 
and Wely are not listed as observers in HS98.

There were certainly even more telescopic observers of sunspots in the 1610s,
e.g. the Benedictine monk Benedetto Antonio Castelli (1576/77 to 1643),
professor for mathematics at U Padua since 1592, suggested to use the
Camera Helioscopica to Galilei to observe and draw sunspots in a letter
dated 1612 May 8, where he attached sunspot drawings (the letter is extant,
but without the drawings), see Reeves \& Van Helden (2010);
Galilei observed and detected the phases of Venus also after a suggestion by Castelli.

\subsection{Astronomical discussions of their time}

In the first few years after the invention of the telescope,
Marius found himself in the midst of major debates:
are the phenomena described as spots before or on the Sun, i.e. on the solar surface or
due to transits of previously unknown small bodies in the solar system --
or even something else?
While Galilei and Scheiner fought for being the first to have discovered
sunspots, it was later noticed that Harriot was the first to have drawn them 
with the help of a telescope (Dec 1610)
and that Johann Fabricius (1611) was the first to publish about them --
not to mention many Chinese, Korean, Japanese, Arabic, and also a few European
scholars, who saw spots with the unaided eye 
for centuries already before the invention of the telescope.
Medieval European and Arabic scholars tried to observe Venus and Mercury transits
in order to find out which of these two planets is closer to the Sun, which was
hard to tell from their very similar synodic periods (e.g. Goldstein 1969, 
Neuh\"auser \& Neuh\"auser 2015); most or all such presumable transit 
observations were probably sunspots (with the only possible exception being an
observation by Ibn S{\={\i}}n\={a}, see Goldstein 1969, Kapoor 2013);
also Kepler detected a sunspot a few years before the telescopic era (1607 May 28),
but thought that it was a transit of Mercury (e.g. Wittmann \& Xu 1987).
Given the recent discovery of Venus phases,
the search for Venus and Mercury transits, 
and the discovery of moons around Jupiter,
the interpretation of telescopic sunspot observations
as transits of unknown small solar system bodies appears to be understandable.

What is the sequence of the inner planets,
what is the architecture of the solar system, 
i.e. does the Sun orbit around the Earth or the other way round?
Marius understood that the data available at his time -- neither stellar abberation
nor stellar parallaxes were measured -- did not 
prove
the heliocentric model.
One of several otherwise insufficient arguments given by Galilei against 
the geocentric model and in favour of the heliocentric model, 
namely the phases of Venus, in particular the
full phase of Venus (i.e. behind, but slightly above or below the Sun, upper conjunction),
was in contradiction with the geocentric model -- given the largest possible
elongation of Venus from the Sun known since long ago; 
it could still not proof the heliocentric model, 
but it was fully consistent also with the geo-heliocentric model
of Theon of Smyrna (2nd century AD) and Tycho Brahe.
(In Mundus Iovialis, Marius (1614) wrote that he found Jupiter together
with its four large moons to orbit the Sun, but not Earth (page 84 in Schl\"or 1988).)
Marius and other observers at that time saw stars as round circles through their
(very small) telescopes, 
today known as Airy disks, 
but thought 
that they had resolved
the stars; by assuming that stars have the same diameter as the Sun, they could then
estimate their presumable (but far too small) distances; from the non-detection of
their parallaxes they rejected the heliocentric model and preferred the geo-heliocentric model.  
See Graney (2009, 2010, 2015) and Graney \& Grayson (2011) for details.

See also Vaquero \& Vazquez (2009) for more about early sunspot observations and related debates.

\subsection{Calendar issues}

The time period studied here is shortly after the calendar reform,
when the new Gregorian calendar replaced the previous Julian calendar:
1582 Oct 4 was immediately followed by Oct 15, i.e. the ten days Oct 5-14 were left out,
while the sequence of weekdays was uninterrupted.
This reform was initiated by Pope Gregory XIII (AD 1502-1585) modifying slightly
the previous calendar by Gaius Julius Caesar (100-44 BC).
The implementation of the reform was slow and took place at different
times depending on region and religion (protestant or catholic),
e.g. in most protestant German states, the reform was implemented by
jumping from 1700 Feb 18 to 1700 Mar 1 (see von den Brincken 2000).
HS98 presumably have transformed all dates in their catalogue to the Gregorian calendar (new style).

Simon Marius himself, working as state astronomer in a protestant country,
used the old Julian calendar according to Zinner (1942):
in his appendix to Mundus Iovialis (Marius 1614, but with the appendix written in 1615),
Marius gave both the Julian and Gregorian dates (e.g. {\em 17./27. Nov.}, see Sect. 4.4).
That he used the Julian calendar in the main text of Mundus Iovialis (Marius 1614)
was specified explicitely by him (page 96 in the edition of Schl\"or 1988) and was
also checked and confirmed by Prickard (1917) by comparison of the Jupiter moon
observations from Marius and Galilei (see also footnote 2).
Also in his book about the 1618 comet (Marius 1619), 
he used the Julian calendar, see Sect. 3.3, footnote 5.
David Fabricius and his son Johann Fabricius, 
who observed sunspots early in 1611, also used the Julian calendar -- and this also applies to
Harriot in England, who obtained the first datable telescopic sunspot detection
in December 1610 (HS98) as well as to, e.g., Jungius, Schmidnerus, and Saxonius.

The other observers from 1611 to 1619, who are studied here in comparison to
Marius, namely Malapert, Scheiner, Tanner, Riccioli, Tard\'e, Argoli, Wely, Perovius, 
and Galilei, either worked in Italy (Galilei) and/or were catholic 
priests (mostly Jesuits), i.e. all used the Gregorian calendar.

We list calendar dates either in Gregorian new style or give the date for
both the Julian and Gregorian style in the form {\em year month x/y} with
x being the Julian date (day) and y being the Gregorian date (day), 
e.g. 1582 Oct 5/15; 
this does not indicate a date range, but the two different dates in the Julian and the Gregorian calendar
for the very same day.
Dates given in direct quotations from Marius, Fabricius, Saxonius,
Harriot, and Jungius have to be considered Julian dates.

\section{Observations from 1617 to 1619}

For the years 1617 and 1618, HS98 name Marius as one of few observers.
According to HS98, 
Simon Marius would have observed all days from 1617 June 7 to 1618 Dec 31 --
except three short intervals between 1618 Mar 8 and Jul 18,
presumably all Gregorian dates.
However, he would never have detected any spot at all (HS98).

Only for meteorological reasons, it appears extremely unlikely that Marius
would have been able to exclude spots on 333 days in 1618
including the last few months of the year,
when the sky in Germany is often overcast for several days to weeks.
E.g., in his work on the comet of 1618, Marius (1619) does mention a few nights,
where the weather was bad, including several subsequent nights, for Dec 1618.

It may not be necessary to observe the sun each and every day to detect
most spots -- large sunspots have life-times of more than a day;
it is possible to miss certain spots which either disappear or rotate out of
view too soon, e.g. during a bad-weather period or a monitoring break of a few days, 
or that spots disappear too soon after they have formed or rotated into view.

Let us consider the data from HS98 for 1617 and 1618 in detail,
we present first their data, then naked-eye spots, 
then the text by Marius, part of which was cited by Wolf (1857) and HS98.

No observations by Marius for 1619 are listed in HS98.

\subsection{Hoyt \& Schatten (1998) for 1617 and 1618}

For 1617, Marius would have observed all days from Jun 7 to Dec 31,
without any spot detection (HS98).
In the very same year, Tard\'e in Sarlat, France, the only other observer,
would have observed from May 27 to Jun 6, and in this period,
he would have detected one spot or group each day (HS98);
all dates presumably being Gregorian.

It appears to be quite surprising that Marius would have started his observations
exactly on the day after the presumable end of the observations of Tard\'e, 
i.e. on the day after the end of solar activity (eleven days with one spot or group).

For 1618, Marius would have observed all days except three short breaks
(Mar 8-18, Jun 21-29, and Jul 7-18), and again, he would never have spotted any spot (HS98).
For 1618, there are three other observers:
\begin{itemize}
\item Malapert (1633) detected one spot or group on Mar 8, 10, 12--15, and 18,
and he also observed from Jun 21-29 with one spot or group each day,
as well as one spot or group on Jul 7, 9, 13, 14, 15, 17, and 18,
and would also not have observed any other day that year.
\item According to HS98, Scheiner also detected a spot or group on Mar 8, 10, 12--15, and 18,
but would not have observed any other day that year.
(However, in fact, Scheiner (1626-1630) reports about the observation of Malapert for March 1618;
Malapert (1633) shows only one group, while both Scheiner (1626-1630) and Malapert (1620) 
show two groups for those dates (see also Sect. 7.2); HS98 give just one group.
Scheiner's drawing of spot A is very simular in resolution to the one spot group shown
in Malapert (1633), while Scheiner's drawing of spot B is less crude than the drawing 
of group B by Malapert (1620) himself, so that there probably existed an original drawing by Malapert,
which is not extant anymore.)
\item Riccioli would have observed all days that year except
Mar 8-18, Jun 21-29, and Jul 7-18 (exactly like Marius) 
and would never have detected a spot (like Marius).
\end{itemize}
All those data are from HS98, their table {\em alldata}.

These numbers again appear to be quite surprising:
Riccioli and Marius would have observed exactly on the very same 333 days that year
without any spots. They would have stopped their otherwise very regular observations
for the same three intervals -- exactly during the time when Malapert
(and partly also Scheiner) had detected spots.

\subsection{Naked-eye spots 1617--1619}

There are several reports about sunspots from naked-eye observations in China
for the relevant period 1617 to 1619, namely as follows
(note that the Chinese texts did not discriminate between singular and plural
forms, so that Wittmann \& Xu (1987) sometimes translate {\em spot(s)}, all dates are Gregorian here):
\begin{itemize}
\item Our summary for 1617: On at least 1 day (Jan 11) several spots in at least one group
\begin{itemize}
\item 1617 Jan 11: {\em at about 9 a.m. there are several black spots} [heizi] 
{\em moving about at the
side of the sun} (Wittmann \& Xu 1987), 
otherwise translated to \\
{\em Between 7 and 11 a.m., on one side of the sun there were several black spots}
[heizi] 
{\em rocking to and fro} (Yau \& Stephenson 1988), or \\
{\em 7 to 11 a.m., there were several black spots} 
[heizi] 
{\em roiling and agitating one side of the sun} (Xu et al. 2000)
\item 1617 (only year given): {\em on the sun there were black spot(s)}
[heizi] 
{\em skimmering about}  (Wittmann \& Xu 1987), otherwise translated to \\
{\em within the sun there was a black spot} 
[heizi] 
{\em rocking to and fro} (Yau \& Stephenson 1988),  or \\
{\em there was a black spot} 
[heizi] 
{\em roiling and agitating the sun} (Xu et al. 2000), \\
probably the same spots as before for 1617 Jan 11
\end{itemize}

\item Our summary for 1618 Apr/May: On at least one day (May 22) one big spot or
several spots in one group ({\em ladle/spot})
\begin{itemize}
\item 1618 May 22: {\em within the sun there was a black ladle} (Willis et al. 2005)
\item 1618 Apr 25 to May 23 (only Chinese lunar month given): 
{\em black spot(s)}
 [heizi] 
{\em on the sun} (Wittmann \& Xu 1987) or \\
{\em within the sun there was a black spot}
[heizi]
(Yau \& Stephenson 1988) or \\
{\em there was a black spot} 
[heizi] 
{\em on the sun} (Xu et al. 2000), \\
probably the same spot(s) as before for 1618 May 22
\end{itemize}
\item Our summary for 1618 May/June: On at least three days (June 20-22) one group ({\em vapour})  
\begin{itemize}
\item 1618 June 20-22: {\em black vapour}
[heiqi]\footnote{The Chinese word 
for {\em vapour}, i.e. {\em qi}, can stand for {\em spot(s)},
in particular when combined with {\em black} 
[hei] 
and {\em sun}, but it can also mean {\em aurora(e)}, 
in particular if combined with a colour like red, blue, or green
(see Chapman et al. 2015), 
while {\em heizi} is always {\em black spot(s)}.}
{\em coming in and out of the sun, moving about,
it was seen until day 25} (Jun 22) (Wittmann \& Xu 1987) or \\
{\em for three days until 25} (Jun 22), {\em on one side of the sun there was a black vapour}
[heiqi]
{\em pulsating in and out that roiled and churned for a long time} (Xu et al. 2000), and then also \\
{\em day 23} (Jun 20), {\em from this day until day 25} (Jun 22) {\em for three days. On one side
of the sun there was a black vapour} 
[heiqi] 
{\em coming in and out of the sun and rocking
to and fro for a long time} (Yau \& Stephenson 1988) or \\
{\em from day 23} (Jun 20) {\em until day 25} (Jun 22) {\em black vapours} 
[heiqi] 
{\em pulsated in and out of the sun and churned about} (Xu et al. 2000)
\item 1618 May 24 to Jun 21 (only Chinese lunar month given): 
{\em black spots}
[heizi] 
{\em on the sun fighting with each other} (Wittmann \& Xu 1987) or \\
{\em within the sun there was a black spot} 
[heizi] 
{\em like a ladle} (Yau \& Stephenson 1988) or \\
{\em black spots} 
[heizi] 
{\em on the sun combated each other} (Xu et al. 2000), and then also \\
{\em within the sun there was again a black spot} 
[heizi]; 
{\em its light was wavering} (Yau \& Stephenson 1988) or \\
{\em there was a double black spot on the sun, its light was roiled and agitated}
(Xu et al. 2000), \\
those spots, for which only the month is given, are probably
the same spots as listed otherwise for certain dates (June 20 and 21)
\item 1618 June 22: {\em within the sun, there was a black vapour} [heiqi] 
(Wittmann \& Xu 1987, Yau \& Stephenson 1988)
\item 1618 (only year given): {\em within the sun there was a black spot}
(Yau \& Stephenson 1988), \\
probably one of the spots listed before for 1618 May 22 or June 20-22.
\end{itemize}

\item 1619: There are no reports known for naked-eye sunspots in 1619.
\end{itemize}

In 1618, we see on at least four days (May 22, June 20-22) one group each, 
possibly the very same group seen again after one rotation;
in 1617, we see at least one spot/group on at least one day.
According to Vaquero et al. (2002, 2004), 
there were two spots reports in 1617 and five in 1618.

Let us compare these reports with the telescopic record in HS98.
On the first date (1617 Jan 11), Marius and Tard\'e would not have observed (HS98).

In May 1618, Marius and Riccioli would have always observed with their telescopes,
but without any spot detections (HS98),
even though the Chinese have detected 
a spot on May 22
with the unaided eye.
The spot report from the Chinese for Jun 20-22 is consistent with the data
from Malapert, who detected a spot or group Jun 21-29 (HS98), see Fig. 2.

\begin{figure}
\begin{center}
{\includegraphics[angle=0,width=7cm]{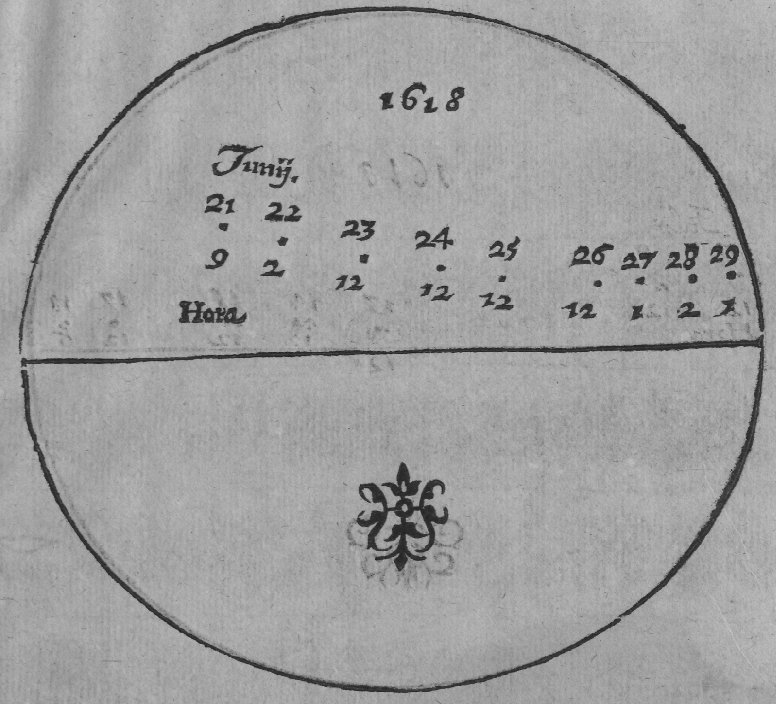}}
\end{center}
\caption{Telescopic sunspot drawing by Malapert (1633) for 1618 June 21-29
(observing hour given below the spot), one spot/group each,
which is consistent with the Chinese report about a naked-eye spot 1618 June 20-22:
{\em for three days until} June 22 {\em on one side of the sun there was a black qi}.
There is another similar coincidence: the Chinese specified 
{\em a star seen on the side of the Sun} for 1625 Sep 2
(Willis et al. 2005),
and Scheiner (1626-30) has drawn one circular large spot for that date
($\sim 1600$ square arc sec, i.e. well visible for the naked eye, Schaefer 1993, Vaquero \& Vazquez 2009).
When the Chinese specify that the {\em qi}, i.e. spot, is located {\em on one/the side},
the meaning could be that this spot was not close to the apparent center of the disk
-- as indeed seen in the drawings by Malapert and Scheiner.
%where the spot is seen relatively far 
%to the upper (left) side for June 21.
We estimated the spot size to be about 160 square arc sec,
so that the spot would not be visible to the naked eye 
(Schaefer 1993, Vaquero \& Vazquez 2009);
we conclude that Malapert did not draw the spot size to scale here:
note that all spots in this drawing are very similar, which is not typical --
since Malapert supported the transit theory, his main interest may have been the
path of the spot(s), shown with correct curvature (B angle) in his drawings;  
see also Fig. 11.
It is believed that the Jesuits (Johannes Terrenz Schreck) brought the telescope to China
not before 1621, so that a telescopic observation in China in 1618 would not be possible.
}
\end{figure}

Shortly before the spot detected on May 22, there also was an aurora on 1618 May 17 seen in China,
for which there are several text variants and translations as follows: 
\begin{itemize}
\item {\em There were two bands of blue-black vapour} [heiqi]\footnote{See previous footnote.}
{\em stretching across the W to E} (Yau et al. 1995, Willis et al. 2005) or
\item {\em This evening there were two bands of blue-black vapour} [qingheiqi] 
{\em stretching across the sky from W to E} 
(Yau et al. 1995, Willis et al. 2005) or
\item {\em This evening there were two bands of bluish-black vapour} [qingheiqi] {\em stretching 
laterally across the sky from W to E} (Xu et al. 2000) or
\item {\em This evening there were two bands of greenish-black vapour} [cingheiqi] {\em stretching 
across the sky from W to E} (Xu et al. 2000).
\end{itemize}
Two of five aurora criteria from Neuh\"auser \& Neuh\"auser (2015) are fulfilled,
auroral colour (blue, green) and {\em W to E},
so that it is a {\em very possible} aurora ({\em This evening} does not neccessarily mean night-time).
If the spot on 1618 May 22 was in a causal connection with the aurora on May 17 (through a CME or flare),
the spot or group should have been present on the solar disk already a few days earlier.
However, Marius and Riccioli would have observed all days in May, 
but would never have detected a spot (HS98).
 
The data in HS98 for 1618 seem to be inconsistent with the reports from China.

\subsection{Reports from Marius for 1617-1619}

HS98 quote Wolf (1857) as source for their data for Marius.

In Wolf (1857), we can read 
(our English translation of the partly German and partly Latin text can be found below,
our additions here and below in square brackets):
\begin{quotation}
Simon Marius, astronomische und astrologi\-sche Beschreibung des Kometen von 1618, N\"urnberg 1619. 4. \\
Die Vorrede dieser Schrift ist {\em Anspach den 6. April 1619}\footnote{Wolf 
(1857)
gave {\em 6. April 1619}, but the original manuscript clearly shows Apr 16;
this is just a typo in Wolf (1857); furthermore, the text about the sunspots 
quoted here is in section V (5), not 4 as given by Wolf. --
Marius did not specify here, whether this date in the dedication is Julian or Gregorian;
if Apr 16 is Julian, the Gregorian date is Apr 26.
For the remainder of the paper, it does not matter much, whether Marius meant here the
Julian or Gregorian date, because his statement about {\em one and a half year prior to
April 1619} may not be precise to better than about one or two months anyway.
For a few dates given in Marius (1619), he adds explicitely that the date is given in the old style
(Julian calendar), and for some other dates, he mentioned the weekday (being consistent with the
given Julian calendar date), so that it is certain that he used the Julian calendar. 
He has detected a comet on 1618 Nov 11 and 21 (Julian) and has then 
obtained positional measurements of a comet 1618 Nov 24 to Dec 19 (Julian),
i.e. Dec 4 to 29 Gregorian (his section III), 
while others have observed comet 1618 W1 from 1618 Nov 23 or 25 (Gregorian) 
until 1619 Jan 22 (Gregorian), the latter by Cysat with a telescope
(according to Kronk (1999), who does not mention Marius for the comets in 1618). 
When Marius mentioned that he saw a comet
on 1618 Nov 11 early in the morning with tail but without the comet head (his section II) 
or since {\em Martini Alten Calendars} (i.e. Nov 11 Julian) (his section IV), 
he refers to the other comet (1618 V1 in Kronk 1999), 
which was observed by others from 1618 Nov 10 or 11 until Nov 29 (Gregorian), or even until Dec 9
(also seen by others with tail but without head);
note that some dates given in Kronk (1999) are incorrectly shifted from a presumable 
Julian date (but truely being Gregorian) by ten more days, namely the last Jesuit 
observation of comet 1618 V1 obviously incorrectly shifted from Nov 29 to Dec 9.
It is quite clear that comet 1618 V1 was discovered earlier
than W1; V1 was observed in China since Nov 14 and W1 since Nov 26,
their dates should not be affected by a wrong Julian-Gregorian calendar conversion. 
It is then clear that Marius first observed comet V1 (1618 Nov 11/21) and then
comet W1 from 1618 Nov 21/Dec 1 until Dec 19/29 -- claiming that all observations
would pertain to only one comet. 
} 
datiert. 
Marius erz\"ahlt, dass er {\em nun \"uber die anderthalb Jahr nicht mehr 
so viel maculas in disco solis habe finden k\"onnen, 
ja gar offt kein einig maculam antroffen, 
das doch vorige Jahr niemals geschehen.} 
Dieser Fleckenarmuth stellt Marius das grosse Kometenjahr 1618 gegen\"uber, 
und f\"ugt dann bei: {\em Ich erinnere es nur, und schliesse nichts.}
\end{quotation}

We translate Wolf's citation from Marius (1619) to English as follows:
\begin{quotation}
Simon Marius, astronomical and astrological description of the comet of 1618, Nuremberg 1619, 4.\\
The foreword of this text is dated and given as {\em foreword 1619 Apr 6}.\footnote{Marius
gave {\em Apr 16}, see previous footnote}
Marius reports that he {\em now, for one and a half year, could not find
as much spots on the solar disk, yet rather often not even a single spot,
as was never the case in the years before.}
Marius compares this dearth of spots with the comet year 1618 and adds
{\em I just recall it, but I do not conclude anything.} 
\end{quotation}

The text in the book by Marius (1619) himself on the comet of 1618, 
dated 1619 Apr 16 (Julian, see footnote 5)
gives even more details (translation below), cited here in German and Latin from Zinner (1942):
\begin{quotation}
..., dieweil ich nun \"uber die anderhalb Jahr nicht mehr so viel maculas in disco Solis 
hab finden k\"onnen, ja gar offt kein einig maculam antroffen, 
das doch vorige Jahr niemals geschehen, 
dahero ich dann in meinen observationibus verzeichnet, Mirum mihi videtur, 
adeo\footnote{Instead 
of {\em adeo}, Zinner (1942) wrote {\em ad eo},
which would have a different meaning; we have consulted the copy of the original
work digitally available at the ETH Z\"urich library and the Marius portal, where it is clearly written as one word;
therefore, the possible meaning of {\em adeo} as {\em rather} is best consistent 
with the context.}
raras vel saepius nullas maculas in disco solis deprehendi, 
quod ante hac nunque est observatum. 
Wie wenn an diesem orth auch etwas verborgen lege. 
Ich erinnere es nur, vnnd schliesse nichts, 
lasse andere hohe, gesunde vnnd scharffe Ingenia den sachen weiters nachdencken,  
Ich thue das meinige, andere thun auch das jhrige, nach deme jhnen Gott gnad verliehen hat, 
man muss der sachen ein anfang machen, vnnd einer dem andern ohne verlesterung die Hand bieten, 
biss man endtlich was gewiesses schliessen kan. Ich hab mich die zeit hero, als von Anno 1611. 
sehr mit gedancken bem\"uhet, was doch solche maculae seyn, oder woher sie entstehen m\"ochten, 
hab aber noch zur zeit keine gedancken gehabt, darauff ich sicherlich beruhen k\"onnte. 
Das sage ich aber: das ich etlichmal maculas caudatas, in disco Solis ausstr\"ucklich gesehen, 
durchauss gleich einem Cometen, darob ich mich offt hoch verwundert hab. 
Wie, wenn solche maculae ein refrigerium weren, summi caloris solis, 
vnnd hernacher per adunationem, vel potius conglobationem zu einem Cometen w\"urden, 
Ich schliesse nichts, kan es auch nicht thun, zeige nur mein gedancken an.
\end{quotation}
Our English translation follows below
(this text is found in a section with considerations about the formation of comets 
and their possible connection to sunspots):
\begin{quotation}
..., while I now, for one and a half year, could not find
as much spots [maculas] on the solar disk, yet rather often not even a single spot [maculam],
as was never the case in the years\footnote{While Usoskin et al. (2015) here translate
{\em Jahr} (in {\em doch vorige Jahr}) 
with {\em year}, we translate with {\em years}: (a) in former times, 
the German word {\em Jahr} was used for both the singular and 
plural meaning for both {\em year} and {\em years}
(Grimm \& Grimm 1854 and Bartz et al. 2004 giving examples from Lessing and Goethe),
(b) if Marius would have meant the singular meaning, he would have said {\em das vorige Jahr},
i.e. with the definite article {\em das} ({\em the}), and
(c) Marius observed spots since 1611 (see below), so that he compared those
one and a half years with several previous years.}
before, I have therefore written this in my observational log books,
[the remaining part of this sentence is in Latin in the original] this appears strange to me, 
that rather few or more often/frequent no spots
could be detected on the disk of the sun, which was never observed before.\footnote{Usoskin et al. (2015)
remarked here that the latter sentence, in the original text by Marius in Latin, would be 
{\em a repetition of what he said before},
but in fact it adds 
information with more precision (see below in this section points (i) to (iii)).
}
As if something would be covered at this location.
I just recall it, but I do not conclude anything,
I let other high, healthy, and sharp-thinking genius (people) think further on those things,
I do my part, others do their parts, given the grace of God,
one must start with it, and should help the other without any hate,
until one can conclude something with more certainty.
I have thought about it a lot since the year 1611, what those spots could be,
and how they would form, but have not come to a conclusion yet, which I could rest on.
But this I say: that I several times have clearly seen tail-like longish spots on the disk of the sun,
indeed somewhat similar to a comet, so that I was often surprised. Like, if those spots would bring
some kind of coolness to the extreme heat of the sun,
and later would become a comet by merging or rather combining,
I do not conclude anything, I cannot do it, but just indicate my thoughts.
\end{quotation}

The {\em observational log books} ({\em observationibus}) mentioned are unfortunately considered lost.
After mentioning his {\em observationibus} ({\em observational log books}), 
he probably had checked them before continuing the writing (and then fell into Latin).
Then, he specified some information with more precision: \\
(i) {\em few} [spots] in addition to {\em could not find as much spots on the solar disk};
the word {\em raras} could mean {\em few} or {\em here and there} or {\em isolated/single}
regarding the number and spacing of spots, or {\em rare} regarding the frequency of spots;
{\em few} might be the best compromise;
{\em raras} is connected grammatically to {\em maculas}. \\
(ii) {\em more often/frequent no spots} in addition to {\em yet rather often not even a single spot};
the Latin {\em saepius} is a comparative degree and translates to {\em more often}
or {\em more frequent},
so that on more than $50\%$ of the observing days the sun was spotless. 
Hence, the active day fraction was below $0.5$, but larger than 0 from fall 1617 to spring 1619,
see also Sect. 6. \\
(iii) {\em never observed before} in addition to {\em never the case in the years before};
with {\em never observed before}, Marius obviously means since the start of sunspot 
observations by himself in August 1611, see also below (Sect. 4.1); this gives
additional evidence that Marius compared those roughly one and a half years with several years before
(see footnote 8). Hence, the active day fraction was $1.0$ from August 1611 to fall 1617
(at least on his observing days, see Sect. 4.9), see also Sect. 6.

If the statement by Marius above that sunspots may provide 
{\em some kind of coolness to the extreme heat of the sun},
should mean that they are cooler than the surrounding photosphere, 
then this would be an acceptable consideration. 
His hypothesis regarding the formation of comets from sunspots was not confirmed,
but indeed, cometary taila are blown by the solar wind;
Marius states the observed fact that tails point away from the sun.
He supports his connection of comets with sunspots by the observational fact
to have observed a large comet (of 1618), but very few spots (in 1.5 yr prior to spring 1619),
i.e. at the same time; in the time before fall 1617, he observed many spots, but no comets.
The statement by Marius about the observed
{\em tail-like longish spots} does not pertain to the period of
1.5 yr prior to spring 1619, but to the observing period since 1611;
such longish spots were apparently considered to have formed comets seen years later
({\em hernacher} for {\em later}, see above).

Marius mentioned a period of (roughly) 1.5 years until some time in the first
3.5 months in 1619: the comet reported by Marius (1619) was detected by him 
until 1618 Dec 19/29 (and he continued to try to observe it until 1618 Dec 25 (Julian),
i.e. 1619 Jan 4 Gregorian, see footnote 5), 
and the dedication of his book is dated 1619 Apr 16 (Julian),
so that the book was written some time in the first 3.5 months of 1619
(the very earliest possible date for the end of those 1.5 yr would be 1619 Jan 4 (Gregorian),
when his comet observations ended, because he connected the reduction of spots with 
the appearance of the comet).
In the remainder of the paper, we assume that the period ended in April 1619,
given that the generic statement about spots is located towards the end 
of the book in section V (of six sections);
the length of the period (1.5 yr) is probably meant to be {\em roughly 1.5 yr}, 
maybe to within one or two months.
(The period of 1.5 yr therefore started roughly between July and October 1617.)
Marius does not mention any evolution of spottedness within those 1.5 years;
it is likely that there was a significant change at the beginning
of those 1.5 years (fall 1617) regarding spottedness: the appearance of spotless days
and/or (much) less spots than before.

From the text by Marius himself, one cannot conclude that he never saw a spot.
Also, the observational period 1617 Jun 7 to 1618 Dec 31 as given in HS98 cannot
be deduced from that text.
(Even if Marius would have reported something
like that he would have observed all of 1618, 
one should consider this as the Julian calendar year of 1618,
while HS98 let him end his monitoring on 31 Dec 1618 on the Gregorian calendar.)
That HS98 let Marius start his successless monitoring on 1617 June 7 is
not justified, neither by some 1.5 years before the end of 1618 nor before 
early 1619 (see next paragraph) nor before April 1619, 
nor by any statement at all from Marius. 

The text by Marius himself as quoted above clearly shows
that he did detect spots, both in those roughly 1.5 years before 1619 Apr,
and even more in the years before those roughly 1.5 years.
In the period before fall 1617, he noticed {\em several times} spots being lengthy like a comet,
obviously describing unresolved groups or double spots.
Marius noticed the decrease from high(er) spot numbers in the years before about fall 1617,
explicitly without any spotless days on observational days,
to much smaller numbers in those roughly 1.5 years before 1619 Apr 
-- in the latter period of roughly 1.5 years, most of the observational days were spotless for him
(as specified in his Latin sentence, i.e. active day fraction below 0.5, but not zero), 
contrary to previous years;
he noticed a decrease in solar activity from the previous Schwabe cycle maximum to a minimum (Sect. 7.3).

HS98 furthermore cite Zinner (1952), where it is just briefly (one line)
mentioned that Marius would have observed spots from 1611 to 1619.
Zinner (1952) gives as reference Zinner (1942), which 
relates to previous years and which we will investigate in Sect. 4.

\subsection{Non-detections by Riccioli (Argoli) in 1618}

According to HS98, Riccioli would have observed in 1618 on the very same 333 days as Marius,
namely all days except the three short breaks when Malapert 
%XXX%and Scheiner 
detected spots.
According to HS98, Riccioli also observed in later years (1632, 1655, 1656, 1657, 1661).
The source given in HS98 is Wolf (1861), where it is specified (our translation from German to English):
\begin{quotation}
[In Almagestum novum, Bononiae 1651 ...] ... on page 96, sources are given. Additionally, it is stated
that there were no spots not only in 1618, when a large comet were shining, 
but also in 1632 from July 12 or 19 until Sep 15, when there was an exceptional dryness
and various observers did not find any spots in the Sun. More generally, when there was bright or
dry weather, no or less spots were seen, while during the coldness in June 1642, there were many spots
on the sun. In the 2nd part, there are citations from Scheiner.
\end{quotation}
HS98 do not list any observations in June 1642.
HS98 also remark for Riccioli as follows: {\em 92 observations for 1655-1661 come from Manfredi (1736) 
and taken as zero spot days based upon comments by other observers}.
Obviously, HS98 concluded from {\em there were no spots not only in 1618}
that Riccioli would have observed all days, but never spotted any spot;
they may have seen this as confirmation of alleged similar observations by Marius.

We have checked the relevant pages (around page 96, the text about sunspots) 
in {\em Almagestum novum} (Riccioli 1651) 
and found only the following relevant statement on page 96:
\begin{quotation}
Itaque anno 1618 quo Trabs, et Cometes fulsit, 
nulla Macula observata fuit, ait Argolus in Pandosio Spherico cap. 44.
Sed neque anno 1632 a die 12. aut 19. Iulii usque ad 15. Septembris, 
quo tempore insignis fuit siccitas, ulla posuit deprehendi 
in Sole Macula Romae a Griembergero Mazorbi ab Argolo et Casenae aut Pisis ab aliis, 
ut testantur Scipio Claramontius et Argolus supra; 
sed et Fortunius Licetus Lib. 6 de novis Astris, 
et Cometis ait a Vincentio Fridiano Patavii observatum saepius, 
suda et sicciori tempestate nullas, aut pauciores Solis maculas 
apparere idemque ait Antonius Maria Reithensis L. 4 c. 2 M. 9 
de anno 1642 addens Iunio finisse frigus ob multitudinem macularum.
\end{quotation}
We translate this to English as follows:
\begin{quotation}
Thus, in the year 1618, when a bar [fiery appearance] and a comet shone brightly,
no spot was observed, says Argolus in his Pandosion Sphaericum chapter 44.
But also in the year 1632 from 12th or 19th July until 15th September, during an unusual dryness, 
not a single spot could be found in the sun 
-- neither by Griembergerus in Rome [Italy] nor by Argolus in Mazorbi [Italy] 
nor by others in Cesena [a town south of Ravenna in Italy] nor Pisa [Italy], 
as confirmed above by Scipio Claramontius and Argolus;
and also Fortunius Licetus said in his {\em six books on new stars and comets} about Vincentius Fridianus,
who more often [or: more frequently]
observed in Padua [Italy] that there appeared none or less sun spots during clear and dry weather,
and the same is also said by Antonius Maria from Reutte in L. 4 c. 2 M. 9 on the year 1642,
where he adds that the coldness [frost] due to the large amount of spots ended in June.
\end{quotation}
The Latin word {\em trabs} originally means just {\em bar}, but was later used also for
{\em fiery appearance} in air or on sky, e.g. for comets.
The mentioned Argolus is the Italian scholar Andrea Argoli (1570-1657),
who is not mentioned in HS98; he published about sunspots in Argoli (1644); see below.
The remaining scholars are not related to our study period of the 1610s:
the Jesuit Christophorus Griembergerus (Christopher Gruintberg) worked as mathematician in Rome, not listed in HS98.
Antonius Maria Reithensis (Anton Maria Schyrleus de Rheita or Johann Burkhard Schyrl or Sch\"urle) from Reutte
is the well-known sunspot observer Rheita (1604-1660), 
see HS98 and Gomez \& Vaquero (2015) about Rheita, who observed spots in 1642.
Scipio Claramontius (Chiaramonti) lived from 1565 to 1652, but is not listed in HS98.
Fortunius Licetus (1577-1657) was an Italian physician, philosopher, and scientist, not listed in HS98.
Vincentius Fridianus is otherwise not known to us. 

The statement from Rheita
({\em coldness [frost] due to the large amount of spots ended in June})
may be interpretable in two different ways,
either coldness due to many spots or coldness ended due to many spots newly appearing.
Given the distribution of spots in HS98, the former possibility is more likely.
Argoli and Licetus about Fridianus (all in Italy) were cited
that there were no or less spots {\em during an unusual dryness}
respectively {\em during clear and dry weather} -- probably indicating a heat wave in summer.
Then, those statements fit together.
Argoli noted (also see next quotation) that there were no spots in 1618 during the comet sightings.
This may be connected to the consideration by Marius (1619) in his book about the 1618 comet (Sect. 3.3) 
that {\em those spots would bring some kind of coolness to the extreme heat of the sun,
and later would become a comet by merging or rather combining}.
All this led to a conjecture in the field of solar-terretrial relations:
more {\em coolness} in the sun, presumably due to many spots, would lead to coldness on Earth.

We will concentrate on 1618 here:
Riccioli cites Argoli (1644) for 1618. In the work {\em Pandosion Sphaericum} by Argoli (1644), 
we found on page 213 in chapter 44 with subheading {\em De solis maculis}:
\begin{quotation}
Anno 1618 tempore quo Trabs, et Cometa affulsit nulla visa est; 
Sic anno 1634 a 19. Iulii usque ad medium Septembris, ut nos Marzobi prope Venetias pluries 
observavimus; hinc admiratione compulsi scripsimus Christophoro Gruintbergero 
Romani Collegii Mathematico, qui eadem literis confirmavit.
\end{quotation}
This is translated to English as follows:
\begin{quotation}
In the year 1618 at a time, when a bar [fiery appearance] and comets [cometa] shone, none [no spot] was observed.
So also in the year 1634 from 19th Jul to mid Sep, as we have observed often in Marzobio near Venice [Italy]. 
Then, moved by admiration, I wrote to Christopher Gruintberg, a mathematician of the Roman College,
who confirmed it in a letter.
\end{quotation}
The last sentence was published before by Vaquero (2003), we quote here his translation.
Parts of the second sentence were also published before by Vaquero (2003), 
but with slightly different Latin and English translation.
Only the first sentence is in relation to 1618.
While Riccioli gives the singular form {\em cometes} probably meaning the main comet of 1618,
Argoli uses the plural form {\em cometa}.

Argoli's statement about spotlessness is restricted to times 
in 1618 {\em when a bar and comets shone},
i.e. 1618 Aug 25 -- Sep 25, 1618 Nov 11 -- 29 (or Dec 9, but see footnote 5), 
and then since 1618 Nov 23 or 25 (Kronk 1999).
It may well be possible that he did detect spots in 1618 before
the first comet appeared, or during the period without a comet.
This is consistent with the dates of known 
telescopic and naked-eye spots in 1618 being all 
from Mar 8 to July 18 (Scheiner and Malapert in HS98), i.e. before August.

Marius (1619) reported about {\em tail-like longish spots} seen since 1611, 
so that he considered whether they {\em would bring
some kind of coolness to the extreme heat of the sun},
and later could become a comet by merging and combining.
When Marius explains the possible connection between comets and spots,
he first argues that comets seem to come from the Sun
(he observed a comet in 1618 very close to the Sun),
for which he gives reference to Cardanus (Marius 1619);
the theory of conneting spots with comets may have been invented by Marius himself, 
because he does not mention references, but says that he indicated his own ideas
(Sect. 3.3).

We stress that Argoli (1644) gives the year 1634 (Jul 19 to mid Sep),
while Riccioli (1651) citing {\em Argolus in his Pandosion Sphaericum} gave
the {\em year 1632 from 12th or 19th Jul until 15th Sep}.\footnote{HS98 used the quotation 
from Wolf (1861) about Riccioli (who gives 1632 citing Argoli) to conclude that Riccioli would have reported
on a sequence of spotless days from 1632 Jul 12 to Sep 15, which would be wrong given the original
quotation from Argoli for 1634.
Vaquero (2003) already published about the observations of Argoli and Gruintberg in 1634,
giving the correct year 1634 based on the work by Argoli himself;
he also argued that non-detection of spots by Argoli for the period 1634 Jul 19 to mid Sep
would be confirmed by Gassendi, who -- according to HS98 -- did not detect any spot from
1634 Jan 1 to Oct 23, even though he would have observed every day from Paris, France (HS98),
which we think is impossible given the typical Paris weather.}

Much later, on page 356 of volume 2 of Almagestum novum (Riccioli 1651), we found:
\begin{quotation}
Nulla macula aut stare in Sole, aut regredi, aut praecipitato visa est hactenus post 
tot annos observationum, id est ab Anno 1611 ad 1627.
\end{quotation}
We translate this to English as follows:
\begin{quotation}
No spots were seen so far, neither standing still in the sun, 
nor coming/moving back or accelerating/falling down --
after so many years of [sunspot] observations; this is from the year 1611 
until the year 1627.\footnote{We are not sure 
whether {\em regredi} here means {\em coming back},
i.e. appearing again due to rotation of the Sun,
or {\em moving backwards}, i.e. retrograde motion of a transiting body,
probably the latter.
Also, we cannot be sure whether {\em praecipitato} means {\em accelerating}
or {\em falling down}, which could both have been considered possible at
that time (nature of spots and comets); it is unlikely that {\em ending} is meant here, because it had
often happened that a spot was seen to disappear before reaching the edge of the disk.}
\end{quotation}

The wording above {\em No spots were seen so far} is clearly restricted to those kinds
of spots which are described next ({\em standing still, coming/moving back or accelerating/falling down}),
so that this is not a generic statement about spotlessness.
We could not find any clear indication that Riccioli would have observed himself in the year 1618,
or even on which days he observed in 1618.

HS98 let Riccioli observe all days in 1618 without three gaps, when others detected spots,
probably again for consistency (HS98: {\em taken as zero spot days based upon comments by other observers}).
This is not justified from the texts by Riccioli himself.

In other parts of his Almagestum novum, Riccioli (1651) wrote about sunspots as follows:
\begin{quotation}
Numerus Macularum varius incertusque est; Aliquando tamen 50. aliquando 33. distincte 
numeratae sunt eodem tempore; sed aliquando una vel altera, et aliquando nulla.
...
Ita nimirum observarunt eas, Ingolstadii Scheinerus, et (P.) Cysatus; 
Patavii et Florentiae Vincentius Fridianus, et Galilaeus; Romae iterum Scheinerus, 
Griembergerus, Paulus Guldinus, Nicolaus Zucchius; Parmae Blancanus, 
Duaci Carolus Malapertius; in Belgio Martinus Hortensius; Bruxellis 
Daniel Antoninus; Conimbricae Gulielmus Velius; In India Orientali Gaspar Ruess.
\end{quotation}
We translate this to English:
\begin{quotation}
The number of spots varies and is uncertain: 
once 50 were seen, once 33 distinct spots at the same time; but once only one or two spots and sometimes smaller.
...
The following men have observed spots: 
In Ingolstadt Scheiner and (P.) Cysat; in Padua and Florence Vincentius Fridianus and Galilei:
in Rome again Scheiner, Grienberger, Paul Guldin, Nicolaus Zucchius; 
in Parma Blancanus; in Douai Carolus Malapertius; 
in Belgium Martinus Hortensius; in Brussels Daniel Antoninus; 
in Coimbra Gulielmus Velius; in Eastern India Gaspar Ruess.
\end{quotation}
Most of them are not listed in HS98.

\subsection{Preliminary results for 1617 to 1619}

The data in HS98 for Marius for 1617 and 1618 are not correct.
Marius clearly says that he did detect spots in those roughly 1.5 years before 1619 Apr,
but much less than before fall 1617; 
and he adds that, more often, there were no spots visible.
which never happened before
(his statement is of course limited by his small telescopes);
this constraints the active day fraction to below 0.5 (but not zero).

The period given in HS98 for Marius, 1617 Jun 7 to 1618 Dec 31, is not given by Marius himself.
That HS98 let Marius start his monitoring on 1617 Jun 7 is probably given by their
(unjustified) conclusions that 
Marius would have observed for some 1.5 years 
until the end of 1618 
and that Marius would have not observed any spot, as well as by the fact that 
Tard\'e detected spots from 1617 May 27 to Jun 6 (see Sect. 4.8). 
Also, the three monitoring breaks in 1618 given in HS98 for Marius are obviously inserted
%XXX% change
in order to be consistent with the spot detections by Malapert (and presumably Scheiner)
in those small periods of time in 1618.

While HS98 lists that Riccioli would have observed the very same 333 days as Marius,
without any spot detection, Riccioli in fact gives incomplete citations from Argoli
that he (Argoli) did not detect any spots while the comets were seen in 1618: 
Aug 25 -- Sep 25, Nov 11 -- 29 (or Dec 9, but see footnote 5), 
and since 1618 Nov 23 or 25 (see Sect. 3.4).

While Clette et al. (2014) and Svalgaard \& Schatten (2016)
(as well as Usoskin et al. (2015) specifically for Marius and Riccioli for 1617/1618)
already suggested to exclude all those data from the solar activity statistics,
where HS98 inserted zeros for many days in a row based on 
(partly and fully mis-interpreted)
generic statements,
we would like to note that omitting undated (unjustified) non-detections 
by Marius (his group numbers in HS98 are always 0) would {\em increase} the average monthly
and yearly (group) sunspot numbers, because of truly non-zero detections by others. 
However, the very important information about a decreasing trend 
given by Marius from the time before fall 1617 to those
one and a half years from fall 1617 to spring 1619 
and that the active day fraction was below 0.5 (but not zero) in that period
would not be reflected when his statements would be omitted.

\begin{table*}
\caption{{\bf Datable sunspot observations by Marius.} 
We summarize here the dates with sunspot observations as reported by Marius.
Dates are given in Julian ([J.]) or both calendars.
We also cite some other important information from his reports here.}
\begin{tabular}{lllll} \hline
Date                   & Spots   & Text (Julian dates)      & Sect. & Remarks \\ \hline
1611 Aug 3/13          & $\ge 1$ & {\em my observations ... from Aug 3, 1611}      & 4.1 & no observers in HS98 (a) \\
before 1611 Dec 29 [J.] &         & {\em spots ... observed in very large numbers}     & 4.1 & many spots \\
before 1611 Dec 29 [J.] &         & {\em always in different form since August [1611]} & 4.1 & not only round \\
1611 Oct 11/21          & $\ge 1$ & {\em implemented on Oct 11 [1611] a different way} (b) & 4.2   & Scheiner for Oct 11/21 (Fig. 3) \\
1611 Oct 3/13          & $\ge 1$ & {\em on Oct 3/13 [1611] ... invented a method} (b) & 4.2   & observed Oct 3/13\\
1611 Nov 17/27         & $\ge 1$ & {\em figure ... I had drawn on 17 Nov 1611}          & 4.4   & Scheiner (Fig. 3) \\
1612 May 30 / June 9   & 14      & {\em May 30 this year [1612], I have seen 14 spots}  & 4.6  & Jungius, Galilei (Figs. 4-6) \\
before 1612 June 30 [J.]  & & {\em see spots clearly ... including their daily motion} & 4.6 & observed often \\
1614 Feb 18/28         &    & {\em my observations ... Aug 3, 1611, to present time} & 4.1 & observed since Aug 1611 \\
before 1614 Feb 18/28  &    & {\em sunspots do not traverse the disk of the sun on the}   &   & general remark \\
                       &    & {\em ecliptic, but build an angle with it} & 4.4 & observed often \\
before 1615 Jul 4/14   & ? & {\em I have shown a figure (from 1611) to (Saxonius)} & 4.7 & see Sect. 4.8 \\
before 1618 fall comet &  & {\em tail-like longish spots on the disk of the sun}         & 3.3 & spot groups \\
before 1619 Apr        &  & {\em for one and a half year, could not find as much spots}  & 3.3 & decreasing activity \\
before 1619 Apr        &  & {\em often not even a single spot, as was never ... before}  & 3.3 & active day frac. =1 before (c) \\
before 1619 Apr        &  & {\em rather few, or more often, no spot} & 3.3 & active day fraction $<0.5$ \\
                       &  & {\em ... which was never observed before ... since 1611}   & 3.3 & active day frac. =1 before (c) \\ \hline
\end{tabular}

Notes:
$^{a}$ Ahasverus Schmidnerus also observed that day (Sect. 4.1), not listed in HS98.
$^{b}$ Change of observing technique on Oct 3/13 and/or 11/21.
$^{c}$ Namely from August 1611 until fall 1617.
\end{table*}

\section{Observations from 1611 to fall 1617}

We will now present additional texts by Marius about his sunspot
observations for the time before fall 1617, 
namely {\em 1611-1619} as indicated by Zinner (1952) by citing Zinner (1942).
Given that a few other observers are also mentioned in the texts
by Marius, we will also discuss them briefly.

In Table 1, we compile datable spot detections by Marius.
He has always observed in Ansbach near Nuremberg in Germany.

There are two East Asian naked-eye sunspot observations 
known for 1611-1616, namely for 1613 Mar 30 and 1616 Oct 10 (e.g. Vaquero 2012),
none of them close to dates discussed below.

\subsection{Marius and Schmidnerus on 1611 Aug 3/13}

In Zinner (1942), a letter from Marius to Maestlin, the teacher of Kepler,
is cited, which is dated to 1611 Dec 29 (Julian): 
\begin{quotation}
Habeo plurimum te quibus ad T. Ex. scriberem, 
utpote de illuminatione veneris et mercurii a Sole in modum lunae, 
et de Maculis in Sole, quas ab Augusto huiusque plurimas semperque diversas observavi.
\end{quotation}
We translate this to English as follows:
\begin{quotation}
I praise You most for those things about which I write to you, His Excellency,
namely the irradiation of Venus and Mercury from the Sun in the same way as the moon,
and about the spots on the sun, which I have observed in very large numbers
and always in different form since August.
\end{quotation}

Hence, Marius has observed sunspots {\em in very large numbers ... since August} 1611.
Unfortunately, 
except the fact that the spots were {\em always in different form},
he does not give exact dates 
here, but just adds that he has
to hurry with finishing the letter, because the courier is waiting and pressing. 

Marius mentions the start of his sunspot observations
in his work {\em Prognosticon for 1613} (Marius 1612),
i.e. the yearly forecast for 1613, finished and dated 1612 June 30 (Julian), he wrote
partly Latin and partly German (also in Klug 1904):
\begin{quotation}
Die maculas in sole belangt, welche von Johann Fabricio und seinem Vattern Herrn Davide Fabricio
erstlich observirt worden, die hab ich voriges Jahr 1611. im Augusto zum erstenmal gesehen,
monstrante Ahasvero Schmidnero Regiomontano Borusso, der damals mich visitiert hat. 
\end{quotation}
We translate this to English as follows:
\begin{quotation}
Regarding the spots in the sun, which were first observed by Johannn Fabricius and his father,
David Fabricius, which I have seen for the first time last year 1611 in August,
as they were shown to me by Ahasverus Schmidnerus from the Preussian K\"onigsberg,
who had visited me at that time. 
\end{quotation}

The person mentioned above, Ahasverus Schmidnerus (called David Schnidner in Klug 1904),
has shown sunspots to Marius;
see Sect. 2.2. (...)
Schmidnerus may well have known about sunspots from Johann Fabricius,
who studied at the same time in Wittenberg and had observed spots in early 1611;
the first publication about telescopic sunspots (Fabricius 1611) also was
printed in Wittenberg.
Marius knew about those early observations: he was in contact with David Fabricius,
whom he got to know during a visit to Brahe years earlier.

In his Latin book on the moons of Jupiter called {\em Mundus Iovialis} (Marius 1614),
dedication dated to 1614 Feb 18/28, 
Marius gives the exact date in the foreword ({\em Praefatio ad Candidum Lectorem});
we cite from pages 42 to 45 from the Latin-German edition by Schl\"or (1988):
\begin{quotation}
Acturus nunc eram de maculis in Sole, uti ante hac proposueram, 
quidquid etiam in eis a 3. Augusti Anno 1611. usque huc observavi manifestare.
Verum non saltem ob causas ab initio indicatas in praesenti nil de eis certo 
determinare volo nec possum, sed quia etiam Doctissimos de iis dissentire, 
et egometipse mihi satisfacere nequeam. Quare relictis iis, 
Quatuor alia nunc subjungam, de quibus in dedicationibus meis 
annuorum prognosticorum hactenus nullam feci mentionem.
\end{quotation}
This was translated to English by Prickard (1916), 
who checked and confirmed that Marius
used the Julian calendar in this work (see also footnote 2), 
as follows:
\begin{quotation}
It had been my intention, according to my former proposal,
to deal now with the spots on the Sun, setting out all my observations
upon them from August 3, 1611, to the present time.
However, I do not wish -- and, indeed, am unable -- to make any definite statement
about them at present, not only from the causes originally pointed out, but for the further
reason that I find the greatest authorities in disagreement, and am unable to
satisfy myself. I therefore pass these matters by, and will take up here
four other points not yet mentioned by me in the dedications of my yearly forecasts.
\end{quotation} 

We saw that Marius started his sunspot observations on 1611 Aug 3/13,
i.e. Aug 3 on the Julian calendar, but Aug 13 on the Gregorian calendar.
We can quite certainly assume that he started the observations with his first positive detection
of at least one spot or group.\footnote{In
the lengthy citation in Sect. 3.3, Marius said that he noticed more often spotless days in the roughly 1.5 years
prior to 1619 Apr, which was never the case in the time before those 1.5 years. If we apply
this statement (never spotless before 1617)
to his observation on 1611 Aug 3/13, the day was definitely not spotless.}
HS98 do not list any observer for 1611 Aug 3/13 (and only one other telescopic detection
before that date, Harriot on 1610 Dec 8/18, see Sect. 7).

Regarding the question, what Marius has {\em mentioned before} about sunspots 
(Mundus Iovialis, Marius 1614),
we can read a few pages earlier at the beginning of the same foreword ({\em Praefatio})
as follows, citing from the Latin-German edition by Schl\"or (1988)
(we also consulted manuscript 180.13 Quod. (3) at 
Herzog August Bibliothek in Wolfenb\"uttel,
Germany, which is available online at the Marius Portal www.simon-marius.net):
\begin{quotation}
Constitueram apud me, Candide Lector, pluribus in hac praefatione tecum agere, 
et de iis omnibus, quae hactenus per instrumentum belgicum, vulgo perspicillum vocatum, 
a me in Sole, Luna, caeterisque sideribus, atque adeo in toto coelo observata sunt, 
longam orationem instituere, prout diversis in locis hujus libelli videre licet. 
Verum cum non tantum adversa valetudo, aliaque negocia intervenientia 
a proposito me detinuerint, sed et nundinae Francofurtenses appropinquarent,
et libellus ipse jam sub praelo versaretur, promissis stare non potui, 
sed in aliud tempus hanc observationum mearum publicationem praeter 
voluntatem meam differre coactus sum.
\end{quotation} 
This was translated to English by Prickard (1916) as follows:
\begin{quotation} 
It had been my intention, Candid Reader, to deal with you at some length
in this preface, and to give a lengthened statement of all the objects which I
have observed to the present time through the Belgian instrument commonly called a spy-glass,
in the Sun, the Moon, the other stars, and in the heavens generally,
as you may see in various passages of this little book.
But, as bad 
health and interruptions caused by other business have kept
me back, and also the Frankfurt fair was close at hand, and the book
was already going through the press, I have been unable to keep
my promise, and find myself unwillingly compelled to reserve for
another time the publication of my observations. 
\end{quotation} 

To summarize, from 
all the texts cited above we can clearly conclude that Marius did detect spots since 1611 Aug 3/13, 
namely {\em in very large numbers and always in different form} 
until at least 1611 Dec 29 Julian (his letter to Maestlin),
yet, even until at least 1614 (Marius 1614),
for further observations until 1619, see Marius (1619) in Sect. 3.3.
Many spot detections in 1611 are
well possible considering the data in HS98:
Scheiner and Harriot have seen one to six spots or groups
on each of 42 different days from October to December 1611.
Furthermore, David Fabricius wrote on 1611 Dec 11 in a letter to 
Maestlin (citing from Reeves \& Van Helden 2010): 
\begin{quotation}
Indeed, this summer [1611] I often observed ten or eleven spots scattered on the Sun's disk
at one time.
\end{quotation} 
This is also fully consistent with {\em very large numbers}
reported for that time by Marius. 

It is possible that some of those spots are connected with
an aurora seen on 1611 Aug 27 in Brasso (formerly Kronstadt) in Romania (Rethly \& Berkes 1963):
\begin{quotation}
27. Augusti sind 12 gro\ss e feurige Strahlen am Himmel gesehen worden, ganze vier Stunden \"uber.
\end{quotation}
(Rethly \& Berkes 1963), which we translate to English as follows:
\begin{quotation}
On 27 August, 12 large fiery rays were seen on the sky, for four full hours.
\end{quotation}
This report fulfils two aurora criteria (Neuh\"auser \& Neuh\"auser 2015), namely
red colour and some motion ({\em fiery}), hence a {\em very possible} aurora;
Rethly \& Berkes (1963) list another, less detailed, 
report from Segesvar in Romania for the same date:
\begin{quotation}
Schreckliche Himmelszeichen waren gegen Sonnenuntergang und Mitternacht gesehen.
\end{quotation}
which we translate to English as follows:
\begin{quotation}
Horrible celestial signs were seen around sunset and midnight,
\end{quotation}
this fulfils only one criterion (night-time).
Hungary had implemented the Gregorian calendar already in 1587 (see von den Brincken 2000),
which is also used in Rethly \& Berkes (1963).
New moon was on 1611 Aug 9 and Sep 7, 
so that Aug 27 was a few days after full moon and, 
hence, dark in the first few hours of the night.

\subsection{Marius on 1611 Oct 3/13 and/or 11/21}

In his work {\em Prognosticon for 1613}, cited above, Marius (1612) continues
(again partly Latin and partly German, also in Klug 1904):
\begin{quotation}
Als mir aber solcher 
modus nicht genug gethan, nemlich durch den radium obscura camera acceptum, 
adhibitio instrumento belgico, als hab ich den 11. October 
einen anderen Weg erdacht, dass ich die Sonnen durch das benannte Instrument 
ohn alle verletzung dess Gesichts bey hellem Himmel sehen, vnnd die maculas 
gar distincte, sampt jhrem t\"aglichen motu observirn kan. 
Aber hiervon zu anderer zeit mehr. 
\end{quotation}
We translate this to English as follows:
\begin{quotation}
When that [original] way of observing them [spots] was not
sufficient any more for me, namely through the light ray in a dark room
[Camera Helioscopica] by using the Belgian instrument, 
I have thought and implemented on Oct 11 [1611] a different way, 
so that I could see the sun and its spots clearly 
through the mentioned instrument during the bright day without harm for the face,
including their daily motion. But later more about this.
\end{quotation}
Here, we see that Marius observed regularly and on subsequent days ({\em daily motion}).
The text says that Marius already used the Camera Helioscopica before 1611 Oct 11/21,
the improvement on Oct 11/21 was probably a better way to 
see {\em spots clearly} and their {\em daily motion};
he explains the projection method with the telescope to a white screen in a dark room
in more detail in the foreword of Mundus Iovialis (Marius 1614);
the improvement may have been regarding the placement of the white screen perpendicular
to the telescope ({\em spots clearly}) and then by drawing the spots day to day 
({\em daily motion}) onto some paper.

\begin{figure*}
\begin{center}
{\includegraphics[angle=0,width=12cm]{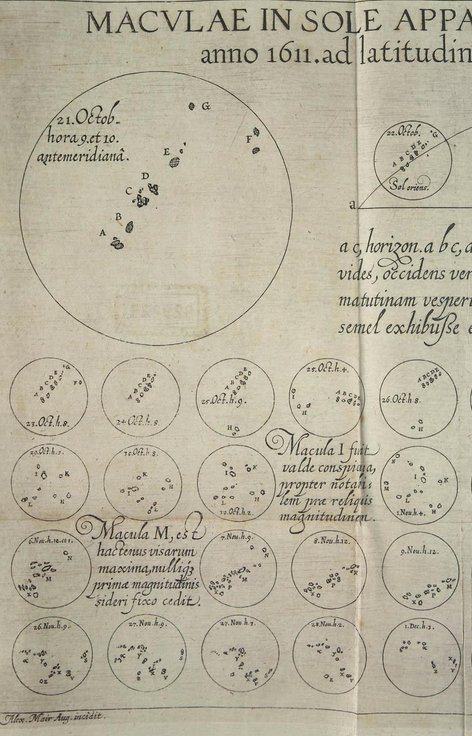}}
\end{center}
\caption{These drawings of sunspots are from Scheiner (1612) from 1611 Oct 21 to Dec 2 
(all his dates are Gregorian).
The drawing for 1611 Oct 21 is shown in the upper left; Marius has changed his sunspot
observing technique on this date, i.e. Oct 11 Julian (and/or on Oct 3/13), 
so that he may have seen the spots as drawn here (by Scheiner).
Also Tanner had observed spots on 1611 Oct 21 (see Sect. 4.5).
Marius has then drawn himself the spots as seen on 1611 Nov 17/27, as was also done by Scheiner.
Note the large range of heliographic latitudes of the spots seen on Nov 27 (the 2nd and 3rd
drawing in the bottom with text inside being {\em 27 Nov h 9} for 9h in the morning and 
{\em 27 Nov h 3} for 3h in the afternoon, spots are labbeled with letters),
the day when also Marius observed spots and produced a drawing of the spots (considered lost).
Marius discussed his drawing for that date with Petrus Saxo (Saxonius) on 1615 Jul 4/14
(see Sect. 4.7).
These drawings were first published as cupper plate in Jan 1612, then also appeared in 
{\em Apelles} (Scheiner 1612), from where we took them.
Marius did not know these drawings by 1615 (Sect. 4.2).
}
\end{figure*}

Since Marius states that he changed his observing technique on 1611 Oct 11/21,
he detected at least one spot group on that day, too
(we again apply the consideration in footnote 12).

After the appearance of {\em Mundus Iovialis}, whose main part was finished and published in 1614
(cited here as Marius 1614), Marius wrote an appendix or addendum {\em Ad candidum lectorem}, dated 1615,
in Latin and translated to German recently by Gaab \& Leich (2014),
where we can read (Latin taken from manuscript 180.13 Quod. (3) at Herzog August Bibliothek in Wolfenb\"uttel,
Germany, which is available online at the Marius Portal www.simon-marius.net):
\begin{quotation}
(Folio G4v)
Hoc saltem addo et sancte affirmo me praeter nuncium siderium, nihil habere, 
a Galilaeo nec etiam legisse. Quinimo nec etiam Apellis librum nancisci hactenus potui, 
nescio quo fato hoc acciderit, cum tamen diligentissime Noribergae de eo inquisierim.
Primi inventores et observatores macularum solarium sunt duo Fabricii Pater et Filius, 
verum quia haeretici putantur, nomina illorum supprimuntur.

(Folio H1r)
Modum observandi colores Astrorum adjnveni Anno 1611. 
Sicut et eodem anno excogitavi rationem 3./13. Octobris, 
per tubum observandi maculas solares in ipso sole, 
absque ullo damno oculorum; Sicut et id ipsum, quod maculae solares 
non ad Eclipticae ductum, discum Solis transeant, sed angulum cum ea faciant, ...
\end{quotation}
Note that Marius really gave both the Julian and the corresponding Gregorian date
in this text, as clearly seen in the manuscript. Marius also gave both dates in his
yearly Calendars, where he listed each day with calculated astronomical 
and expected astrological events.

We translate this to English as follows (considering the German translation by Gaab \& Leich 2014):
\begin{quotation}
[page G4v] I add at least this and confirm it with holy emphasis that I do not poses
anything else from Galilei than Sidereus Nuncius and that I also did not read anything else.
Also, I could not yet get hold of the book by Apelles [by Scheiner as pseudonymous author on spots]; 
I do not know why,
even though I have searched for it carefully in Nuremberg. The first discoverers and observers
of sunspots are the two Fabricius, father and son, but because they are considered heretics,
their names are not cited.

[page H1r] In the year 1611, I have found a method to observe the colours of the stars.
Also in the same year on October 3/13, I have invented a method to observe sunspots on
the sun itself through a tube, without any harm for the eye; in addition, [I add] that sunspots
do not traverse the disk of the sun along the ecliptic, but that they
build an angle with it, ...
\end{quotation}

With the {\em tube} Marius may have meant what he described in the foreword to
{\em Mundus Iovialis} (Marius 1614), namely that he observed the sun and its spots
with the naked eye at low altitude (hence, {\em without any harm for the eye})
through some {\em tube}: {\em if the sun stood low, I used a black paper arranged 
as narrow tube, whose narrow opening hole
was put to the eye, but its wider opening hole towards the Sun} (Marius 1614).

Marius mentioned two slightly different dates for implementing new observing techniques,
namely 1611 Oct 11/21 in Marius (1612) and 1611 Oct 3/13 in the 
1615 appendix to Mundus Iovialis (Marius 1614).
Either the two different dates indicate two steps in the implementation of the
new observing technique, or one of them is given by mistake.
Hence, Marius has observed sunspots on 1611 Oct 3/13 and/or 11/21.
In Oct 1611, also Scheiner observed regularly, but not daily:
on 1611 Oct 21 (Gregorian), he detected four groups (HS98),
but there were no observations on 1611 Oct 13 (Gregorian) according to HS98.
We can see in Fig. 3 that Scheiner marked seven spots or groups (A to G) on Oct 11/21,
which we think form at least five groups, but the grouping of spots is somewhat subjective.

Scheiner's drawing 
of the spots of 1611 Oct 21 appeared in {\em Apelles}, where it is the drawing with the
earliest date (see Fig. 3);
Marius mentioned above that he did not had available a copy of {\em Apelles}
at the time of writing the {\em Mundus Iovalis} appendix in 1615 --
it may well be that Marius did not know at this time (1615) that Scheiner was
the author of those letters of {\em Apelles}, see also Sect. 4.7.

There may have been an aurora seen in Europe in 1611 October (Link 1964),
but we could not check the textual description.

In the 1615 appendix to {\em Mundus Iovialis}, Marius gives an important result
from his observations: {\em sunspots do not traverse the disk of the sun along the ecliptic, 
but ... they build an angle with it.}
Both for spots moving on the solar surface and especially for small solar system bodies
transiting the sun, it might have been expected at that time that they would
traverse the disk of the sun on the ecliptic 
(or maybe parallel to it according to Tard\'e, see Sect. 4.8).
To notice the inclined path of the spots, it may have been neccessary to
draw a spot day to day into the same drawing with, e.g., the Camera Helioscopica.
The statement that the spots (observed almost daily, see above) 
form an angle with the ecliptic includes the notion that 
the solar equator is inclined to the ecliptic.
The amount of this effect, or whether it changes with time, e.g. within a year,
is not reported by Marius.

Marius gives some more details about his solar observations in 
the foreword of {\em Mundus Iovialis}, Marius (1614), 
continuing the citation from Sect. 4.1:
\begin{quotation}
Quartum est, peculiaris quaedam observatio in Sole, praeter maculas, 
de qua inter me et Dominum Davidem Fabricium Theologum in Frisia orientali, 
et Astronomum excellentissimum Amicum meum singularem, per literas aliquoties disceptatum est. ...
Interdum enim quasi stare videtur radius, quoad motu illum, qui alias diurnus vocatur, 
interdum vero quasi in momento saltu quodam facto in consequentia ferri. 
Eidem motui inaequali etiam obnoxiae sunt maculae Solares.
...
Hic igitur motus aut inest Soli, aut terrae, aut denique aeri. 
Ab aere existere non posse puto, ...
\end{quotation}
We translate this to English as follows:
\begin{quotation}
The forth observation is a very special one on the sun in connection with the spots; I and Mr. David Fabricius,
a theology scholar from eastern Frisia, a very excellent astronomer and my dearest friend, have written 
a few times about them. ...  
Sometimes, the ray [from the sun]
seems to stand almost still in its motion, which is otherwise the usual daily motion. But sometimes
it [the ray] seems to move further like jumping. The same uneven motion also applied to sunspots.
... This motion either originates from the sun or from the Earth or from the air.
I think it is not due to the air, ...
\end{quotation}

Hence, Marius seems to have noticed effects of the sun and its spots,
which we now call {\em seeing};
Marius (1614) also mentioned that David Fabricius considered the air to cause this effect,
which indeed is correct.
This effect was also reported by the Chinese naked-eye observers, e.g.
{\em several black spots rocking to and fro} for 1617 Jan 11 (more in Sect. 3.2).

\subsection{Excursus: David and Johann Fabricius in March 1611 in Dornum}

Given that Marius mentioned the observations by Fabricius above,
see Sects. 4.1 and 4.2,
we will now discuss them briefly.

David and Johann Faber (or Goldschmidt or Goldsmid, Lat.: Fabricius) from Dornum in East Friesland,
north-western Germany, had detected their first spot on 1611 Feb 27 and 28 (Julian),
hence 1611 Mar 9 and 10 (Gregorian), as published by Johann Fabricius (1611),
the first publication on telescopic sunspots (Fabricius 1611) --
published there on folios C2v-C4r, as cited by, e.g., Casanovas (1997), 
Vaquero \& Vazquez (2009), or Reeves \& Van Helden 2010).
The exact observing dates were not given in Fabricius (1611),
but were mentioned in the {\em Prognosticon for 1615} by David Fabricius, 
cited in Reeves \& Van Helden (2010):
\begin{quotation}
... spots ... that are at present found and observed on the Sun,
such as were observed for the very first time, in my presence, by my son
Johannes Fabricius, a medical student, in the year 1611, on 27 Feb, old style,
through the Dutch spectacles.
\end{quotation} 
They speak for those first two days clearly about only one spot (or group):
\begin{quotation}
We then saw the spot more distinctly and certainly.
\end{quotation}
It is clear from Fabricius (1611) that they observed one spot 
on two subsequent days in early 1611, to be dated 1611 Feb 27 and 28 (Julian),
hence 1611 Mar 9 and 10 (Gregorian) according to the {\em Prognosticon for 1615}. 
Casanovas (1997) add: {\em The next day, they saw with great
pleasure and with excitement that the spot had moved a little from east to west}.
Casanovas (1997) continue quoting that {\em after a few days, a second spot appeared
at the limb; then a third one, which they could follow until they disappeared on the
west limb and reappeared on the east}, citing and summarizing Fabricius (1611).

From those observational data, we can almost certainly conclude that the first three spots 
or groups were
all seen in March 1611 (Gregorian), while the spot that, 
as reported, 
{\em reappeared on the east} 
was probably already in April 1611. For March 1611, the reported daily spot (or group ?) 
number was 1 to 3. There was no mention about spotless days.

It is interesting to note that there was an aurora sighting on the days of the
sunspot observations by father and son Fabricius (1611 Feb 27 \& 28, Julian), 
namely on 1611 Mar 10 (Gregorian), as far south as Korea:
\begin{quotation}
During the first watch of the night, in the three directions E, W, and N
there were scarlet vapours [qi], five of which were shaped like torches.
After a long time they were extinguished.
\end{quotation}
as given in Xu et al. (2000), with a very similar translation in Yau et al. (1995)
with {\em red} instead of {\em scarlet}, also listed as {\em R} for {\em red} in Lee et al. (2004);
this description fulfils three aurorae criteria (Neuh\"auser \& Neuh\"auser 2015),
northern direction, night-time, and red color, i.e. a {\em probable} aurora.
The telescopic sunspots detected by Fabricius (1611) are not listed in HS98.

\subsection{Marius on 1611 Nov 17/27}

Directly after the above quotation (Sect. 4.2) from Marius in his 1615 appendix to {\em Mundus Iovialis},
where he listed various observations and inventions, 
we can read (Latin taken from manuscript 180.13 Quod. (3) at Herzog August Bibliothek in Wolfenb\"uttel,
Germany, which is available online at the Marius Portal www.simon-marius.net):
\begin{quotation}
(Folio H1r) 
..., prout etiam figuram die 17./27. Novembris Anni 1611. delineatam priusdicto Holsato monstravi, 
qui cum admiratione illam intuitus est, et addidit hoc secreto sibi concreditum esse a Scheinero.
\end{quotation}
We translate this to English as follows (considering the German translation by Gaab \& Leich 2014):
\begin{quotation}
[page H1r] ..., and that I have shown a
figure, which I had drawn on the 17th/27th day of November of the year 1611, to the 
previously mentioned Holsteinian, who looked at it with admiration and
added that this would have been shared with him in secret by Scheiner.
\end{quotation}

It is known that the person from Holstein (Saxonius) had visited Marius on 1615 Jul 4/14 (Gaab \& Leich 2014). 
We will discuss his possible observation on 1615 Jul 4/14 with the visitor from {\em Holstein} in Sect. 4.7 below.

In the 1615 appendix to {\em Mundus Iovialis},
Marius clearly states that he has drawn sunspots for 1611 Nov 17/27 and that he
has shown the drawing to the visitor from Holstein (Saxonius), who was in contact with Scheiner and who
told him (Marius) that he (Saxonius) had seen such (a) drawing(s) from Scheiner.
It seems that Saxonius did not mention the {\em Apelles}, where such drawings
were published (pseudonymous by Scheiner).
As mentioned in Sect. 4.2, Marius did not have available a copy of {\em Apelles}
with the drawings from Nov 17/27 by now (1615).
Unfortunately, the drawing from Marius has not been found, yet.
We can judge what Marius has seen and drawn that day,
namely from the drawings by Scheiner on this very date, see Fig. 3.

It is interesting to note that there was a possible aurora sighting two days after
the sunspot drawing (1611 Nov 17/27) as far south as Korea, namely on
1611 Nov 29 (Gregorian), for which Lee et al. (2004) list a {\em red} ({\em R}) 
aurora.\footnote{In occidental records, 
words like {\em fire} or {\em burning} indicate both red colour and some (apparent) motion
for aurorae, {\em flame(s)} do(es) not necessarily indicate red colour;
in oriental reports, the phrases {\em like fire} and {\em like flame(s)} do not mean red colour nor motion;
most oriental reports are written by astronomers, they mention explicitly colour and changes,
see Neuh\"auser \& Neuh\"auser (2015) and Chapman et al. (2015) for more discussion about aurora criteria.
For the possible aurora on 1611 Nov 29, we know only that Lee et al. (2004) describe them as {\em red},
but we cannot exclude that the original text reports only {\em flames}, i.e. no colour.}

Why did Marius show a figure with sunspot(s) drawn by him to Saxonius from Holstein?
One could imagine that Marius and his visitor would have at least tried
to observe spots together during the visit. If the weather was too overcast on that or those day(s),
then he would instead have shown a drawing to him.
On the other hand, it is also well possible that the drawing shows a particularly large
number of spots and/or spots of unusual form and/or on unusual location(s), 
so that Marius did show it to his visitor anyway 
(possibly in addition to the collaborative observation that day).
Two drawings of sunspots by Scheiner for that day, 1611 Nov 17/27, shown here in Fig. 3,
indicate a special situation around that day:
Many spots are distributed over a large range of heliographic latitudes,
about half of them near the equator and all others only on one hemisphere.
This particular sunspot distribution may have been discussed in connection
to the nature of spots, as such a large range of heliographic latitudes may not
be consistent with one of the theories discussed
(namely that they are transits of small solar system bodies).
(See also Sect. 4.7 on Marius and Saxonius.)

It is well possible that Marius often produced drawings
(given that the technique was available to him, namely the Camera Helioscopica,
see Marius 1614):
he observed the {\em daily motion} of spots (Marius 1612) 
and that their path is inclined to the ecliptic (in the 1615 appendix to Marius 1614).

\begin{figure*}
\begin{center}
{\includegraphics[angle=0,width=16cm]{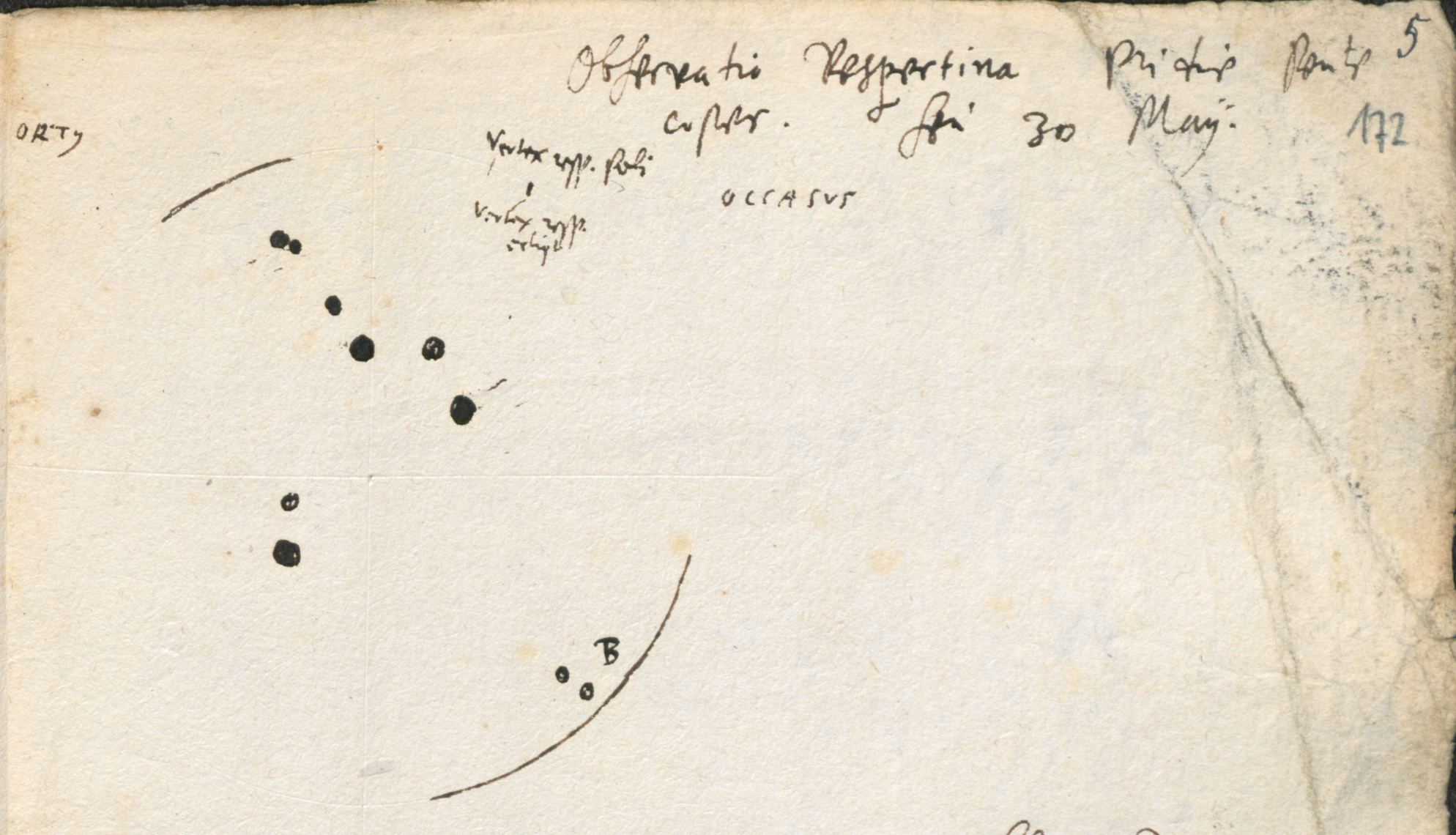}}
\end{center}
\caption{This drawing of sunspots is from Jungius from Giessen, Germany, for 1612 May 30 (Julian),
i.e. June 9 (Gregorian). We can clearly see ten spots in five pairs or groups (HS98 gave 5 groups).
This drawing is taken from the observational log {\em Maculae Solares 1612/13} by
Jungius, folio/page 172, obtained in digital form from the University library of U Hamburg.
The text written by Jungius says: upper right {\em Observatio Vespertina pridie Pente/ costen. seu 30. Maii},
i.e. {\em Evening observation on the day before pentecost. or 30 May};
in 1612, May 30 [Julian] was indeed the day before pentecost sunday;
slightly to the upper right of the solar disk there are two small tick marks or arrows
with the following caption: {\em Vertex resp. Poli}, i.e. direction towards {\em celestial pole},
and below of it {\em Vertex resp. eclipsis}, i.e. direction towards the {\em ecliptic pole};
to the very left, we can read {\em ortus} for {\em east} and to the upper right
of the solar disk {\em occasus}
for {\em west}; one of the spots (or a group or pair) is labelled {\em B};
the numbers to the upper right ({\em 5} and {\em 172}) are page or folio numbers.
Marius reported to have seen 14 spots that day, i.e. more than Jungius.}
\end{figure*}

\subsection{Excursus: Scheiner, Cysat, Tanner, and other Jesuits in March 1611 and since October 1611 in Ingolstadt}

Given that we compare the observations by Marius (Sects. 4.2 \& 4.4)
with observations by Scheiner in 1611,
we will now discuss him briefly.

Scheiner and his assistant Cysat saw sunspots in March 1611.
Casanovas (1997) wrote: {\em One day in March 1611, he [Scheiner] and
his assistant observer Johann B. Cysat were looking at the Sun through the smoke
rising from the university's tower which had caught fire, both saw spots in the Sun.}

While we did not find a clear indication of the source for this statement in Casanovas (1997),
in particular for the {\em fire}, there is a similar statement by Scheiner in his
{\em Rosa Ursina} (Scheiner 1626-30), which was partly also quoted by Reeves \& Van Helden (2010).
Scheiner wrote in the preface {\em Ad Lectorem} on the 2nd (unnumbered) page as follows:
\begin{quotation}
Anno igitur 1611 in Universitate Ingolstadiana Matheseos Professor Mense Martio, conscensa Templi nostri turri, 
Telioscopio per Nebulam moderatam in Solem proportionate hebetatum directo, non ex ullo rumore praevio, 
sed Solis explorandi studio spontaneo ductus, Maculas solares prima vice deprehendi, socio Io. Baptista Cysato, 
Theologiae tunc studioso, qui ex illo tempore me pro vitris coloratis parandis vehementer incitavit, quod 
dum perago, mense Octobri anni eiusdem supervenerunt iterum more suo aliqui dies nebulis temperatis respersi, 
quibus ad observandum Solem invitantibus Tubum opticum in eundem ex meo cubiculo direxi (quemadmodum 
lib. I. pag. 63. ex Apellis Tabula expressum habes) die videlicet 21. hora Astronomica 21. Germanica seu a media 
nocte 9. antemeridiana sive matutina, in eoque Maculas secundum vidi, multisque aliis Patribus, 
et studiosis ad horam 10. usque ostendis circa quod idem tempus, P. Adamus Tanner earundem Macularum aspectu primo 
e suo cubiculo potitus est, (id quod tom. I. Disp. 6. de Creat. Mundi quaest. 3. paragraph 5. 
de Maculis solaribus, num. 69. ipse enarrat) ...
\end{quotation} 
We translate this to English as follows (partly following Reeves \& Van Helden 2010): 
\begin{quotation}
In the year 1611, when I was professor for mathematics at U Ingolstadt, in the month of March,
having ascended the tower of our church, 
and having directed the telescope through moderate clouds [Lat. nebulam: clouds or nebulae or fog] to the Sun,
led not by an advance rumour but by free desire
to investigate the Sun, I saw sunspots for the first time with my associate Joh. Baptist Cysat,
then a student of theology, who had strongly urged me at that time to obtain coloured glasses,
which I am still doing, and in the month of October of the same year, there were again, as usual, 
certain days with temperate clouds [Lat. nebulis: clouds or nebulae or fog], 
and after we had invited a few people for observations of the Sun,
I directed an optical tube from my small room to it [the Sun] (as you can find on page 63 of book I 
from Apelles), it was during the day in the 21st astronomical and 21st German hour, or 
in the 9th hour after midnight before noon [between 8 and 9h local time in the morning], 
and in this [hour] I have seen for the second time
the spots, and while we have shown them to many other [Jesuit] fathers and students until the
10th hour, it was around the same time that Father Adam Tanner for the first time enjoyed the
view of these spots from his small room (this is what he himself reports in {\em Tomus I Disput. 
6 de Creatione Mundi quaest. 3, paragraph 5} about the sunspots).
\end{quotation}

Hence, as we can see, there is no mention of a fire, but clearly of normal moderate clouds,
and also no mention of a university tower, but a church tower (in contrast to the
quotation from Casanovas 1997).
Scheiner (1626-30) mentioned that he has shown the spots in October 1611 to many other Jesuits and students,
and he also mentions that Adam Tanner (also a Jesuit) observed and detected spots. 
The drawing of sunspots for 1611 Oct 21 is the first drawing shown in Scheiner's Apelles
(our Fig. 3), so he started his monitoring campaign on that day.

The observations reported for March 1611 by Scheiner and the two Fabricius are
fully consistent with each other. Scheiner uses the plural for spots for both March and October 1611.
Given their reports, there is no evidence for spotless days in those months.
As noticed above, there was an aurora sighting on 1611 Mar 10 (Gregorian), as far south as Korea.

Given that Scheiner mentioned Tanner for observations in 1611,
we will also discuss him briefly;
see Sect. 2.2. (...)
Tanner is an additional, new sunspot observer for October 1611, not listed in HS98.

In {\em Rosa Ursina}, Scheiner (1626-30) mentioned that Tanner would have written about sunspots in the part 
given as {\em Disp. VI. De Creatione Mundi, Quaest. III. Dub. III., paragraph V.}, which can
be found in the work by Tanner (1626) called {\em Universa Theologia Scholastica},
where we can read as follows on page 1726:
\begin{quotation}
Igitur cum eius rei incertus quidam rumor iam antea aliunde ad nos Ingolstadium fuisset allatus, 
accidit Anno MDCXI, die 21. Octob. hora fere dimidia ante decimam nostram antemeridianam, hoc est, 
duabus horis cum dimidia antemeridiem, ut aptissima eius rei explorandum se offerret occasio. 
Cum nam eo tempore coelum tenui quadam nebula ita obductum esset, ut tamen Solis discus, 
radiis vehementioribus undique recisis, plane ac sincere spectabilis esset, admoto tubo optico, non 
difficulter maculas quaesitas in eo deprehendo. Et mox admiratione simul et hilaritate perfusus, ad 
insolitum spectaculum plures advoco et invito; qui pari affectu diu multumque una mecum rem 
eandem contemplati sunt. Inde vero totius anni sequentis curriculo, variis modis, ac praesertim 
directa per tubum inspectione (quae ad physicam considerationem plus habere videbatur momenti) pene quotidie eas observavi.
\end{quotation}
We translate this to English as follows:
\begin{quotation}
After a certain dubious rumour about this matter [sunspot observations?]
had arrived here with us in Ingolstadt from elsewhere,
it happened in the year 1611 on 21 October, about half an hour before the tenth hour in the morning,
that is two and a half hours before noon [around 9:30h local time in the morning], 
that there was a very useful moment, to investigate the matter.
Namely, at that time the sky became covered by a thin clouds [Lat. nebula: clouds or nebulae or fog] 
in a way, so that the disk of the sun was clearly
and well visible with less strong rays on all sides, and I discovered without difficulty the wanted spots on it
after I had directed an optical tube to it. ... And, indeed, from this time on in the course of the following year,
I have almost daily observed them in different ways, but mostly through the directed tube (which appeared to
be more important for the physical consideration).
\end{quotation}

Hence, Tanner says that he observed sunspots (plural, hence more than one spot, i.e. one or more groups) 
on 1611 Oct 21 (Gregorian date, as he was a Jesuit).
Tanner then observed spots almost every day
in the rest of the year 1611 and in the following year 1612.
The statement by Tanner ({\em I have almost daily observed}) is not restricted to
fully clear days, but Tanner and Scheiner (also Harriot, see Sect. 7.1) tell us that they did observe the sun
well when there were thin clouds, which are not atypical for the winter half year
October to March in Ingolstadt and elsewhere in Germany (and England).
From his report, there is no evidence for spotless days.
According to HS98, there would be 242 active and 10 inactive days in 1612.
His observation for 1611 Oct 21 is consistent with Scheiner (Fig. 3).
The fact that Tanner detected spots on 1611 Oct 21 
is also mentioned in Sharratt (1996) and Reeves \& Van Helden (2010), but not in HS98.

We would like to note that Riccioli also mentioned the early observations by Scheiner,
as he wrote in his Almagestum novum (Riccioli 1651):
\begin{quotation}
P. Christophorus Scheinerus e Soc. IESU, qui anno 1611. Ingolstadii mense Maio, per occasionem rimandi Solis apparentem diametrum ope Telescopii 
corpuscula haec animadvertit, eaque coram P P. Iacobo Gretsero, Adamo Tannero ...
\end{quotation}
We translate this to Emglish:
\begin{quotation}
Pater Christopherus Scheiner from the Jesuit community, who had in the year 1611 in the month of May the opportunity to
investigate the visible disk [diametrum] of the sun and noticed the spots [Lat.: {\em corpuscula} literary meaning
{\em small bodies}] with the help of the 
telescope in the presence of Pater Jacob Gretser and Pater Adam Tanner ...
\end{quotation}
Hence, Riccioli mentioned an additional early sunspot observer, Jesuit Pater Jacob Gretser 
(born 1578 in Markdorf, died in 1625 in Ingolstadt, both Germany);
what is dated to May 1611 in Riccioli 
may well be the observations of March and/or October 1611 reported above by Scheiner himself.

\subsection{Marius on 1612 May 30 (Julian): 14 spots}

In his {\em Prognosticon for 1613}, finished and dated to 1612 June 30 (Julian), 
Marius wrote, again partly Latin and partly German
(also in Klug 1904):
\begin{quotation} 
Den 30. May diss Jahrs, hab ich 14. solcher auff einmal gesehen. 
Es seyn aber nicht in ipso corpore solari, sondern seyn corpora, quae circa Solem feruntur. 
\end{quotation}
We translate this to English as follows:
\begin{quotation}
On May 30 [Julian] of this year [1612], I have seen 14 such [spots] at once.
They were [would be], however, not on the solar body themself, 
but they were [would be] bodies orbiting the sun. 
\end{quotation}

Given that the Prognosticon for 1613 was dated to 1612 June 30 [Julian],
this text about those 14 sunspots on {\em May 30} (Julian)
were clearly observed on 1612 Jun 9 (Gregorian),
i.e. only shortly before the text was written.

On that very same day, also Galilei and Jungius observed:
Galilei had seen seven to ten spot groups (HS98), his largest daily number,
and Jungius in Giessen had seen five groups (HS98).
Those large numbers appear to be consistent with Marius giving {\em 14 spots}.
It is of course somewhat subjective how many groups there are and how many
individual spots are present inside or outside of groups.
We show the drawings by both Galilei and Jungius in Figs. 4-6, respectively.
Marius has seen those spots in this way (same day) or slightly different (different instrumentation).

The drawing by Jungius shows five spot pairs or groups with a total of ten spots,
the one by Galilei shows seven to nine groups, partly resolved into smaller structures
with a total of some 25 to 30 spots. 
In the drawing by Jungius, only the smallest spots seen by Galilei are missing.

According to HS98, also Harriot would have observed spots that day,
HS98 gave five groups on his drawing.
However, the catalogue of the drawings of Harriot (digilib.mpiwg-berlin.mpg.de)
does not contain a drawing for 1612 May 30 (Julian), June 9 (Gregorian);
see Table 3 for the corrected group sunspot numbers for those days.
Harriot did observe on both the day before and after that date and had detected
basically the same five groups as Jungius on May 30 and 31 (Julian),
but a few more spots (13 or 14 for Harriot, 10 for Jungius).
Galilei saw additional smaller spots, both inside the groups detected
also by the others, but also a few more weak groups with two weak spots each.
Marius reported 14 spots for May 30 / June 9. 

It is well possible and understandable that several European observers monitored the Sun
closely those days because of the solar eclipse visible in Europe on 1612 May 30 (Gregorian).
According to HS98, Galilei, Harriot and Jungius reported sunspots for that day
(Table 3).

It is quite obvious that Marius gave this particular day (1612 May 30/June 9) as example,
because he never had 
seen so many spots 
on any others days until the date of this
statement (1612 June 30/July 10).
We can then assume 13 spots/groups as upper limit for Marius for the time before 1612 June 30/July 10.
For 1612 May 30 / June 9, also Galilei reported his largest spot/group number
for this period.

There are no naked-eye sunspots known for 1612 May/June (e.g. Vaquero 2012), 
but from AD 1612 Aug 19-21:
Galilei saw sunspots with both the telescope and the unaided eye,
and he has drawn the telescopic sunspots for 1612 Aug 19 (Vaquero 2004).
Tanner (1626) reported that he has detected spots {\em almost daily} in 1612
from Ingolstadt, see Sect 4.5.

There was an aurora observation on 1612 Aug 4 in Kolozsvar (formerly Klausenburg) in Romania:
\begin{quotation}
Es wurde diese Nacht ein gro\ss es Himmels\-wunder im Norden gesehen.
\end{quotation}
which we translate to English as follows:
\begin{quotation}
This night, a large celestial wonder was seen in the north.
\end{quotation}
This fulfils two criteria (Neuh\"auser \& Neuh\"auser 2015),
namely night-time and northern direction.
There were also some aurora observations on 1612 Aug 6 (greg.) in Z\"urich, Switzerland 
({\em Streitendes Heer am Himmel} (Fritz 1873), i.e. {\em fighting war army on sky}),
which could have been an aurora
(after new moon on 1612 Jul 28, it was partly moon-less in the nights of Aug 4 and 6).
There were more aurora observations on Aug 28 in Kolozsvar (formerly Klausenburg) in Romania, 
(Rethly \& Berkes 1963): 
\begin{quotation}
Gro\ss e Himmelswunder entstanden die ganze Nacht hindurch am n\"ordlichen Himmel,
\end{quotation}
which we translate to English as follows:
\begin{quotation}
Large celestial wonders formed the whole night through on the northern sky.
\end{quotation}
i.e. at night and in the north, a {\em very possible} aurora,
also reported from some other locations (Rethly \& Berkes 1963);
indeed, it was a dark night: new moon was on 1612 Aug 26.

\subsection{Marius and Saxonius on 1615 Jul 4/14?}

We had read above (Sect. 4.4) from Marius written 1615 in the appendix to {\em Mundus Iovialis}
(see Marius 1614):
\begin{quotation}
..., and that I have shown a
figure, which I had drawn on the 17th/27th day of November of the year 1611, to the 
previously mentioned Holsteinian, who looked at it with admiration and
added that this would have been shared with him in secret by Scheiner.
\end{quotation}

Regarding the person from Holstein, Marius had mentioned before
(our translation to English):
\begin{quotation}
Namely on [1615] July 4/14 there was a highly educated man here, Mr. Petrus Saxo from Holstein, 
student of mathematics,
who undertook a travel from Ingolstadt [southern Germany] from Scheiner directly to me.
\end{quotation}

Petrus Saxonius (1591-1625) was from Husum in northern Germany; 
he was travelling in southern Germany in 1614, also visiting Scheiner in Ingolstadt;
according to HS98, he had observed sunspots in Feb and Mar 1616, but see next Section;
since September 1617, he was professor for mathematics in Altdorf (Gaab 2011).
Petrus Saxonius visited Marius on (or since) 1615 Jul 4/14.
It is quite likely that Marius and Saxonius observed sunspots together that day,
but we have no firm statement about it.

Tard\'e (1620) has drawn some 30 spots for 1615 Aug 25 (image reprinted in 
Vaquero \& Vazquez 2009),\footnote{In Tard\'e (1620), page 23 (figure caption) \& 24 (drawing), 
it is clearly stated that the observation was done on 1615 Aug 25 (which is a Gregorian date,
since Tard\'e was a Catholic priest in France), this date was also given in Baumgartner (1987),
while HS98 listed ten groups for Mar 25 and five groups for Aug 15,
possibly partly following Wolf (1859) -- none of these numbers can be found in Tard\'e (1620).
On the drawing, one can identify 30 spots in some 7 to 9 groups with one or more spots each
for 1615 Aug 25 (one group being very large). Tard\'e himself remarked to have seen
30 spots clearly resolved from each other.}
Since they (also) are spread over a large range of heliographic latitudes,
a similar pattern one month earlier during the visit of Saxonius to Marius
may have motivated their discussion of the nature of spots given their
large range of heliographic latitudes, so that Marius has shown him another
example (1611 Nov 17/27).

There are no naked-eye sunspots known for 1615 (e.g. Vaquero 2012).

\subsection{Saxonius in Altdorf and Tard\'e in Sarlat in Feb/Mar 1616}

We discuss Saxonius here, because he visited Marius in 1615 and is listed
as observer for 1616 in HS98 (together with Tard\'e).
According to both Wolf (1857) and HS98, Petrus Saxonius would have observed sunspots on
1616 Feb 24 \& 26 as well as Mar 4, 6-9, 11, 12, 14, 16, and 17 (with 2-8 groups per day),
see Figs. 7-9.
There is some overlap only with Tard\'e, who would have recorded one spot or group 1616 Mar 3-14
(HS98), see Figs. 9 \& 10.

These observations by Saxonius
are based on the reproduction or copy on 
a copper plate
based on drawings by Saxonius, available in the
Germanisches Nationalmuseum (German National Museum) in Nuremberg,
from where we obtained a digital copy.
The caption specifies that {\em sunspots [were] observed by Petrus Saxonius 
from Holstein at the academy of Altdorf} near Nuremberg (Fig. 7). 

The dates given by Wolf (1857) and HS98 are those on the drawings.
In HS98, all dates are supposedly Gregorian.

Since Saxonius was from protestant northern Germany, 
where his father worked as protestant pastor,
and since Saxonius now worked and published at U Altdorf in the protestant area of Nuremberg,
he used the old Julian calendar in his writings, as did Marius.
Hence, the dates of his drawings have to be considered Julian dates.
Then, the dates in Wolf (1857) and HS98 are off by 10 days.

\begin{figure}[t!]
\begin{center}
{\includegraphics[angle=0,width=8cm]{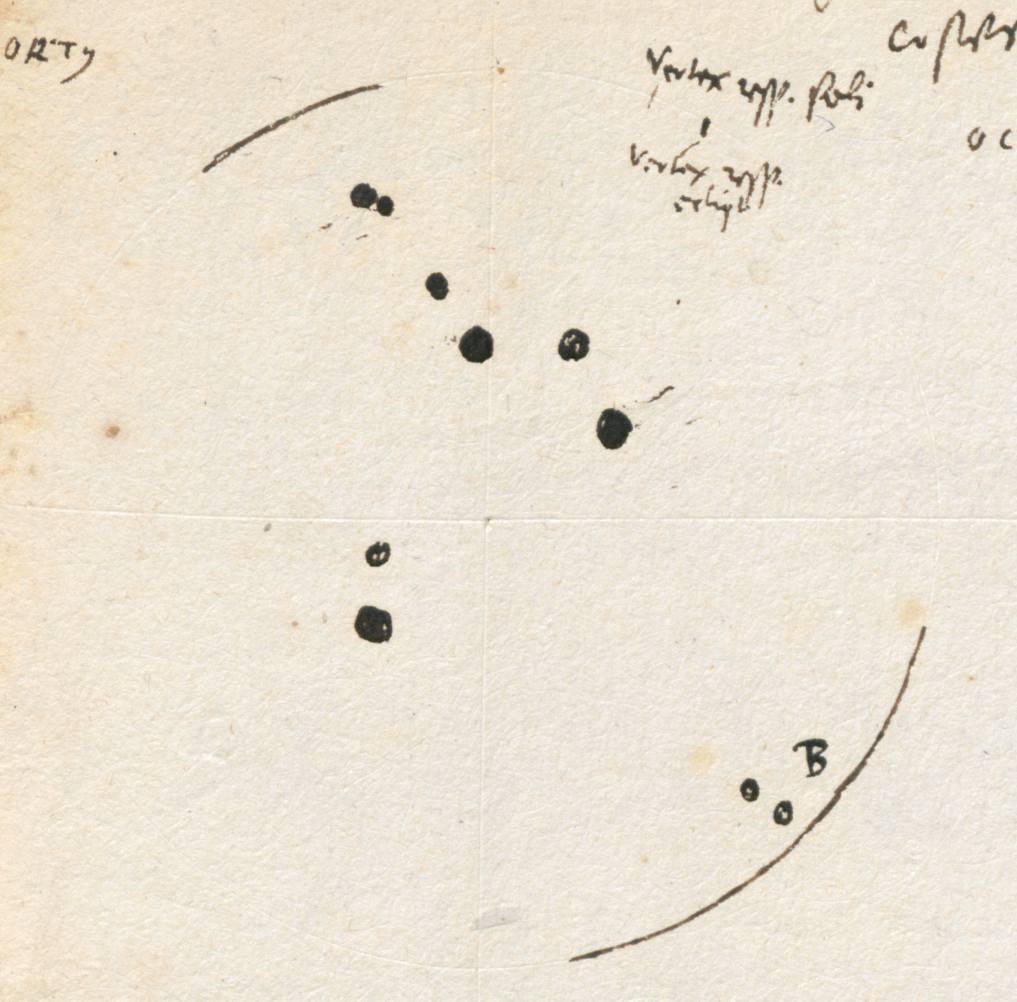}}
\end{center}
\caption{This is again the drawing of sunspots from Jungius,
as shown in Fig. 4 (1612 June 9 evening, Gregorian),
but just the drawing itself without some parts of the caption, so that it can better be
compared to the drawing from Galileo Galilei on the very same day shown to the right in Fig. 6.
The spots as drawn by Jungius are obviously larger than the same spots drawn for the same day by Galilei (Fig. 6),
but both drawings show similar relative sizes within each spot pair.
Some differences in spot location in the drawings by Jungius (evening of June 9) and Galilei (Fig. 6, June 9)
cannot be explained only by solar rotation during that day,
so that the heliographic coordinates in the Jungius drawing may be somewhat uncertain.}
\end{figure}

\begin{figure}[t!]
\begin{center}
{\includegraphics[angle=0,width=8cm]{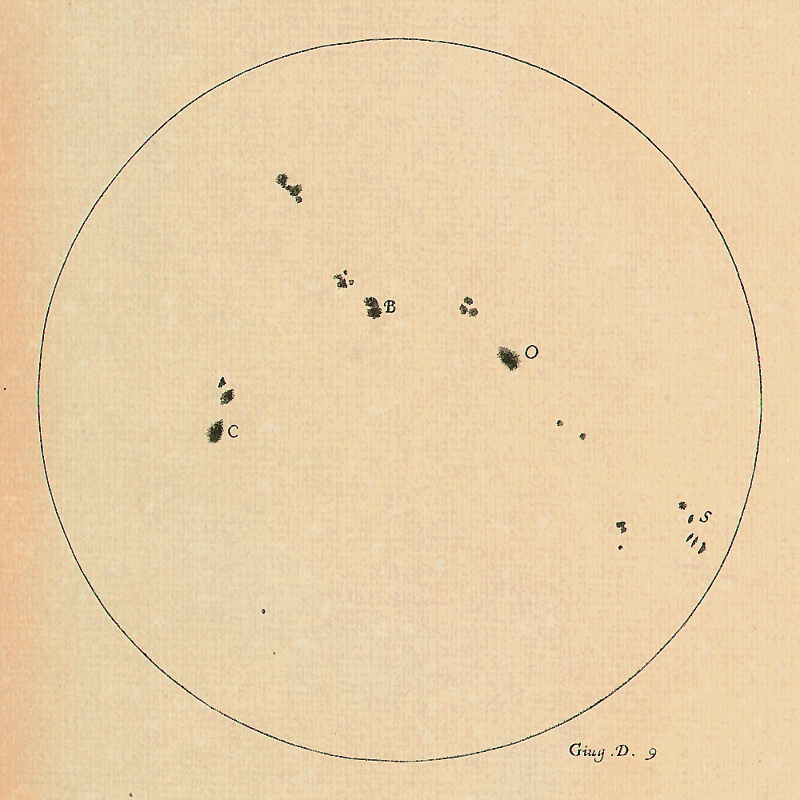}}
\end{center}
\caption{This drawing of sunspots is from Galileo Galilei from Italy for 1612 Jun 9 (Gregorian).
We can see seven to nine groups, partly resolved into smaller structures (about 25 to 30 spots).
Marius reported 14 spots that day.
HS98 list either eight (from Wolf) or ten groups (Sakurai's counting) for that day for Galilei.
The caption text to the lower right says {\em Giug. D. 9} for June day 9 (Giugno is Italian for June).
This drawing is taken from the Galileo Project (galileo.rice.edu, copyright Albert Van Helden)
and was shown before in Casas et al. (2006).
}
\end{figure}

We list corrected observing dates for Saxonius in Table 2 and compare his observations with
those by Tard\'e.
After the correction from Julian to Gregorian calendar, all observations by
Saxonius lie in March 1616, and since no one (else) has observed in Feb 1616 (HS98),
there is no monthly group sunspot number available for Feb 1616 (any more).

On three days, both Saxonius and Tard\'e observed.
Wolf (1859)\footnote{This paper is dated Wolf 1850 on ADS,
but it is clearly dated {\em Februar 1859} at the bottom of its first page.}
gives the spot numbers from Tard\'e (1620),
which are also listed in HS98 (who remarked that {\em these observations are poor}):
Tard\'e (1620) shows a drawing of one spot for 1616 March 3-14, i.e. partly for the
same days as the drawing by Saxonius. While the drawing by Tard\'e for 1615 Aug 25 shows
30 spots, all other drawings by Tard\'e for 1616 and 1617 show one spot each.

It may appear surprising that Tard\'e saw always one spot or group,
while Saxonius saw three on March 5, two on March 7, and even seven on March 14.
The drawing by Saxonius clearly shows three spot pairs
(in three groups) for Feb 24 / Mar 5, two such pairs/groups for Feb 26 / Mar 7
and seven such pairs in three groups on Mar 4/14. 
One can count here either seven pairs or groups or three groups (same on Mar 6/16).
(Zolotova \& Ponyavin (2015) neither noticed the dating problem (HS98 did not
correct Saxonius from Julian to Gregorian) nor the apparent contradiction
in group numbers (being apparent for both cases, with or without date correction),
even though discussing Tard\'e and Saxonius.)

After the drawing of 30 spots for 1615 Aug 25,  
Tard\'e (1620) presents five drawings with one spot each for the periods
1615 Nov 17-27, 1616 Mar 3-14, 1616 Apr 16-27, 1616 May 17-28, and 1617 May 27 to June 6
(all dates being Gregorian).\footnote{HS98 list the latter spot correctly for 1617 May 27 to June 6,
but they list one spot for 1616 May 17-26, then two spots for May 27 \& 28,
and then again one spot for May 29 to June 6; we suppose that HS98 incorrectly 
copied the spot report for 1617 also to 1616 and added it to the spot seen 1616 May 17-28
getting two spots on 1616 May 27 \& 28 
and one spot for 1616 May 29 to June 6.
}
We consulted both the Latin (Tard\'e 1620) and the French translation by Tard\'e himself from 1623 (Tard\'e 1627).
We should keep in mind that Tard\'e (1620) advocated the hypothesis that sunspots are due to
transits of solar system bodies (planets) called {\em Bourbon stars}, see Baumgartner (1987).

Did Tard\'e (1620) miss the other spots seen by Saxonius or did he purposely omit them
in his drawings~? \\
(i) Tard\'e (1620) wants to show here that a spot can cross the solar disk in some 10 or 13 days,
as he explains in his book:
spots move regularly in a straight line from limb to limb, from East to West, 
all move parallel to the {\em ecliptic} or on it, some move through the center of disk,
spots appearing on the same day need a different amount of days to path through, 10-13 days,
he also acknowledges that several spots appear almost at once near the center of the disk,
while others disappear before reaching the western limb
(points 8-14 in his chapter IV).
In all five drawings, he follows one
particular spot from its appearance in the East to its last sighting in the West -- adding other spots
would complicate the drawing. 
Furthermore, it may appear surprising that, whenever Tard\'e reports 
a spot (in 1616 and 1617, but not so on 1615 Aug 25), he saw it
continuously for 11 to 12 days without breaks (e.g. due to weather).
He also mentioned that he could never detect
retrograde motion of spots, even though he may have missed certain days due to bad weather.
It may appear possible that Tard\'e has interpolated the spot location for days with bad weather.
His drawings are quite realistic in regard of the daily motion of the spot
and also show foreshortening, 
but Tard\'e's spots are not drawn to scale, and their
path does not show the curvature due to the B-angle (see Fig. 10). \\
(ii) For the last spot shown, 1617 May 27 to June 6 (Fig. 8), he explicitly adds
that in this particular case, there was only one single spot alone on the disk
(Tard\'e 1620)\footnote{Tard\'e translated and extended
his Latin book to French (Tard\'e 1627), as quoted in Wolf (1859):
{\em sa promenade fut d'onze jours, tousiours seul, sans compagnon:
car pendant ces onze jours autre que luy ne fut veu dans l'aire 
du soleil}, which is consistent with the Latin text.} 
\begin{quotation}
Hic Solem aggreditur 27. Maii 1617. \& ultra 6. Junii non apparuit,
peregrinatus est in facie Solis undecim dies sine comite, nullus enim praeter 
ipsum his diebus in Sole visus est.
\end{quotation}
We translate this to English:
\begin{quotation}
As the sun is concerned, from 27 May and not longer than until 6 June,
it [a spot] wandered on the surface of the sun on 11 days without a companion,
really, none [no spot] appeared -- except this one -- seen in the sun these days.
\end{quotation}
We can implicitly conclude that, on the other occasions,
there may have been additional spots on the disk, as he mentions the exception only
for the last case. 
From the comparison between Tard\'e and Saxonius for March 1616 (Figs. 7-10),
we know for that case that there were indeed more spots on the disk than drawn
by Tard\'e. Also his explanatory texts support that he has often seen several
spots on the disk at once.  \\
(iii) The drawing for 1615 Aug 25 with 30 spots shows that he admits
that many spots can be on the sun at once. He would not withhold such evidence;
he also wrote explicitly that several spots can be on the disk at once (his points 13, 14, 19, and 20). \\
(iv) Because of his transit theory, it is possible that Tard\'e 
was interested especially in round spots.
Indeed, all spot drawings shown in Tard\'e (1620) display round spots
(but see Sect. 7.1 on his theory for non-circular spots). 
However, he acknowledged that spots have uneven edges with whitish and blackish fringes.
While he has drawn only round spots, he did describe also 
non-circular spots,
so that his numbers 
-- when obtained from drawings --
have to be regarded as lower limits. 
Furthermore, he may have opted to show these particular five cases of spots
seen for 11-12 days from their first appearance in the very East 
until the last sighting in the very West, because that would be expected for
transiting bodies, while other spots appear and/or disappear somewhere on the disk. 

In sum, we can conclude that Tard\'e (1620) did see other spots --
not only during the listed periods.
His numbers have to be regarded as lower limits (except for 1617 May 27 to June 6 
and 1615 Aug 25).

The observations by Saxonius and Tard\'e in March 1616 are not contradictory at all:
The spot seen in the drawing by Tard\'e (1620) can be identified in the drawings
by Saxonius as one particular double-spot, see Figs. 7-10.
The spots in the drawing by Tard\'e are all circular and all about the same size
(except the foreshortening effect), hence probably not drawn to scale,
but he did notice (his book, point 17) that spots have different sizes.
The spots drawn by Saxonius have different sizes and forms,
so that they may well be to scale (but his drawings are very small).

Tard\'e (1627) also remarked that he often had spotless days even for subsequent days
(for his observing period from Feb 1615 to 1619, 
shortly before his book appeared 1620 in the original Latin), point 2 in his chapter 4.
This is also consistent with information from Marius, see Sect. 3.

\begin{table}
\caption{{\bf Corrected observing dates by Saxonius in 1616.} 
While Saxonius gave his dates on the Julian calendar, Wolf (1857) and HS98 assumed they would have
been on the Gregorian calendar. Therefore, we correct the dates from Saxonius by shifting them
from the Julian to the Gregorian calendar by ten days (taking into account the leap day in 1616).
{\em n/o} for {\em not observed}.
The values by Tard\'e are lower limits (see Sect. 4.8).
The number of groups (in fact pairs) given for Saxonius follow HS98.
}
\begin{center}
\begin{tabular}{ll|ll} \hline
\multicolumn{2}{c}{1616}& \multicolumn{2}{c}{number of groups}  \\
Julian & Gregorian & Saxonius & Tard\'e   \\ \hline
Feb 22 & Mar 3    & n/o & $\ge 1$    \\
Feb 23 & Mar 4    & n/o & $\ge 1$    \\
Feb 24 & Mar 5    & 3   & $\ge 1$    \\
Feb 25 & Mar 6    & n/o & $\ge 1$    \\
Feb 26 & Mar 7    & 2   & $\ge 1$    \\ 
Feb 27 & Mar 8    & n/o & $\ge 1$    \\
Feb 28 & Mar 9    & n/o & $\ge 1$    \\
Feb 29 & Mar 10   & n/o & $\ge 1$    \\
Mar 1 & Mar 11   & n/o & $\ge 1$    \\
Mar 2 & Mar 12   & n/o & $\ge 1$    \\
Mar 3 & Mar 13   & n/o & $\ge 1$    \\
Mar 4 & Mar 14   & 7   & $\ge 1$    \\
Mar 5 & Mar 15   & n/o & n/o  \\
Mar 6 & Mar 16   & 7   & n/o  \\
Mar 7 & Mar 17   & 6   & n/o  \\
Mar 8 & Mar 18   & 7   & n/o  \\
Mar 9 & Mar 19   & 8   & n/o  \\
Mar 10 & Mar 20   & n/o & n/o  \\
Mar 11 & Mar 21   & 3   & n/o  \\
Mar 12 & Mar 22   & 2   & n/o  \\
Mar 13 & Mar 23   & n/o & n/o  \\
Mar 14 & Mar 24   & 4   & n/o  \\
Mar 15 & Mar 25   & n/o & n/o  \\
Mar 16 & Mar 26   & 4   & n/o  \\
Mar 17 & Mar 27   & 3   & n/o  \\ \hline
\end{tabular}
\end{center}
\end{table}

\subsection{Spotless days}

From the text cited above 
in Sect. 3.3 
from Marius (1619) that there were more often 
spotless days in those roughly 1.5 years before 1619 Apr,
but that there were no spotless days before those roughly 1.5 years, 
i.e. until fall 1617,
we can conclude that Marius 
either did not observe on those days
in the period 
1611 Aug 3/13 
to fall 1616,
when the sun was spotless --
or that he detected spots when others did not detect any.
There were 16 days in that period,
when other observers noticed a spotless sun, namely as follows 
(according to HS98, all dates Gregorian --
with reservations, because we noticed some shortcomings in HS98,
we did not check the sources of these observations):
\begin{itemize}
\item 1611 Dec 29 (Harriot: no spots)
\item 1612 Mar 2, 4, 5, 6, Apr 13-17, 23 (Harriot: no spots)
\item 1612 Mar 2, 4 (Cigoli: no spots)
\item 1616 Nov 13-15, 22, 23 (Scheiner: no spots)
\end{itemize}
On these dates, there are no naked-eye sunspots known (e.g. Vaquero 2012).

As shown in Figs. 4-6 for 1612 June 9 (Gregorian), Jungius and Harriot saw the same five groups, 
while Marius detected slightly more spots (14) than Jungius,
a similar number as Harriot (up to 14), but less than Galileo.
The fact that Harriot reported spotless days for ten days in 1612 does
not need to be a contradiction to the statement by Marius that there were no
spotless days before fall 1617: even though Marius and Harriot saw about the
same number of spots around 1612 June 9, it is possible that 
Marius saw on other days one (or a few) more spots,
which were not spotted by Harriot -- or, Marius may not have observed those days.
Also, from Tanner's record, there is no evidence for spotless days in 1612, 
even though of {\em almost daily} observations (Sect. 4.5).

We can conclude that all essential elements in the statements by Marius can be confirmed,
while no parts were falsified.

\section{Group sunspot numbers}

We could now use the new and/or revised data found above in the work by Marius 
and others to compute new
daily, monthly, and yearly group sunspot numbers for the relevant periods,
i.e. to revise HS98.

As far as the periods before and inside the Maunder Minimum are concerned,
telescopic group sunspot numbers (HS98) were previously revised
for a few days, months, and years (e.g. by additional, newly found old
sunspot observations, newly found after 1998, i.e. since HS98), 
see publications by Vaquero (2003), 
Casas et al. (2006),
Vaquero et al. (2007, 2011), 
Vaquero \& Trigo (2014), Carrasco et al. (2015), Gomez \& Vaquero (2015), 
and Neuh\"auser et al. (2015).

\begin{figure*}
\begin{center}
{\includegraphics[angle=0,width=16cm]{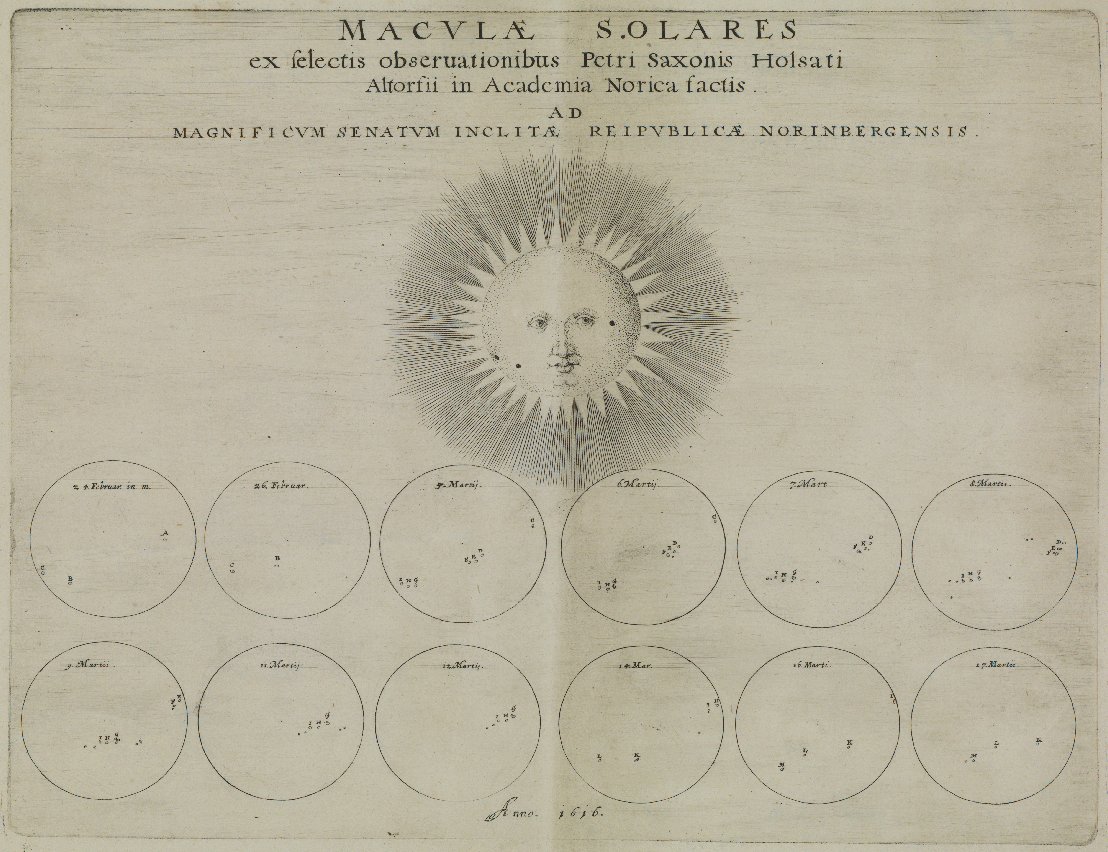}}
%{\includegraphics[angle=0,width=16cm]{saxo-full.tiff}}
\end{center}
\caption{These are the reproductions based on drawings by Saxonius for 1616 Feb 24/Mar 5 to Mar 17/27,
which are available in the Germanisches Nationalmuseum (German National Museum) in Nuremberg (Inv. nr. HB 12704)
on a cupper plate (25.8 cm $\times$ 32.2 cm) produced in Altdorf near Nuremberg,
shown here with their permission.
The caption on the top reads: {\em Maculae Solares ex selectis obseruationibus Petri Saxonis Holsati
Altorfii in Academia Norica factis ad Magnificvm senatvm inclitae reipvblicae Norinbergensis},
i.e. on {\em sunspots observed by Petrus Saxonius from Holstein at the academy of Altdorf,
dedicated to the senate of the republic of Nuremberg} [Reichsstadt].
The twelfe drawings in the bottom two rows are dated ({\em Anno 1616}, as given in the bottom row);
while Wolf and HS98 assumed that the dates were given on the Gregorian calendar,
they are in fact given on the Julian calendar, see Table 2.
We see for each day one (or more) spot(s) plus a letter.
The larger image of the sun in the top center also shows the three spot pairs
(combined to one large unresolved spot each) 
from the first observing date (1616 Feb 24/Mar 5) as in the leftmost drawing in the 2nd-to-bottom row.
The 2nd-to-left spot on that day is the one also drawn by Tard\'e, see also Figs. 8-10.
On the first three dates, also Tard\'e has observed (Fig. 9, Table 2).
It may well be that, historically, drawing a face in the sun with eyes, nose, and mouth 
may have been motivated by sunspots.}
\end{figure*}

HS98 have estimated daily, monthly, and yearly group sunspot numbers for all the
observations listed by them.
We see a number of problems in the system applied by HS98 and in Equ. (2), namely as follows:
\begin{itemize}
\item Similar unjustified assumption, as made by HS98 for Marius (and Riccioli) for the years 1617 and 1618,
namely that they would have observed (almost) the whole year without detections,
also affect other European observers of the 17th and 18th century listed in HS98,
where there are observers who appear to have observed and excluded spots for all
365 or 366 days of particular years:
Zahn, Hevelius, Picard, Fogel, Weigel, Weickmann, Siverus, Agerholm, Wurzelbaur, Derham,
and Adelburner. Only for Fogel, Weigel, Siverus, und Weickmann,
HS98 remark that {\em original observations are probably lost so we do not know
exactly what days [they were] observing}. Zeros in the HS98 tables, even though the observers
either did not observe or even did detect spots, can underestimate the sunspot numbers.
\item The factor $12.08$ in Equ. (2) comes from the mean number of individual spots per group
and a scaling of the previous centuries to the Greenwich observatory monitoring period (HS98), but
it is a priori not known whether solar activity was indeed similar on average
during those two epochs -- during and before the Greenwich observatory monitoring period.
\item It is not clear whether the number of spots per group is constant,
e.g., inside and outside of Grand Maxima and Minima.
\item In case of the observation by Marius on 1612 May 30 (Julian), where he reported 14 spots,
are those 14 spots or 14 groups? Without any other evidence, in the system of HS98 we would have
to assume that they form 14 groups, so that his daily group sunspot number would be 212
(namely $14 \cdot 12.08 \cdot 1.255 = 212.25$), 
which would probably be an overestimate of solar activity with Equ. 2.
\item It is often somewhat subjective as to how many groups are formed by the spots, see e.g. Figs. 3-6.
\item When Marius reported 14 spots for 30 May 1612 (Julian), they were probably grouped
in five groups -- by comparison with the drawings 
by Jungius, Galilei, and Harriot on and around that date (Figs. 4-6).
\item We obtained some lower limits for the numbers of spots for certain observers for certain dates,
e.g. for Marius and Schmidnerus for 1611 Aug 3/13. 
However, Equ. (2) does not allow to take them into account when calculating daily means.
\item HS98 recommend that data for years with less than 20 observational days per year should not 
be considered (nor plotted) due to too 
low statistical significance (for monthly and yearly means) -- why exactly 20?
\item HS98 interpolate for days without observations which lie within days with detections,
assuming that the detected spot(s) were also present on the days in between the detections.
Such data are listed in their table {\em filldata}. It may in some cases have happened that the
number of groups changed between detections. 
\item Given the problems with Marius and other observers as found here and listed above,
the correction factors $k^{\prime}$ in Equ. (2) are questionable for Marius and the other observers
affected in a similar way. 
\item Then, if the correction factors $k^{\prime}$ and observing days 
of some observers listed above would need to be revised,
the correction factors of other observers may also have to be updated, namely those, to whom the
above listed observers were compared.
\item For observers, who have observed on days when no one else has observed (e.g. Fabricius), 
it is not possible to obtain such a correction factor by comparison to other observers,
but HS98 have just assigned $k^{\prime}=1.255 \pm 0.112$ to them.
\item HS98 (in the first table of their appendix) list the number of comparison observers
(used to calculate $k^{\prime}$ for all observers) and give zero for, e.g., Scheiner and Malapert,
but according to their table {\em alldata}, there are many comparison days for comparing those observers
%XXX%
to others, e.g. Scheiner seems to overlap with, e.g., Malapert (see Sect. 3.1) and Smogulecz. 
\item HS98 also had to choose {\em primary} observers, to whom the others are compared;
e.g. they give {\em for the observations before 1730 ... Plantade} and for the early 1600s,
they picked Galilei; 
neither Scheiner nor Malapert overlap directly on any days with Galilei,
but Scheiner has overlaps with, e.g., Harriot 
(e.g. 1611 Dec 10, 13, 14 according to HS98),
and Harriot has overlaps with Galilei (e.g. June 1612, see Table 3),
%XXX% next line weg
%and Malapert has overlaps with Scheiner, so that both can be linked and compared with Galilei; 
HS98 assign corrections factors $k^{\prime}=1.250$ for Galilei, 
$k^{\prime}=1.990$ for Harriot, and $k^{\prime}=1.255$ for Scheiner, Malapert, and Marius,
which is not transparent;
for calculating the correction factor $k^{\prime}$ for those primary observers,
HS98 had to choose some interpolation law.
\item HS98 state that, when taking a mean, one should use only those observers with
$k^{\prime}$ between 0.6 and 1.4 for the time since 1848 -- why exactly since 1848,
why exactly 0.6 and why exactly 1.4?
\item For the observations by Galilei, when two different group countings are available, one from HS98 themselves
and one from Sakurai (with two different correction factors $k^{\prime}$), should one consider both when taking
the mean (even though they are not independent of each other as based on the same drawings) or the mean of the
two countings?  
\item While HS98 take the weighted mean for days, for which more than one observer obtained data,
one could instead use only the observer with the most spots or groups, or -- even better -- the 
observer, if this observer was found to credible regarding all the reported observations (including trends).
\item Both the averaging (means) and the error estimate in HS98 appear to be highly complicated and
are based on many subjective decisions.
\end{itemize}
Some of those and other problems were noticed before (see e.g. Clette et al. 2014); 
e.g. Vaquero et al. (2011) noticed the problems with Crabtree's data in 1638 and 1639.

We will refrain from estimating new daily, monthly, and yearly group sunspot numbers here
for the following reasons: \\
(i) We noticed a number of small problems in the HS98 data base (Sect. 3 \& 4), not only for
Marius, whom he have checked in detail, but also for several other observers,
whom we did not even check systematically, but whom we studied only partially and
only in comparison with Marius. Only after a careful study of all the original
texts and drawings, one should consider to reestimate new (group) sunspot numbers. \\
(ii) Important trends as those mentioned by Marius 
(much less spots in the last 1.5 years than before, Sect. 3.3)
cannot be taken into account in Equ. (2). \\
(iii) We noticed that several early observers are missing, like
D. and J. Fabricius, Cysat, Schmidnerus, Tanner, Argoli, Wely, and Perovius. 
It is likely that a number of additional sunspot observers is still missing
(e.g. Sect. 3.4). \\
(iv) The problems in the group sunspot number system listed above.

Furthermore, the group sunspot numbers for 1617, 1618, 1619 and also until 1623
are all 15 (in HS98), so that the Schwabe cycle minimum around that time can hardly be dated precisely.
The trend mentioned by Marius is also not taken into account, namely that spot numbers
and solar activity decrease from before fall 1617 to the time from fall 1617 to spring 1619.

Instead of Equ. (2), one should consider to use the spot area (e.g. Arlt 2011, 2013),
or the group number, i.e. without the factor 12.08 (e.g. Svalgaard \& Schatten 2016),
or the (international or Wolf) sunspot number, or maybe the active day fraction,
or even generic statements or trends;
depending on the particular problem and the availability of sources,
one should select and use certain kinds of quantitative and/or qualitative evidence.

\begin{figure}[t!]
\begin{center}
{\includegraphics[angle=0,width=8cm]{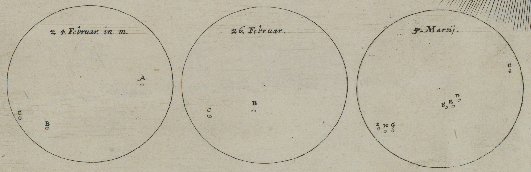}}
\end{center}
\caption{These are the drawings by Saxonius for his first 
three observing days 1616 Feb 24/Mar 5, Feb 26/Mar 7, and Mar 4/14
(part of Fig. 7). On the same days, 1616 Mar 5, 7, and 14, Tard\'e has drawn only his main spot (Fig. 10, Table 2).
This main spot as drawn by Tard\'e (East left, North top, Fig. 9)
is the 2nd-to-left spot group on the left (first) image of Saxonius, 
which has moved towards the upper right in the 3rd image.
}
\end{figure}

\begin{figure*}[t!]
\begin{center}
{\includegraphics[angle=0,width=14cm]{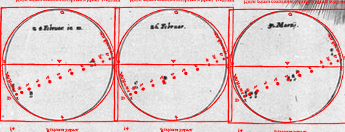}}
\end{center}
\caption{Here, we show the drawings by Saxonius for his first three observing days 
1616 Feb 24/Mar 5, Feb 26/Mar 7, and Mar 4/14
in the background in grey, together with the drawing by Tard\'e for 1616 Mar 3 to 14 in the foreground in red overlaid
(in red, we see for each day one spot and the date from Tard\'e).
While Tard\'e indicated that he placed North to the top and East to the left (Fig. 10),
we see that Saxonius placed East to the left, too, and we assume that he placed North to the bottom
(probably due to his observing technique);
in our overlay figure, North is to the bottom.
We can thereby identify the spot group recorded by both: 
On the left image, the third spot from the left by Tard\'e (Mar 5)
is very close to a spot (group) drawn by Saxonius for Feb 24/Mar 5;
then, in the center image, the fifth spot from the left by Tard\'e (Mar 7) is very close to a 
spot (group) drawn by Saxonius for Feb 26/Mar 7;
and in the right image, the last spot on the right by Tard\'e (Mar 14) is very close to 
a spot (group) drawn by Saxonius for Mar 4/14.}
\end{figure*}

\begin{figure}[t!]
\begin{center}
{\includegraphics[angle=0,width=7cm]{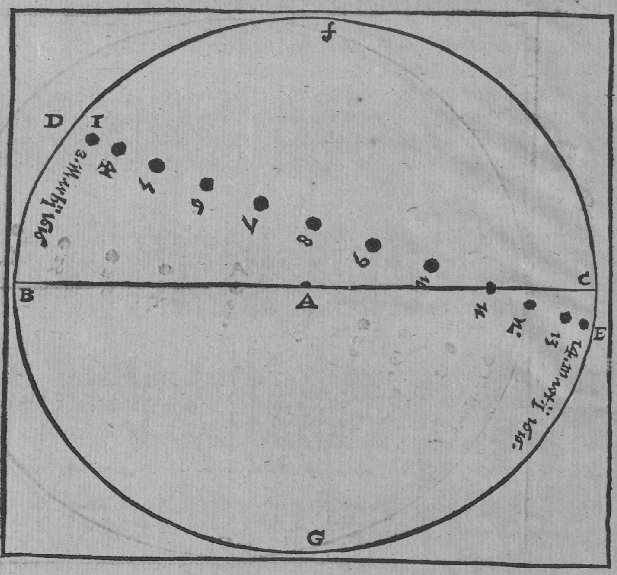}}
\end{center}
\caption{This is the drawing by Tard\'e for 1616 Mar 3 to 14
obtained from his book Tard\'e (1620) on {\em Borbonian Stars}, figure 7 on page 34.
The letters indicate as follows, according to Tard\'e (1620), his figure caption on page 32: 
{\em A - centrum disci Solis}, i.e. center of the disk,
{\em B - pars orientalis}, i.e. East (left),
{\em C - Occidua}, i.e. West (right),
{\em F - Septentrionalis}, i.e. North (top),
{\em G - Australis}, i.e. South (bottom),
{\em BC - linea aequatori parallela}, i.e. a (full) line parallel to the equator,
and {\em lineam DE describens initio facio a D plaga orientali},
i.e. line DE describes the path of the spot, which first appeared in the East.
Each spot shown is labelled with the observing date, the first with {\em 3. Martii 1616} (1616 Mar 3),
the last with {\em 14. Martii 1616} (1616 Mar 14).
His figure caption reads: {\em Huius primus contuitus accidit tertia Martii, \& ultimus
14. eiusdem 1616. Mora fuit undecim dierum}, i.e.: {\em The first appearance of this [spot] was on the 3rd of March
and the last on the 14th of the same month in the year 1616. Its presence was for 11 days}
(from the 3rd to the 14th of March, we have 12 days (inclusive counting,
and 12 spot days are seen in the drawing), the spot was seen from some time
during the (bright) day on the 3rd until about the same time on the 14th, hence indeed for 11 days).
The path of the spot crossing the disk as drawn by Tard\'e does not show a significant curvature,
even though the solar B-angle is large at this time of the year (March).
}
\end{figure}

\begin{table*}
\caption{{\bf Correction of values from Harriot for 1612 June 8-13.}
For the days 1612 June 8-13 (Gregorian), we list the number of groups for Harriot
as given in HS98 (alldata) and the corrected (just shifted) numbers according to the
drawings by Harriot himself (see http://digilib.mpiwg-berlin.mpg.de).
{\em n/o} for {\em not observed} (on that day). 
We did not change the number of groups, i.e. did
not question the listing by HS98, but we just correct the dating. There is a simple shift error in
the table alldata in HS98. The monthly number does not change, because the correction
is just a permutation. Days before Jun 8 and after Jun 13 are not affected by the mistake.
We list daily means from HS98 and our calculation (both without the correction),
and then in the last column the daily means after the correction --
where the two other observers Galilei and Jungius were of course also taken into account.
For none of the possible combination for Galilei (see Note (a) below), we can reproduce the exact HS98 values,
possibly at least partly due to their rounding (the same problem happens for both the
{\em alldata} table from HS98 as listed below and the {\em filldata} table from HS98,
where they interpolated for days without observations -- similar problems with HS98 were
noticed for other years in Neuh\"auser et al. 2015 and Svalgaard \& Schatten 2016).
In the 2nd column, we list Wolf's numbers (Wolf 1858), which compare
well with our corrected estimate from Harriot's drawings.
}
\begin{tabular}{l|c|cc|l|lll|l} \hline
Date & Wolf(e) & \multicolumn{3}{c}{Number Groups} & \multicolumn{4}{c}{Daily Mean Group Sunspot Number} \\
June  & (1858)   & HS98   & Harriot & HS98         & \multicolumn{3}{c}{Our Mean (a) }         & Corrected \\
1612     & gr/sp    & Value  & Drawing & Mean         & w/ Gal/HS    & w/ Gal/Sak  & w/ 2 $\times$ Gal  & Mean (b)\\ \hline
 & & & & & & & & \\[-8pt]
~~8      & 5g12s & n/o  & 5       & $117 \pm 17$ & $122.5 \pm 23.7$     & $134.6 \pm 6.4$  & $125.0 \pm 17.3$ & $123.8 \pm 14.3$  \\
~~9$^{\,c}$ & n/o  & 5    & n/o     & $123 \pm 5$  & $126.7 \pm 10.8$(d) & $129.8 \pm 9.5$  & $127.6 \pm 9$    & $130.0 \pm 9.2$(c) \\
10     & 6g14s & n/o  & 6       & $108 \pm 4$  & $108.6 \pm 4.0$      & $101.3 \pm 14.4$ & $102.7 \pm 10.5$ & $113.1 \pm 22.4$ \\
11     & n/o  & 6    & n/o     & $116 \pm 4$  & $125.5 \pm 16.9$(d) & $124.2 \pm 17.5$ & $123.4 \pm 14.4$ & $116.4 \pm 4.7$ \\
12     & 5g14s & n/o  & 5       & $109 \pm 8$  & $122.5 \pm 23.7$     & $121.6 \pm 24.8$ & $116.3 \pm 19.8$ & $117.3 \pm 16.3$ \\
13     & n/o  & 5    & n/o     & $112 \pm 11$ & $116.7 \pm 24.5$     & $121.6 \pm 24.8$ & $113.5 \pm 21$   & $111.3 \pm 25.1$ \\ \hline
\end{tabular}

\smallskip

Notes: 
$^{a}$ {\em w/ Gal/HS} means that we exclude Sakurai's group number counting for Galilei, 
i.e. we use the HS98 group number counting for Galilei only, together with all other observers
that day listed in HS98 ({\em w/} for {\em with}),
{\em w/ Gal/Sak} for using Sakurai's group number counting for Galilei (as listed in HS98)
and the other observers, but not the counting by HS98 themselves, 
and {\em w/ 2 x Gal} for using both -- in all cases with the wrong (permuted) entries for Harriot according to HS98.
$^{b}$ using both HS98's and Sakurai's group numbers for Galilei and the 
corrected entries for Harriot according to his drawings.
$^{c}$ On June 9, Marius has observed 14 spots, not considered in the means in this table;
with the 14 spots as 14 groups reported by Marius, 
his daily group sunspot number would be $14 \cdot 12.08 \cdot (1.255 \pm 0.112) = 212 \pm 19$;
then, we obtain a daily {\em mean} group sunspot number of $151 \pm 42$;
assuming that those 14 spots were assembled in five groups (by comparison with the drawings 
by Jungius, Galilei, and Harriot on and around that date, Fig. 4-6),
the daily group sunspot number of Marius would be $5 \cdot 12.08 \cdot (1.255 \pm 0.112) = 75.8 \pm 6.8$,
and the daily mean group sunspot number would be $116 \pm 28$.
$^{d}$ $120.5 \pm 0.4$ and $116.1 \pm 6.4$ for June 9 and 11, respectively, 
when omitting $2\sigma$ deviations as in HS98. 
$^{e}$ In the usual notation of Wolf, e.g. {\em 5,12} for June 8 for 12 spots in 5 groups (here: {\em 5g12s}) as measured for
Harriot by Carrington for Wolf.
\end{table*}

\section{Active day fractions 1611-1620}

Let us now consider the 
active day frations F$_{\rm a}$): number of days (N$_{\rm a}$) 
with at least one sunspot divided by the number (N) of observing days in a given period
(e.g. Maunder 1922, Kovaltsov et al. 2004).
Active day fractions F$_{\rm a}$ compared to group sunspot numbers (from HS98) are listed in Table 4.

In the latest work on solar activity with the active day fraction in the Maunder Minimum,
Vaquero et al. (2015) studied three cases: \\
(i) In the {\em loose} model, they left out 
all periods with generic statements (zero in HS98) being {\em longer than a month}. 
If applied to the data in HS98 for 1618, they would leave out all alleged zeros from Riccioli
and Marius except the period from 30 June to 6 July 1618 (which is not longer than a month);
then, the active day fraction would be 1.0 for March, June, and July in 1618 and also for 1618 in total;
%XXX%
this is, because -- according to HS98 -- Malapert (and presumably also Scheiner) reported only those 21
days in 1618 when they detected at least one spot, but they did not report any spotless day;
the active day fraction with 21 active days and those presumable (HS98) seven inactive days from 30 June to 6 July
would be 0.75; if those seven alleged spotless days from Riccioli and Marius would be left out, 
the active day fraction would be 1.0. 
An active day fraction of 0.75 or even 1.0 for 1618 would be in strong contradiction to the statement
from Marius that he saw more often spotless days from fall 1617 to spring 1619.\\
(ii) In the {\em optimum} model, Vaquero et al. (2015) left out all generic statements (long periods
with zeros in HS98) for observers, who did not detect any spot that year;
for 1618, this model would leave out Riccioli and Marius completely, so that the active day fraction
would be 1.0. \\
(iii) In the {\em strict} model, Vaquero et al. (2015) left out all generic statements (long periods
with zeros in HS98) except when two observers independently reported a spotless day;
for 1618, both Riccioli and Marius allegedly report spotless days from June 30 to July 6 (HS98),
so that those data would be left in; the active day fraction would then be 0.75. \\
Hence, the active day fractions from all three models are problematic.\footnote{Kovaltsov et al. (2004)
suggest to use a correlation between the active day fraction and group sunspot numbers (found for
the time since 1850) to estimate the group sunspot numbers for problematic years with low group 
sunspot numbers -- however, this works only for years with low active day fraction, i.e. again
poor statistics; furthermore, the years with such low active day fractions after 1850 are the
Schwabe cycle minima years, which may not be typical for periods before 1850, e.g. in a Grand
Minimum. Comparing active day fractions and (group) sunspot numbers may still be useful to
find problematic years (e.g. Vaquero et al. 2012).} 
While Vaquero et al. (2015) did not apply these three models to the 1610s,
similar problems with generic statements interpreted by HS98 as zero spot numbers for
each and every day for months to years may apply to several other observers after the 1610s.

We stress again that the generic statements from Marius were clearly mis-interpreted by HS98 
as meaning zero spots for several months, while only from the quotation of the sentences from Marius
given in Wolf (1857) it was clear that the interpretation by HS98 is not correct.
In Table 4, we apply the statement from Marius that the active day fraction
is below 0.5 (but not zero) from fall 1617 to spring 1619 to three separate
periods: fall 1617, 1618, and Jan-Apr 1619.
It is possible that the active day fraction changed with time within those 1.5 years,
e.g. to could have decreased: Argoli (Sect. 3.4) reported that there were no spots during
the periods when comets were seen in 1618 (all in the 2nd half of 1618).

\begin{table}
\caption{{\bf Group sunspot numbers and active day fractions}
from data in HS98 (alldata) and our revised values.
{\em n/a} means {\em no datable observations available}.
We list only those months, where the values are to be revised
and/or when Marius reported {\em datable} observations 
(and all other years, so that variations can be seen).}
\begin{tabular}{lccc} \hline
Month    & R$_{\rm G}$ HS98 & \multicolumn{2}{c}{active day fraction $F$} \\
year     & Equ. 2          & (HS98) & this work \\ \hline
Mar  & n/a             & n/a              & 1         \\ 
Aug  & n/a             & n/a              & 1 (a)     \\ 
Oct  & $57.3 \pm 11.3$ & 1                & 1 (a)     \\ 
Nov  & $60.7 \pm 14.3$ & 1                & 1 (a)     \\ 
\multicolumn{2}{l}{since Aug} & 1                & 1 (a)     \\
1611 & $54.7 \pm 5.9$  & 0.98 (b)         & 0.98 (b)  \\ \hline
Jun  & $99.6 \pm 16.3$ & 1                & 1 (a)     \\ 
1612 & $92.1 \pm 2.7$   & 0.96 (e)        & 0.96      \\ \hline
1613 & $92.3 \pm 7.8$   & 1 & 1 (a) \\ \hline
1614 & $121.0 \pm 15.5$ & 1 & 1 (a,c) \\ \hline
Mar  & $151.0 \pm 0.0$  & 1 & n/a (g,s) \\
1615 & $30.1 \pm 3.7$   & 1 & 1 (g) \\ \hline
Feb  & $29.8 \pm 8.0$   & 1 & n/a (l,s)  \\
Mar  & $43.6 \pm 18.1$  & 1 & 1 (l)  \\
Jun  & $15.0 \pm 0.0$   & 1 & n/a (k,s) \\
1616 & $21.6 \pm 2.3$   & 0.91 (d) & 0.89 (d)   \\ \hline
Jun  & $3.0 \pm 6.1$    & 0.2   & 1 (t)    \\
Jul-Dec & $0 \pm 0$    & 0 (j) & (r,t) \\
since fall &           & 0 & $0 < F < 0.5$ (r) \\
1617 & $0.8 \pm 0.0$   & 0.05 (f) & \hspace{-0.3cm} $0.03 < F \le 0.92$ (r,u)     \\ \hline
Jan  & $0.0 \pm 0.0$   & 0 & n/a (r) \\
Feb  & $0.0 \pm 0.0$   & 0 & n/a (r) \\
Mar  & $5.3 \pm 7.3$   & 0.23 & (r,h) \\
Apr  & $0.0 \pm 0.0$   & 0 & n/a (r) \\
May  & $0.0 \pm 0.0$   & 0 & n/a (r) \\
Jun  & $4.5 \pm 7.0$   & 0.30 & (r,h) \\
Jul  & $5.8 \pm 7.4$   & 0.23 & (r,h) \\
Aug-Dec & $0 \pm 0$   & 0 (j) & (r,q) \\
1618     & $1.3 \pm 0.0$   & 0.06 (p) & $0.08 < F < 0.5$ (r)   \\  \hline
1-4/1619     & \hspace{-0.3cm} $15.0 \pm 1.8$ (v)  & 1 (n) & $0.06 < F < 0.5$ (r,t)   \\ \hline
1620     & $15.0 \pm 1.7$  & 1 (m) & \hspace{-0.2cm} $0.11 \le F \le 0.99$ (m) \\ \hline
\end{tabular}

Notes:
$^{a}$ Marius: never spotless before fall 1617,
i.e. $F=1$ whenever he observed;
also 1 from datable observations --
not necessarily valid for each month.
$^{b}$ 45 active and 1 inactive day (HS98), 
we added active days in Mar 1611 (Sect. 4.3) and 1 in Aug 1611.
$^{c}$ 1614 only one observing day with 8 groups (HS98).
$^{d}$ 49 active and 5 inactive days in 1616 (HS98),
but only 40 active and 5 inactive days (Table 2 and footnote 16).
$^{e}$ 240 active and 10 inactive days (HS98).
$^{f}$ 11 active and 219 inactive days (HS98, Sect. 3.1).
$^{g}$ Footnote 14; 12 active days in 1615.
$^{h}$ Additional active days in Table 5.
$^{j}$ Valid for each month given. 
$^{k}$ Footnote 16.
$^{l}$ Table 2.
$^{m}$ 35 active days (HS98)
plus 4 more active and 2 inactive days according to Malapert (Sect. 7.2),
not in HS98 ($F \ge 0.11$ if days without records were inactive).
$^{n}$ 7 active and no inactive days for 1 Jan -- 26 Apr 1618 (HS98).
$^{p}$ 23 active days in HS98, 5 more in Table 5. 
$^{q}$ Argoli: no spots during comets, mainly in Sep, Nov, Dec (Sect. 3.4).
$^{r}$ Marius: more often spotless days in 1.5 yr before spring 1619
(i.e. $F < 0.5$, but not 0) -- not necessarily valid for each month.
$^{s}$ If note (a) is valid for each month, then $F=1$. 
$^{t}$ Not clear, when exactly the period of 1.5 yr started, 
it could be anytime from July to end of Oct 1617 (Sect. 3.3),
we call it {\em fall}.
$^{u}$~Upper limit 0.92, if note (r) applies only to Nov and Dec 1617.
$^{v}$~$15.0 \pm 1.8$ for the whole year 1619, $15.0 \pm 0.0$ for Jan,
no values for Feb-Apr in HS98.
\end{table}

The active day fractions given above for 1611 to 1616, according to the values in HS98
and our corrections and additions (Table 4),
all being close to 1, are not inconsistent with the statement by Marius that he did not
find any spotless days before fall 1617 -- he certainly did not observe on all days
(see Sect. 4.9),
it could have been too overcast, and he mentioned that he was sometimes ill or travelling.

While we have an active day fraction of 0.96 to 1 in 1611 to 1615,
it then drops to 0.89 in 1616 and further to below $0.5$ from fall 1617 to spring 1619
(given the statement by Marius).
The active day fraction may be revised slightly by additional observations
by Galilei from 1612 Feb 12 to early May (see Vaquero \& Vazquez 2009, Reeves \& Van Helden 2010), 
which were not listed in HS98
(nor considered in Table 4); Harriot (also) observed a lot in those months (HS98).

Let us try to constrain the active day fraction for the period from
fall 1617 to spring 1619 even further:
According to Marius, it is lower than $0.5$.
%XXX%  next 2 lines
Malapert stated for 21 days in 1618 that he saw at least one spot. 
He did not report spotless days (HS98). If we assume that less than 337 days of 1618 would
have been spotless, then the active day fraction would be larger than $0.08$.
Hence, both limits together constrain the active day fraction to $> 0.08$ and $< 0.5$ for 1618.
This is indeed lower than 0.89 in 1616 (Table 4), i.e. consistent with decreasing activity.
(We note that some of the numbers above were obtained from the compilations by
HS98, which were otherwise shown to have problems -- one should check all reports
from all observers before obtaining final results.)

A limiting factor in using the active day fraction is the fact that it does not reflect
the number and sizes of spots on active days.
Furthermore, in the early telescopic era, observers concentrated on reporting spots, 
and have probably left out dates of spotless days in their records;
therefore, in particular for 1617-1620, we cannot constrain the active day fraction
better than in Table 4.
Active day fractions from generic statements (like the one from Marius studied here)
can be quite useful.

\section{The Schwabe cycle minimum around 1620}

We can now use all the data presented above to investigate the Schwabe cycle minimum
near the turn from the 1610s to the 1620s, 
which is dated 1619.0 in HS98.
The minima in the yearly group sunspot numbers lie in 1617 and 1618 (Table 4),
they are then given as 15.0 for all years from 1619 to 1623 in HS98.
The timing of this minimum was also discussed recently in Zolotova \& Ponyavin (2015), 
also in regard of Marius, so that we will first consider that paper.
Then, we will present the drawings by Malapert from 1618 into the 1620s,
which yield the heliographic latitudes -- to be used to date the Schwabe cycle minimum.

\subsection{Discussion of Zolotova \& Ponyavin (2015)}

We discuss some issues from sections 3 and 4 in Zolotova \& Ponyavin (2015)
on an alleged dominant world view and the timing of the Schwabe cycle minimum around 1618,
both in relation to Marius.

Zolotova \& Ponyavin (2015) discussed the sunspot observations of Marius.
They wrote:
\begin{quotation}
According to Hoyt \& Schatten (1998) ... Marius and Riccioli ... 
did not even register a single spot ... 
In March 1618 Scheiner and Malapert synchronously observed a sunspot group. 
It is noteworthy that when the Sun became active, 
Marius and Riccioli immediately stopped observations.
\end{quotation}
The conclusion or assumption that Marius and Riccioli would have stopped their observations,
%XXX%
exactly when the sun became active, i.e. when Malapert detected a spot or group,
is clearly not justified, according to the text from Marius quoted in previous sections:
The preface of his work on the large comet of 1618 finished in April 1619 (Marius 1619, our Sect. 3.3)
clearly shows that he observed regularly 
(also) in the period fall 1617 to spring 1619, he did notice more often spotless days.
(As mentioned before, Sect. 3.1, the drawing in Scheiner (1626-30) is from an observation by
Malapert, so that they did not {\em synchronously observe a sunspot group}.)

Furthermore, we gave more than one example where Marius reported spots for the very same day,
when others have reported (and drawn) spots, so that the claim by Zolotova \& Ponyavin (2015),
he (and others) would stop observing and/or reporting when others observe spots and/or
when the sun gets active, is not supported by the evidence of the historic transmission.

The key argument in Zolotova \& Ponyavin (2015) is that some observers did not report certain
spots, in particular non-circular spots would not have been reported
by certain observers, in particular Marius,
\begin{quotation}
caused by the dominant world view of the seventeenth century 
that spots (Sun's planets) are shadows from a transit of unknown celestial bodies. 
Hence, an object on the solar surface with an irregular shape or consisting 
of a set of small spots could have been omitted in a textual report 
because it was impossible to recognize that this object is a celestial body.
\end{quotation}
First, the conclusion made by Zolotova \& Ponyavin (2015) cannot be supported for Marius:
he reported {\em tail-like longish spots}, i.e. non-circular spots (Sect. 3.3), or
{\em spots ... observed in very large numbers ... always in different form since August}
reported by Marius for the time from Aug to Dec 1611 in his letter to Maestlin from
late December 1611 (Julian dates), see Sect. 4.1.

Second, even though both Tard\'e (Sect. 4.8) and Malapert (Sect. 7.2) supported the transit theory, 
so that they may have been mostly interested in roundish spots, 
Tard\'e (1620) acknowledges explicitly that non-circular / non-roundish spots are sometimes seen
(see also Baumgartner 1987): spots and their surroundings change getting longish, 
or split into 2 or 3 or more, or merge (points 19 and 20 in his chapter IV).
Tard\'e (1620) argued that it was difficult -- due to the large
brightness of the sun -- to get accurate images of spots (so that some may only appear to be
non-circular, but may in fact be circular),
and that spherical bodies -- when transiting before/across the sun -- would be seen
first in crescent phase, then in full phase, and then again in crescent phase,
hence the different forms seen (inspired by Venus phases, which were just discovered) --
but he obviously overestimated that effect strongly.

Third, what is called a {\em dominant world view} in Zolotova \& Ponyavin (2015),
spots as transiting small solar system bodies or their shadows,
was just one of several theories.
Marius mentioned at least three of them: \\
(i) spots are something like clouds on the surface or in the atmosphere
of the sun possibly coming as emission from the interior of the sun, \\
(ii) spots are transiting bodies orbiting close to the sun or their shaddows, or \\
(iii) spots are some kind of evaporations from the sun that cool the sun
and then form comets. \\
Marius himself mentioned the first two possibilities in 1612 (Sect. 4.6),
but in 1619 (Sect. 3.3) he seems to tend to the latter, partly based
on his own ideas.\footnote{Usoskin et al. (2015) 
discuss in their section 2 only the two extremes, (i) and (ii), but not the third possibility,
which can be seen as some kind of a compromise between the former two,
based on a similar theory for the formation of comets as evaporations from
Earth in Aristotle's {\em Meteorology}.
Marius mentioned this possibility in his work on the large comet of 1618 (published
April 1619), see Sect. 3.3, where Marius may have considered that the currently low number of spots
were connected to the large comet of 1618.
Such a connection was also considered by Riccioli,
see Sect. 3.4.}

Zolotova \& Ponyavin (2015) may want to indicate that most observers with a 
Catholic background, in particular the Jesuits, 
favoured the {\em asteroid transit} hypothesis.
Apart from the fact that Galilei was confirmed by the catholic church in early years
that spots on the sun do not contradict directly any statements in the Bible,
opinions 
differed among different observers,
e.g. David Fabricius thought that they are transits of small bodies,
while his son Johann thought that they are on the surface of the sun,
both were protestants.
Hence, neither among the Catholics (e.g. Jesuits and Galilei)
nor among the protestants, the opinion was homogeneous:
in addition to the two opinions among the two
Fabricius, Marius added a third possibility, see above.
Independent of the confession, there was a scholarly dispute about the
nature of sunspots, as also mentioned by Marius (1614):
{\em I let other high, healthy, and sharp-thinking genius (people) think further on those things,
I do my part, others do their parts, given the grace of God,
one must start with it, and should help the other without any hate,
until one can conclude something with more certainty}, see Sect. 3.3,
and {\em I find the greatest authorities in disagreement}, see Sect. 4.1.
In particular, there was no {\em dominant world view} against non-circular spots
and no {\em crucial difference} between sunspots that were drawn and sunspots that
were reported (only) in texts.
Several observers have reported and/or drawn non-circular spots like Scheiner and Marius.

The only example given for the claim by Zolotova \& Ponyavin (2015),
that drawings and textual descriptions would differ, is the
case of Harriot's observation on 1610 Dec 8/18\footnote{We note that there was an aurora close in time,
namely on 1610 Dec 17, as observed in Kolozsvar in Romania (formerly Klausenburg):
{\em Den 17. Dezember wurde gegen Norden ein gro\ss es feuriges Kriegsheer gesehen von Abends bis 
gegen Mitternacht ...} (Rethly \& Berkes 1963),
i.e. {\em On 17 Dec, a large fiery war army was seen towards the north from the evening until about midnight},
which fulfils four aurora criteria (Neuh\"auser \& Neuh\"auser 2015), namely red color and some motion ({\em (fiery})
as well as northern direction and night-time, i.e. it is a {\em very probable} aurora.}
(the very first dated and documented telescopic sunspot observation), 
when he did draw three spots
and wrote in addition (cited after Reeves \& Van Helden 2010)
\begin{quotation}
1610 Syon, Decemb. 8, mane [Saturday]. 
The altitude of the Sonne being 7 or 8 degrees. 
It being a frost \& a mist. 
I saw the sonne in this manner. \\
Instrument. 10/1. B. I saw it twice or thrice, 
once with the right ey \& other time with the left. 
In the space of a minute time, after the Sonne was to cleare.
\end{quotation}
The wording used here ({\em to cleare}) means that the sun became too bright due to
increasing altitude and/or decreasing mist. This meaning becomes clear from the content.\footnote{In the
online {\em Oxford English Dictionary} on www.oed.com, it is given that {\em cleare} can
mean {\em fully light, bright} as opposed to dusk or twilight, as e.g. used in R. Grafton (1569) Chron. II. 100
{\em It was done in the cleare day light}, or {\em full of sunshine, bright} or {\em free from cloud, mists, and haze},
or in the sense of {\em clear weather ... in which the air is transparent 
so that distant objects are distinctly seen ... a sky void of cloud}.}
Zolotova \& Ponyavin (2015) want to construct a contradiction:
Harriot wrote {\em ''the Sun was clear'',  but it is accompanied by a sketch of three spots}.
The wording {\em to cleare} does not mean {\em spotless}, but just too bright to be observed.
Given that Harriot did draw the spots, there was no need to mention them in the caption text;
when he wrote {\em I saw the sonne in this manner} (with a larger than usual line break
afterwards in his hand-writing) he probably wanted to point to his drawing ({\em in this manner})
which was located to the right of those lines.
Vaquero \& Vazquez (2009), who also quote the text from Harriot and show his drawing,
do not see a contradiction here at all. This conclusion was also drawn by Usoskin et al. (2015).

\subsection{Observations reported by Malapert since 1618}

Some of the observations listed by HS98 for Malapert
were actually obtained 
not only by him:
e.g., Malapert indeed detected spot(s) on 1618 Mar 8, 10, 12-15, and 18 (his drawings in Malapert 1620 and 1633),
but Cysat in Ingolstadt, Germany, also detected spot(s) on 1618 Mar 8-11 and 17,
and Perovius in Kalisz, Poland, also detected spot(s) on 1618 Mar 9 --
consistent drawings for both the observers in Ingolstadt and Kalisz are given in Malapert (1633).
This adds spot detections on Mar 9, 11, and 17 not listed in HS98.
See Table 5 for an overview of dates, where the listing in HS98 has to be revised.

Malapert himself preferred the planetary transit hypothesis as explanations for sunspots --
his book (Malapert 1633) is entitled {\em The Austrian planets circling the Sun} 
(our English translation). 
He shows in his drawings mostly one spot each traversing the solar disk
(sometimes, some intermediate days are missing).
An important feature of his drawing is the fact
that the path of the spots is seen with reliable curvature
(given the B-angle on those dates).
In a few drawings, Malapert shows some variations like groups with several spots
(e.g. March 1618); his drawings remain vague regarding the size and form of the spots.
(We did not consult all texts from Malapert, but found a detailed description of
the spot of June 1620, see below.) 

We have used the curvature of the spot paths on the disk as drawn by Malapert (1633)
together with the given dates (month and day) to estimate the heliographic latitude. 
The paths are always 
aligned to the solar equator. 
(The curvature in the drawings for Malapert's own observations
are consistent with north being towards the upper part of the pages in Malapert (1633),
as he has indicated in the first few of his drawings.)
In Dec 1620, 
the path of the spot is very close to the equator and without curvature
(B-angle close to zero).
From 1618 to Dec 1620, the drawn spots tend to appear at lower and lower latitude -- 
with one exception: in Oct 1620, the path appears at relatively high latitude --
which was immediately noticed by Malapert, see Fig. 12
(similar also in Sep and Nov 1621).
Wely also measured the high southern latitude of the spot in Oct 1620,
according to the figure caption in Malapert (1633).

Malapert's observations are credible: 
(a) Reports, drawings, and measurements (of the separation
of spots to the solar limb) by others (as given in Malapert 1633) are all consistent
with his own observations. 
%XXX%
(b) His detections in May 1625 are on the very same days as obtained by Scheiner in Rome (HS98). 
(c) His detections in Nov 1621 are partly on the same days as obtained by
Scheiner in Rome. 
(d) His detections on 1618 June 21 \& 22 are obtained on the same day as naked-eye detections in China.
(e) For the time, when Argoli reported spotlessness, namely during the comet sightings in 1618
(Sect. 3.4), Malapert did not report spots.

Malapert has shown two partly different drawings for 1618 Mar 8-18 in his books from 1620 and 1633:
while we can see at least one spot group close to the ecliptic in Malapert (1633) for Mar 8, 10, 12-15, and 18,
we can see not only the same spot group close to the ecliptic in Malapert (1620), but also one more group (made up
by one to two spots) further north, the latter group labelled B, the former A.
Spot group B was seen by him on Mar 8, 10, and 12-15, i.e. on the same days as group A except on Mar 18,
when group B was not seen any more (Western edge).
In the drawings shown in Malapert (1633) for the observations of Cysat in Ingolstadt 
and Perovius in Kalisz, we can also see only one spot group, the one labelled A in Malapert (1620).
The spot drawing for 1618 Mar 8-18 is the only one shown in Malapert (1620), where he also discussed
stars in the Orion nebula, the four large moons of Jupiter, the ring around Saturn, and craters on the Moon.
The fact that Malapert (1620) shows an additional spot group for March 1618, which is not shown in Malapert (1633),
may indicate that he -- at least in this case (March 1618 in Malapert 1633) -- 
shows only one spot group at a time, i.e. one group
per drawing, e.g. in order not to make the drawing too complicated or because he wanted to show
how one group moved from the East to the West.
For 1618 Mar 8, 10, and 12-15, the group count has to be changed from 1 in HS98 to at least 2,
the second group is shown in Malapert (1620); even in Malapert (1633), 
one could count one or two groups in the drawing for March 1618 --
for most days, one (or two) group were resolved into several spots.

The detection of two sunspot group by Malapert in March 1618 (Malapert 1620 drawing)
is not inconsistent with Marius, who reported that there were more often spotless days
from fall 1617 to spring 1619, but who also did mention explicitely that there were spots
in that period (Sect. 3.3).

On the very last page in Malapert (1633), 
in his last chapter, a detailed discussion of peculiar (otherwise undated) observations,
he mentioned that he saw a special spot in 1620 June 6 \& 7:
\begin{quotation}
Quomodo evenisse arbitror ut anno 1620. die 6. Iunii macula quaedam circa longitudinem 50, 
latitudinem vero Australem 30 comparens, die sequenti retrocessisse videretur; 
neque ultra hoc biduum macula illa mihi conspecta est.
\end{quotation}
We translate this to English as follows:
\begin{quotation}
I suppose it happened in a way as in the year 1620 on 6th of June, 
when a certain spot at about 50 [degrees] longitude and
indeed at a southern [Australem] latitude of 30 [degrees] appeared,
on the following day it was seen to have weakened [retrocessisse];
and that spot was observed by me not longer than for those two days.
\end{quotation}
We can conclude that there was a spot (or group) on June 6 \& 7, 
and that at least June 5 and 8 may have been spotless (Malapert would not report appearance or
disappearance only due to clouds). 

Since the solar equator is near the apparent center of the solar disk in June, 
we can conclude that those $30^{\circ}$ south is 
definitely far south and belongs to the new Schwabe cycle.
Given that the first high-latitude spots drawn for this new cycle, 
1620 Oct as well as 1621 Sep and Nov,
are also in the south, we may conclude that the southern hemisphere was leading
the spot production in that new Schwabe cycle.

The first spot at high latitude was seen 
on 1620 June 6 and 7, 
and is not listed in HS98.
Since all the spots shown in the drawings based on his observations are seen to move from the 
Eastern to the Western edge, we may be able to conclude
that Malapert obviously wanted to show in his drawings only spots seen to move 
all the way from one edge to the other --
because he favored the transit theory for spots.

Since Malapert found the very southern spot in June 1620 to be unususal, he obviously had started
his observations not long ago (1618), namely since there were no high-latitude spots anymore.
From Marius and Argoli, we know that solar activity was weak and decreasing since about 1617 or 1618.
Therefore, there were probably not many spots visible to move all the way from the eastern to
western edge anymore. His reports on such spots to move all the way may be rather complete
for the time until 1620. 

\begin{table}
\caption{{\bf Observations by Malapert}  
(1620, 1633), but for 1618 and 1620 only,
where the data base by HS98 has to be modified$^{a}$.
The group sunspot number on active days is always 1 in HS98.
Our checkmarks show only that the observer has deetcted at least one spot
or group, but we do not specify or confirm the number.
}
\begin{tabular}{r|c|cccc|c} \hline
Date       & HS & Mal- & Pero-  & Wely & Cy- & No \\
           & 98  & apert & vius &      & sat & te   \\ \hline
1618 Mar 8 & 1 & \checkmark & - & - & \checkmark & b,c \\
         9 & - & - & \checkmark & - & \checkmark & c \\
        10 & 1 & \checkmark & - & - & \checkmark & b,c \\
        11 & - & - & - & - & \checkmark & c \\
     12-15 & 1 & \checkmark & - & - & - & f \\
        16 & - & - & - & - & - & f \\
        17 & - & - & - & - & \checkmark & c \\ 
        18 & 1 & \checkmark & - & - & - & f  \\ \hline
1618 Jun 21-29 & 1 & \checkmark & - & - & - & f \\ \hline
1618 Jul 7 & 1 & \checkmark & - & - & - & f  \\
         8 & - & - & \checkmark & - & - & c \\
         9 & 1 & \checkmark & - & - & - & f  \\
     10-12 & - & - & - & - & - &  f \\
        13 & 1 & \checkmark & \checkmark & - & - & c \\
  14 \& 15 & 1 & \checkmark & - & - & - & f  \\
        16 & - & - & - & - & - & f  \\
        17 & 1 & \checkmark & - & - & - &  f \\
        18 & 1 & \checkmark & \checkmark & - & - & c \\
        19 & - & - & \checkmark & - & - & c \\ \hline
1620 Feb 17-20 & 1 & \checkmark & - & - & - & f  \\
            21 & - & - & - & - & - &  f \\
            22 & 1 & \checkmark & - & - & - & f  \\
            23 & - & - & - & - & - &  f \\
         24-28 & 1 & \checkmark & - & - & - & f  \\ \hline
1620 Apr 11 & 1 & \checkmark & - & - & - & f  \\
   12 \& 13 & - & - & - & - & - &  f \\
      14-21 & 1 & \checkmark & - & - & - & f  \\ \hline    
1620 Jun 6 \& 7 & - & \checkmark & - & - & - & g \\ \hline
1620 Oct 21 & 1 & - & - & \checkmark & - & d \\
         22 &1 & \checkmark & - & \checkmark & - & d \\
         23 &  - & - & - & - & - &  f \\
         24 &1 & - & - & \checkmark & - & d \\
         25 & 1 & \checkmark & - & - & - & f  \\
         26 &1 & - & - & \checkmark & - & d \\
         27 &1 & \checkmark & - & \checkmark & - & d \\
   28 \& 29 &1 & - & - & \checkmark & - & d \\
         30 &1 & \checkmark & - & \checkmark & - & d \\ 
         31 &- & - & - & \checkmark & - & d \\ \hline
1620 Dec  2 &1 & \checkmark & - & - & - & f  \\
       3-5 & 1 & - & - & \checkmark & - & d \\
          6 & - & - & - & - & - &  f \\
          7 &1 & \checkmark & - & \checkmark & - & d \\
       8-10 & - & - & - & - & - &  f \\
         11 &- & - & - & \checkmark & - & d \\
         12 &1 & \checkmark & - & \checkmark & - & d \\
         13 &1 & - & - & \checkmark & - & d \\ \hline
\end{tabular}

Notes: 
$^{a}$ Plus new spots by Wely on 1621 Sep 13, 14, 15, \& 16,
and on 1621 Nov 30 by Malapert, not in HS98.
$^{b}$~Malapert (1620): one more group on 1618 Mar 8, 10, and 12-15.
$^{c}$~Drawings from Perovius, Cysat, and Malapert.
$^{d}$~Separation of spot from solar limb from Wely given 
in figure caption in Malapert (1633). 
$^{f}$~No changes compared to HS98.
$^{g}$~Reported in the last sentences in Malapert (1633), spotless on June 5 \& 8.
\end{table}

\subsection{Dating the Schwabe cycle minimum}

In the context of discussing the period of 1616 to 1623 with a Schwabe cycle minimum,
Zolotova \& Ponyavin (2015) argue {\em that the reports of that period reflect a sunspot transit,
but not the exact number of spots};
as example, they give 1626 with Malapert (always 1-2 groups according to HS98)
and Scheiner (drawing several groups). Zolotova \& Ponyavin (2015):
{\em this finding supports our idea that group sunspot number extracted
from the text sources without drawings are underestimated};
Zolotova \& Ponyavin (2015) probably assume that HS98 got the number (1-2 groups) for Malapert
from his text, but in fact Malapert (1633) shows drawings for almost each month in 1626.
On the contrast, we have shown that Tard\'e (1620) has drawn several
times one spot only, but wrote that there were several spots on the disk
(details in Sect. 4.8). When there was really one spot only, he mentioned
this fact explicitly. Images are not necessarily the better evidence.

Zolotova \& Ponyavin (2015) argue that the presumable spot minimum in
1617-1618 (in HS98) would not be present 
partly because Marius would 
have stopped observing whenever the sun became active.
Regardless of the timing of the minimum, the argument cannot be supported: 
that Marius did see spots in 1617/1618(/1619) and
did not stop observing when the sun became active, was already shown above.

Wolf (1856) dated the first telescopic Schwabe cycle minima to be
1611.11, 1622.22, and 1633.33;
later on, Wolf (1858) dated the first to
1610.8 or 1611.0 with the data from Harriot,
HS98 dated the first Schwabe cycle minima as follows: 1610.8, 1619.0, 1634.0.
The minima in the yearly group sunspot numbers in HS98 lie in 1617 and 1618 (Table 4),
so that the minimum would lie close to 1618.0.
From the active day fractions calculated with the values in table alldata from HS98,
one would also get a minimum in 1617 and/or 1618, see Table 4. 
HS98 then give group sunspot numbers being 15.0 for all years from 1619 to 1623.

Let us now present the timing of the Schwabe cycle minimum 
(we compile all known spots and aurorae around the minimum in Table 6):
the information from the historic report from China (Sect. 3.2) and from
Europe are fully consistent with each other, e.g. Malapert 
confirms a Chinese naked-eye spot in June 1618 by telescopic observations (Fig. 2).
Marius gives detailed information for the time span fall 1617 to spring 1619
that there were much less spots and more often no spots (compared to before fall 1617);
after the naked-eye spot sighting in May/June 1618,
there are no reports about naked-eye sightings in 1619, 
the next one is then in October 1620,
again simultaneous with a telescopic spot by Malapert.
Malapert observed one spot (group) each 1619 Jan 13-20
(a {\em very probable} aurora was seen also that month in Korea, Jan 4-7)
and then again in Aug 1619.
In 1620, telescopic spots were recorded for Feb 17-28, Apr 11-21, Jun 6 \& 7, Oct 21-31, and Dec 2-13, 
(at least) one group each by Malapert (HS98).
Tard\'e (1620) reported spotless days, even on subsequent days, for his observing
period, i.e. 1615 to 1619, see Sect. 4.8 -- also Scheiner for 1616 (HS98).
Long breaks without 
dated spot sightings were from Aug-Dec 1618 (mostly with comets, Sect. 3.4), 
Feb-Jul 1619, and Sep 1619-Jan 1620.
There is stronger activity since 1621 (see Table 6).

The active day fraction shows a general decrease from $\simeq 1$ in 1611-1615
to much lower values in 1618/1619;
from the qualitative report by Marius the active days fraction
was 1.0 until fall 1617, and for the period from fall 1617 to spring 1619,
the active day fraction was below 0.5 (but not zero),
consistent with all (but few) observations by others;
for the time after spring 1619, we do not have any generic statements like those from Marius;
for 1620, the value lies anywhere from 0.11 to 0.99, see Table 4,
based on reports and drawings by Malapert 
(at least 38 active days and at least two inactive days giving a lower and an upper limit, respectively).
For the years 1621 to 1623, the number is 1 according to scattered data in HS98.

As seen in Table 6 and Fig. 11-13, the last (known, drawn) spots at low latitude (previous cycle)
were seen in Apr and Dec 1620, while the first {\em drawn} (and reported) spots at high latitude 
were seen in Oct 1620
(partly simultaneous with a naked-eye spot in China, i.e. a very large spot/group); 
in addition, Malapert (1618) also reported a high-latitude
southern spot ($30^{\circ}$ {\em south}) for 1620 June 6 \& 7 (without drawing).
It is well-known that there is an overlap in
time between the last low-latitude and first high-latitude spots, i.e. an overlap in time
between two neighbouring cycles, see e.g. butterfly diagram in Hathaway (2010).

\begin{figure}
\begin{center}
{\includegraphics[angle=0,width=7cm]{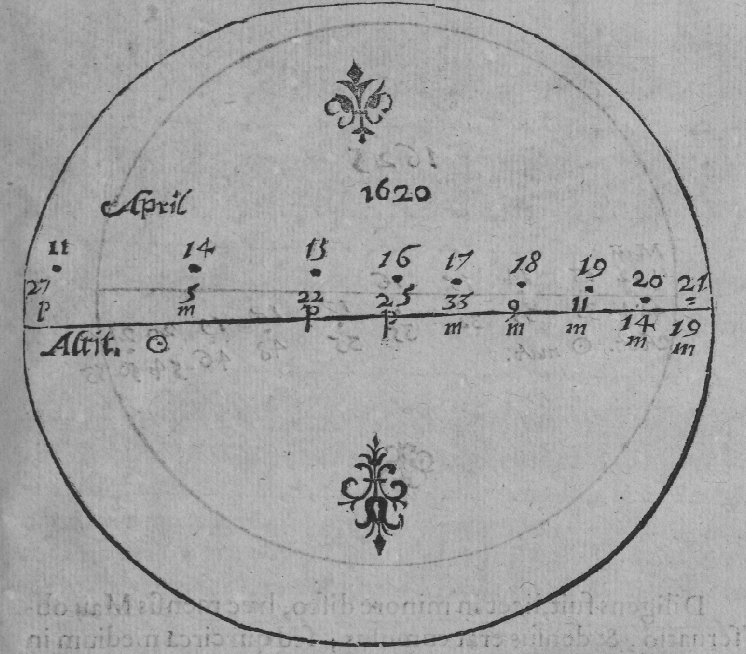}}
\end{center}
\caption{Telescopic sunspot drawing by Malapert (1633) for 1620 April 
(observing dates are given above the spots, observing time in minutes before noon
({\em m} for {\em mattuttina}) or after noon ({\em p} for {\em po[st]meridianam}) below the spots,
labbeled as {\em Altit. [of the] sun}) 
with low-latitude spots
belonging to the ending Schwabe cycle. These are the last drawn spots before
the first spots at higher latitude (Fig. 11).
(As it often happens in old books, some ink from the back page is seen.)}
\end{figure}

\begin{figure}
\begin{center}
{\includegraphics[angle=0,width=7cm]{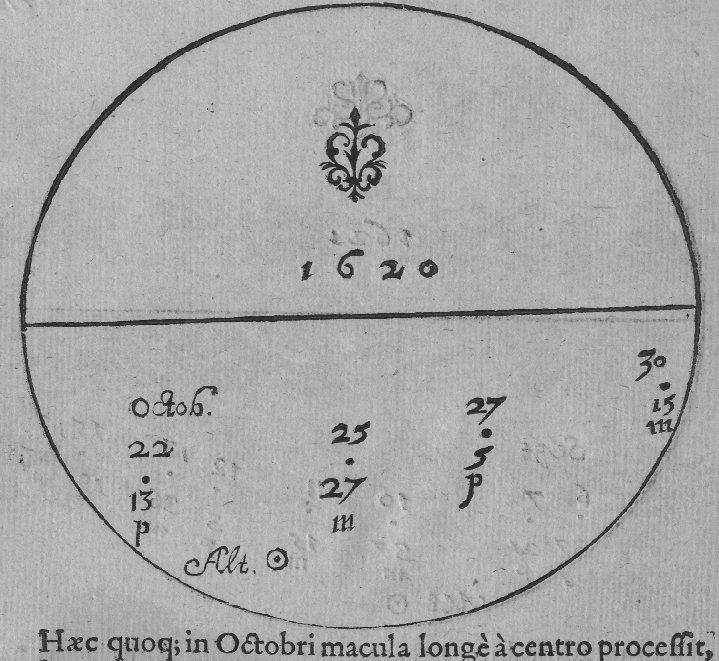}}
\end{center}
\caption{Telescopic sunspot drawing by Malapert (1633) for 1620 Oct 
(labbeled by him as in the previous figure) with high-latitude spots
belonging to the new Schwabe cycle. The spots in 1621 -- as drawn by
Malapert (1633) -- are all at high latitude. For the spot seen in Oct 1620, Malapert (1633)
wrote in his figure caption {\em in Octobri macula longe a centro processit},
i.e. that {\em in October, a spot travelled at large separation from the center},
indeed at high heliographic latitude indicating the start of a new Schwabe cycle.
Malapert noticed the same effect on the spot in Sep 1621.
While this is the first high-latitude spot {\em drawn} by Malapert (1633), he did mention a spot 
for two days at {\em $30^{\circ}$ southern} latitude for 1620 June 6 \& 7 (only text, no drawing) --
this is the first spot reported from the new cycle.
The Chinese reported a spot for some day(s) during Oct 15-24, i.e. probably simultaneous.
See also Fig. 2. In his figure caption, Malapert (1633) mentions that Wely could still
detect the spot on Oct 31, and gives his separation measurement.
We note that Wely also measured the high southern latitude of this spot.
}
\end{figure}

\begin{figure}
\begin{center}
{\includegraphics[angle=0,width=7cm]{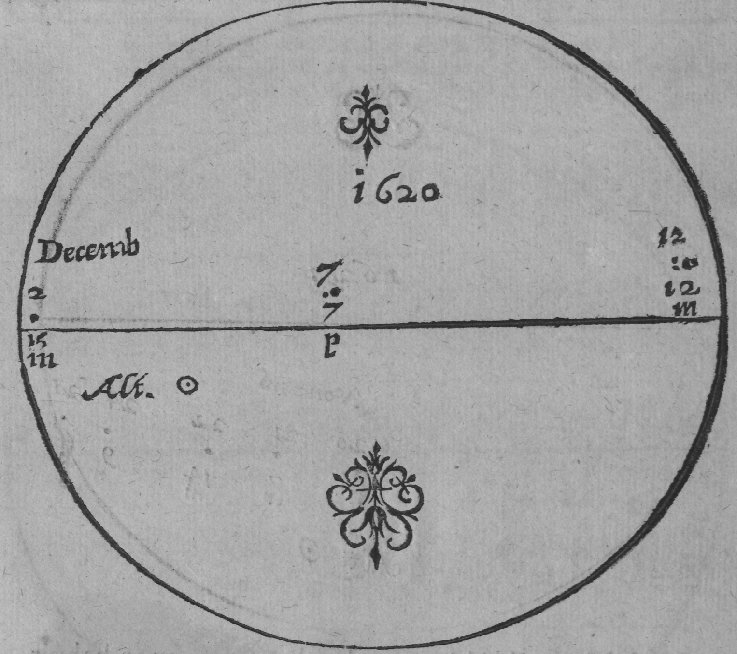}}
\end{center}
\caption{Telescopic sunspot drawing by Malapert (1633) for 1620 Dec 
(labbeled by him as in the previous figures)
with the last known low-latitude spot
belonging to the ending Schwabe cycle, 
shortly after the first reported and/or drawn new spots at high latitude from the new cycle.}
\end{figure}

Taking all evidence together, 
we can date the transition from one Schwabe cycle to the next one (minimum)
to somewhere between June and Dec 1620.
(The active day fraction for 1620 in not well constrained (0.11 to 0.99) due
to only 41 observing days, so that it is not inconsistent with the occurence of a minimum.
It is interesting to note that the first sunspot minimum in the telescopic era
is datable by the transition in spot latitude and that this latitude change 
was noticed by Malapert (Figs. 11-13).
This minimum was not particularly deep: there was no separation in time between
the last low-latitude and the first high-latitude spot, and there was not a full year
without spots.

\section{Summary}

Hoyt \& Schatten (1998) list Marius only for 1617 and 1618, 
but without any spot detections (Sect. 3.1).
They cite, but mis-interprete Wolf (1858) and Zinner (1942, 1952).
Zinner (1942, 1952) in fact 
wrote that Marius observed sunspots from August 1611 
until at least 1619\footnote{Zinner (1942):
{\em Marius hat die Sonne nach Flecken von August 1611 bis mindestens 1619 nachgesehen}, i.e.
{\em Marius has searched for sunspots from August 1611 until at least 1619}.
Vaquero \& Vazquez (2009) also mention that Marius {\em was also an observer of sunspots in 1611}.},
and Wolf (1857) gave explicitely the essential quotation from Marius 
regarding the time 1617-1619 (Sect. 3.3).

\begin{table*}
\caption{{\bf Sunspots and aurorae 1617--1621 around the Schwabe cycle minimum.}
We list here the sunspot and aurora observations from 1617 to 1621
in order to date the Schwabe cycle minimum at that time. For aurorae, we give the number
of aurora criteria fulfilled for the text found in the reference listed -- according to the
five aurora criteria given in Neuh\"auser \& Neuh\"auser (2015): night-time, colour, dynamics,
northern direction, and repetition. Sightings, which were listed by others as potential aurorae,
but which do not fulfil any of the criteria, are not listed (except in Note a).
In the last row, we indicate whether there was
or may have been a (quasi-)simultaneous sighting of both aurorae and sunspots, 
i.e. whether aurorae were reported 
within a few days 
of sunspots sightings (see Willis et al. 2005 for other such cases),
or whether there were simultaneous naked-eye and telescopic sightings. 
For aurorae possibly simultaneous with sunspots, we list the source text in the Notes.} 
\begin{tabular}{l|llll|c} \hline
Date (Gregorian)   & sunspots                & aurora   & observer    & Remarks, Ref. & quasi- \\ 
                   &                         & criteria & or location & or Section    & simultan \\ \hline
before 1617 fall   & many spots              &          & Marius      & Sect. 3.3 & \\  \hline
1617 Jan 11        & several naked-eye spots &          & China       & Sect. 3.2 & \\
1617 May 27-June 6 & 1 spot group            &          & Tard\'e     & Sect. 4.8 & \\ 
1617 fall-1619 spring & few or no spots      &          & Marius      & Sect. 3.3 & \\ \hline
1618 Mar 8-18      & spots and groups        &          & Malapert et al. & Sect. 7.2 & \\
1618 May 17        &                         & 2  (e)   & China       & Yau et al. \& Xu et al. & sim \\
1618 May 22        & 1 naked-eye spot group  &          & China       & Sect. 3.2 & sim \\
1618 Jun 20-22     & 1 naked-eye spot group  &          & China       & Sect. 3.2 & sim \\
1618 Jun 21-29     & 1 spot group            &          & Malapert    & Fig. 2 & sim \\
1618 Jul 7-19      & 1 spot group            &          & Malapert et al. & Sect. 7.2 & sim \\ 
1618 Jul 19        &                         & 2  (f)   & China       & Yau et al. &  sim \\
1618 Nov 17        &                         & 2        & Korea       & Yau et al. & cor. hole? (m) \\
1618 Dec 14        &                         & 1        & Korea       & Yau et al. & cor. hole? (m) \\
1618 Aug-Dec (k)   & no spots                &          & Argoli      & Sect. 3.4 & \\
1617 fall-1619 spring & few or no spots      &          & Marius   & Sect. 3.3 & \\ \hline  
1619 Jan 4-7       &                         & 4        & Korea    & Yau et al. \& Xu et al. & \\
1619 Jan 13-20     & 1 spot group            &          & Malapert & low latitude, Sect. 7.2 & \\
1619 Aug 16-28     & 1 spot group            &          & Malapert & low latitude & \\ \hline
1620 Feb 3         &                         & 2        & China    & Xu et al. & \\
1620 Feb 17-28     & 1 spot group            & (a)      & Malapert & low latitude, Sect. 7.2 & (a) \\
1620 Apr 11-21     & 1-2 spots               &          & Malapert & low latitude, Fig. 11 & \\
1620 Jun 6 \& 7    & 1 spot                  &          & Malapert & $30^{\circ}$ {\em south}, Sect. 7.2 & \\
1620 Aug 19        &                         & 1        & China    & Xu et al. & \\
1620 Oct 19        &                         & 1 (g)    & China    & Yau et al. & sim \\ 
1620 Oct 15-24     & naked-eye spot          &          & China    & Yau \& Stephenson 1988 & sim (i) \\
1620 Oct 22-31 (b) & 1 spot group            &          & Malapert et al. & high latitude, Fig. 12 & sim (d,i) \\
1620 Dec 2-13  (c) & 1-3 spots               &          & Malapert et al. & low latitude, Fig. 13 & (d) \\ \hline
1621 Jan 9-11      & 1 group                 &          & Schickard & Ref. HS98 & \\
1621 May 2         &                         & 3        & China     & Xu et al. & \\
1621 May 23        & naked-eye spot(s)       &          & China    & Wittmann \& Xu 1987 & \\
1621 Sep 6-16      & 1 spot group            &          & Malapert & high latitude & sim \\
1621 Sep 12        &                         & (h)      & Europe/Syria & Fritz 1873 & sim \\
1621 Sep 26-30     & 1 spot group            &          & Scheiner & Ref. HS98 & \\
1621 Oct 1-15      & 1 spot group            &          & Scheiner \& Smoguleczz & Ref. HS98 & \\
1621 Oct 5-Nov 1   & 1 spot group            &          & Smogulez & Ref. HS98 & sim \\
1621 Oct 25-31     & 1 spot group            &          & Scheiner & Ref. HS98 & sim \\
1621 Oct 15-Nov 12 & naked-eye spots         &          & China    & Wittmann \& Xu 1987 & sim (l) \\
1621 Nov 20-25     & 1 spot group            &          & Malapert & high latitude & \\ 
1621 Nov 16-25     & 1 spot group            &          & Scheiner & Ref. HS98 & \\  \hline
\end{tabular}

Notes:
$^{a}$ A potential aurora is given by Fritz 1873 for Istanbul for 1620 Feb 22 without text, which could be simultaneous;
according to the drawing by Malapert (1633), a spot (group) was near the center of the solar disk on Feb 22.
$^{b}$ Different in HS98 (Oct 21, 22, 24-30), but see our Fig. 12 from Malapert (1633) and his caption. 
$^{c}$ Different in HS98 (Dec 2-5, 7, 12, 13), but see Malapert (1633). 
$^{d}$ The Schwabe cycle minimum lies here.
$^{e}$ {\em evening 2 bands green/blue-black vapour E to W}.
$^{f}$ {\em 3 vapours red, green, white hanging down ... they wavered}.
$^{g}$ {\em red vapour across sky}.
$^{h}$ Strong aurorae seen in Western Europe, Venice, Italy (by Galilei), and Aleppo, Syria (Fritz 1873).
$^{i}$ The Chinese spot observation gives the {\em decade} Oct 15-24, i.e. a 10-day-period, 
so that their observation could have been on any one or more days in that period;
if it refers to the same spot (group) as observed by Malapert and Wely Oct 22-31, see Fig. 12, 
then the Chinese probably saw the spot simultaneous with Malapert.
$^{k}$ 1618 Aug 25 - Sep 25 and since Nov 11, while comets were seen.
$^{l}$ Seen sometime during the given month.
$^{m}$ The difference of one solar rotation period could indicate a coronal hole aurora,
typical for the declining phase.
\end{table*}

With the original texts written and published by Marius,
we could find the following (often well datable) information on his sunspot observations:
\begin{itemize}
\item Simon Marius and Ahasverus Schmidnerus together saw at least one spot or group on 1611 Aug 3/13 (Sect. 4.1).
\item Marius observed spots {\em in very large numbers} from 1611 Aug 3 to Dec 29 (Julian),
consistent with David Fabricius, Scheiner, and Harriot (Sect. 4.1 \& 4.5);
Marius saw spots {\em always in different form}, with {\em their daily motion},
and that they {\em do not cross the disk of the sun on the
ecliptic, but build an angle with it}.
This indicates regular observations.
\item Marius improved his observing technique
on 1611 Oct 3/13 and/or Oct 11/21, and he detected at least one spot on that (or those) day(s) (Sect. 4.2).
\item Marius draw sunspots at least once, namely on 1611 Nov 17/27 (also drawn by Scheiner for that day) (Sect. 4.4).
\item Marius reported 14 spots for 1612 May 30 (Julian), which is probably his largest
daily number until 1612 June 30 (same for Galilei) (Sect. 4.6).
\item Simon Marius may have observed spot(s) together with Petrus Saxonius on 1615 Jul 4/14 (Sect. 4.7).
\item Marius observed not only many spots since 1611 Aug 3/13,
but had no spotless days before fall 1617, 
i.e. before the period of 1.5 years ending 1619 Apr (Sect. 3.3 \& 4.9);
this implies an active day fraction of 1.0 before fall 1617.
\item Marius observed spots in a period of (roughly) 1.5 years 
before 1619 Apr (Sect. 3.3), but much less than before.
\item With his statement that there were more often spotless days in those
(roughly) 1.5 years before 1619 Apr, together with limits on the active day fractions
from Malapert et al. (Sect. 7.2) and Scheiner (HS98), we could
constrain the active day fraction to $> 0.08$ and $< 0.5$ for 1618 (Sect. 6, Table 4).
\item The generic statement by Marius also constrains Schwabe cycle minima
to lie before August 1611 (no spotless days seen Aug 1611 to fall 1617)
and in or after 1619, consistent with other observers and a typical cycle length;
the maximum was somewhere 1612 (14 spots by Marius and others) to 1615 (30 spots by Tard\'e).
\end{itemize}

Given that Marius reported in 1619 that there were no spotless days before about fall 1617
(Sect. 3.3), we could conclude that he did not observe on those days from 1611 Aug 3/13
until at least end of 1616, when others noticed a spotless sun, namely on 16 days --
or that he could detect spot(s) when others missed them (Sect. 4.9).

The observations by Marius since 1611 Aug 3/13 are among the very first telescopic sunspot records.
Harriot has detected his first three telescopic spots in England on 1610 Dec 8/18 (HS98, 
see also our Sect. 7.1),
and Johann and David Fabricius in northern Germany had detected their first spot on 1611 Feb 27 and 28 (Julian),
hence 1611 Mar 9 and 10 (Gregorian), Sect. 4.3;
also, Scheiner and Cysat observed spots in March 1611 (Sect. 4.5).

All essential elements in the statements on spots by Marius for 1611--1619 can be confirmed,
while no parts were falsified (Sects. 3, 4, 7.2, 7.3), so that his texts are highly credible.

We presented the following additional correction and additions with respect to HS98:
\begin{itemize}
\item David and Johann Fabricius observed 1-3 spots in March 1611 (starting on Mar 9 and 10 Gregorian), Sect. 4.3.
\item Scheiner and Cysat detected spots on one day in March 1611 from Ingolstadt (Sect. 4.5).
\item Scheiner and other Jesuits detected spots on 21 Oct 1611 from Ingolstadt (Sect. 4.5).
\item Tanner detected spots on 1611 Oct 21 and also on most of the remaining days that year 
and throughout 1612, also from Ingolstadt:
{\em from this time [1611 Oct 21] on in the course of the following year,
I have almost daily observed [spots] in different ways};
this generic statement means that his active day fraction was 1.0 from 1611 Oct 21 until
the end of 1612 on the days he observed (Sect. 4.5).
\item Dates for observations by Harriot in June 1612 as listed in HS98
do not correspond to his drawings (Table 3, Sect. 4.6).
\item The observation of 30 spots in up to some ten groups by Tard\'e (1620) 
was neither on 1615 Mar 25 nor Aug 15 (as in HS98),
but on 1615 Aug 25, as clearly given in Tard\'e (1620), Sect. 4.8. 
\item Group numbers for Tard\'e reported by HS98 for May and June 1616 are not correct (footnote 16):
There was one spot/group for 1616 May 17-28 -- just one on May 27 \& 28 and none 
from 1616 May 19 to June 6 (Sect. 4.8).
\item The dates given by Saxonius for 1616 are Julian dates, which were not transformed
to the Gregorian calendar in HS98 (Sect. 4.8).
\item Group sunspot numbers from Tard\'e have to be regarded as lower limits,
except for 1617 May 27 to June 6, where he specifies to have seen only the one spot drawn
(and of course 1615 Aug 25, when he saw 30 spots), Sect. 4.8 \& 7.1.
\item The period of a spotless sun reported by HS98 for 1618 based on Riccioli is based
on observations by Argoli and has to be restricted to periods when comets were seen (Sect. 3.4),
it is consistent with Marius (Sect. 3.3), Malapert (Sect. 7.2), and Tard\'e (Sect. 4.8), see Table 6.
\item The period of a spotless sun reported by HS98 for 1632 July 12 to Sep 15 (based on Riccioli)
is not correct, because Riccioli gave an incorrect year in his quotation of Argoli, who gave 1634 (Sect. 3.4).
\item The sunspot drawings by Malapert (1633) for 1618 to 1621,  
show several differences compared to the numbers in HS98: sometimes small spots
together with large groups, etc., while HS98 always give 1 (to be studied in detail later);
Malapert (1633) shows three additional drawings for March and July 1618 obtained in Ingolstadt
and Kalisz; Wely has observed 1621 Oct 21, 22, 24, and 26-31, 
while Malapert observed only on Oct 22, 25, 27, and 30 (Fig. 12),
Wely also detected a spot on Oct 31 according to the figure caption in Malapert (1633);
for 1620 Dec, Malapert observed only on Dec 2, 7, and 12 (Fig. 13), 
while Wely observed on Dec 3-5, 7, and 11-13;
in 1621 Sep, there are additional observations by Wely for Sep 13-16;
Malapert (1633) added in the figure caption that he saw one more spot in 1621 Nov 30
not listed in HS98.
Malapert (1633) also mentioned one spot for 
1620 June 6 \& 7, given with very southern heliographic position, 
which he did not draw, 
he also mentioned indirectly that June 5 and 8 were spotless,
the information on these two spots and the spotless days are not listed in HS98 (Sect. 7.2).
Perovius is not listed in HS98 (but reported by Malapert 1633), he has observed
on 1618 Mar 9, Jul 8, and Jul 19 (where HS98 do not list Malapert, which is correct)
and on 1618 Jul 13 and 18 (where HS98 list correctly Malapert).
Cysat is also not listed in HS98 (but reported by Malapert 1633), he has observed
on 1618 Mar 9, 11, and 17 (where HS98 do not list Malapert, which is correct)
and on 1618 Mar 8 and 10 (where HS98 list correctly Malapert).
%XXX$
\item In Malapert (1620), an additional group is seen for 1618 Mar 8, 10, and 12-15,
{\bf
also seen in the drawing in Scheiner (1626-30) based on the observation by Malapert,
but HS98 listed one group for both Malapert and Scheiner (Sects. 3.1 \& 7.2). 
}
\item Most spot/group numbers for Malapert may have to be regarded as lower limits,
but activity was anyway low in his first observing years 1618-1620 (Sect. 7.2).
\end{itemize}

We discussed our strong systematic concern regarding the system used by HS98 and Equ. (2)
in Sect. 5;
furthermore, several early observers for the 1610s are missing in the HS98 data base:
David and Johann Fabricius, Cysat, Schmidnerus, Wely, Perovius, Argoli, and Tanner;
Marius is listed in HS98 only for 1617 and 1618, but observed from 1611 to 1619;
HS98 list Riccioli for 1618 instead of Argoli.
Marius, Tanner, and Argoli provide valuable generic statements;
there are even more telescopic sunspot observers in the 1610s, e.g. Castelli (Sect. 2.2).

The main change resulting from our revision affects the years 1617 and 1618,
where the yearly group sunspot numbers change from 0.8 and 1.3 in HS98, respectively, 
to 15.2 in both years, i.e. they both increase strongly (Sect. 3.3 and Table 4).
While these numbers are all smaller than those for the early and mid 1610s,
the yearly group sunspot numbers remain to be 15 from 1619 to 1623 in HS98.

Marius clearly said that there were (significantly) less spots in the 
those roughly 1.5 years before April 1619, i.e. from about fall 1617 to spring 1619
(Sect. 3.3).
Group sunspot numbers cannot be calculated with trends or limits or undated observations,
as reported by Marius (and others), which demonstrates a major problem
in the group sunspot number system by HS98 and in Equ. (2).

The situation is also problematic for the value of the active day fraction:
When omitting all generic zeros in HS98, the active day fraction in 1617 and 1618
would be 1, while Marius clearly reported that more than half the observing days 
from fall 1617 to spring 1619 were spotless (but also note the small size
of his telescope and, hence, his detection limits). 
With our careful analysis of the historic reports, we could constrain the
active day fraction for 1618 to larger than 0.08, but smaller than 0.5.
This is clearly lower than during the years 1611 to 1615, when is was $\simeq 1$ 
(Sect. 6, Table 4).

We could date the turnover from one Schwabe cycle to the next to about Jun-Dec 1620, 
when the last spots drawn by Malapert at low latitude (Dec) and the first spot 
explicitly mentioned at high latitude (June 6 \& 7) were seen by Malapert (Sects. 7.3);
this is consistent with both apparent breaks in credible aurora and sunspot sightings
and the reports from Marius, Argoli, and Tard\'e (Sects. 3.3, 3.4, 4.8, Table 6).
The Schwabe cycle minimum may be dated around that time (Sect. 7.3).

We could show that some naked-eye spots reported by the Chinese are
consistent with telescopic observations, e.g. 1618 June (Fig. 2, Sect. 3.2).
We also notice that there were cases, where aurorae were detected within a
few days of the sunspot sightings reported here, 
namely in the years 1610, 1611, and 1612 (Sect. 4), 
but also see Table 6.

In addition, we found evidence for the following facts that could also
be relevant for the history of telescopic sunspot observations
(mostly found in the texts by Marius):
\begin{itemize}
\item Marius and Schmidtnerus with their observation
on 1611 Aug 3/13 were among the first known telescopic sunspot observers (Sect. 4.1).
\item Marius changed the observing technique on 1611 Oct 3/13 and/or 11/21 (Sect. 4.2):
He used the Camera Helioscopica (Marius 1612), apparently earlier than 
Castelli, Galilei and Scheiner, 
and improved it by directing the telescope to a (perpendicular) white screen, 
explained in the foreword to {\em Mundus Iovialis} (Marius 1614);
he could then detect and draw the daily motion of spots.
His observing techniques would be worth another investigation
(see also Wolfschmidt 2012a).
\item A theory of conneting spots with comets may have been invented by Marius,
partly based on a writing by Cardanus about a connection of comets with the Sun (Sects. 3.3 \& 3.4). 
\item Marius noticed that the path of spots form an angle with the ecliptic (Sect. 4.2). 
On the contrast, Tard\'e wrote that the path of spots is on the ecliptic or
parallel to it, and he did not draw the path with curvature (Sect. 4.8),
while Malapert has drawn the path with correct curvature (Figs. 11-13).
\item {\em In the year 1611,} Marius {\em have found a method to observe the colours of the stars.}
He added in the Prognosticon for 1616: {\em the significant change in the colours of several fixed stars
... was first noticed with naked eye by Mr. M. Mastlino and Mr. Keplero, and it was rather clearly seen
by myself through the Perspicillum} [telescope].\footnote{In original German: {\em eine merckliche
Abwechsslung der Farben in etlichen Fissternen ... von Herrn M. Maestlino vnd Herrn Keplero erstlich
durch freyes Gesicht ist vermerckt vnd von mir durch das Perspicillum gar deutlich ersehen worden.}
cited after Zinner (1942).} 
\item Malapert noticed in June and Oct 1620 that a spot appeared at large separation from the center of
the solar disk, i.e. at high heliographic latitude. He probably did not see such a high-latitude
spot for quite some time, so that he implicitly noticed the start of a new Schwabe cycle.
Since he mentioned explicitly that the high heliographic latitude is special, we can
conclude that he did not see additional (e.g. earlier) high-latitude spots
(he observed since 1618);
also, if there would have been more low-latitude spots around that time, too, 
he would have mentioned them (Sect. 7.2).
\end{itemize}

The HS98 and Wolf data bases with their large biblio\-graphies are of great value.
For further improvement of sunspot (group) numbers and our understanding of solar activity,
in particular for the time before and during the Maunder Minimum,
it is absolutely essential to check carefully the material from 
all observers in that time -- and also to take into account
lower limits, monthly or yearly averages, and trends mentioned by the observers.

\acknowledgements
In particular, we thank H. Gaab for sharing with us his knowledge about
sunspot observations by Marius.
We are grateful also to Klaus-Dieter Herbst, Pierre Leich, 
and Klaus Matth\"aus for valuable advise on the work of Marius.
We would especially like to acknowledge Pierre Leich, who has motivated us for this study
and who has organized a Marius conference in September 2014, where we could present
some first results. He also build up the Marius portal www.simon-marius.net.
We are grateful to the Simon-Marius high school in Gunzenhausen
for translating the full {\em Mundus Iovialis} from Latin to German
as part of a special school project with eight school students 
under the supervison of their teacher, Joachim Schl\"or (published by Schl\"or 1988).
We thank C.M. Graney for discussing the astronomical debates, in which Marius were involved,
in talks in N\"urnberg and Bamberg in 2014 on {\em Simon Marius -- an astronomer too good}.
We consulted the data base of Hoyt \& Schatten (1998) at
ftp://ftp.ngdc.noaa.gov/STP/space-weather/solar-data/solar-indices/sunspot-numbers/group.
We acknowledge the Moon phase predictions by
Rita Gautschy on www.gautschy.ch/$\sim$rita/archast/mond/Babylonerste.txt.
We thank Regina von Berlepsch (AIP Potsdam) for providing
the publication by Zinner (1952) about the work of Marius --
and Rainer Arlt (AIP Potsdam) for providing some works by Wolf and Tard\'e,
as well as other helpful advise.
We also thank T. Friedli for advise.
The sunspot drawing of Jungius (Figs. 4 \& 5) was obtained in 
digital form from the university library U Hamburg,
we thank Eike Harden (U Hamburg) for his help.
Fig. 6 from Galileo Galilei was taken from
the Galileo Project on galileo.rice.edu (copyright Albert Van Helden).
The drawings of Harriot was consulted on digilib.mpiwg-berlin.mpg.de.
Daniela Luge and Klaus-Dieter Herbst help reading and translating the hand-written Latin text of Jungius.
Daniela Luge (U Jena) supported us in the translation of Latin texts.
We also would like to thank D. Vanderbeke (U Jena) for pointing us to the Oxford English Dictionary on www.oed.com.
We acknowledge the Germanisches Nationalmuseum N\"urnberg for the digital copy of the drawing
by Saxonius (Figs. 7-9).
We thank Brian Mac Cuarta SJ for information about Guilielmus Wely
and Jarek Wlodarczk, the director of the Institute of History of Science of Polish
Academy of Sciences in Warsaw, for information about Simon Perovius.
We also acknowledge Jesse Chapman (U Stanford) for advise on the Chinese wording for aurorae and spots.
We received the drawings from the works by Tard\'e and Malapert 
in digital form from the Thuringian State and University Library Jena.
We would like to thank an anonymous referee for good suggestions 
and encouragement to extend this study to other early sunspot observers.

{}

\end{document}